\documentclass[traditabstract,times]{aa}

\newcommand{\ergps}{erg\thinspace s$^{-1}$}
\newcommand{\ergpspsqcm}{erg\thinspace s$^{-1}$\thinspace cm$^{-2}$}
\newcommand{\psqcm}{cm$^{-2}$}
\newcommand{\nH}{$N_{\rm H}$}

\usepackage{graphicx}
\begin{document}

\title{C-GOALS: Chandra observations of a complete sample of luminous
  infrared galaxies from the IRAS Revised Bright Galaxy Survey}

\author{K. Iwasawa\inst{1}\thanks{Email: kazushi.iwasawa@icc.ub.edu} 
\and 
D.B. Sanders \inst{2}
\and
Stacy H. Teng \inst{3}
\and
Vivian U \inst{2,4,5}
\and
L. Armus \inst{6}
\and
A.S. Evans \inst{7}
\and
J.H.~Howell \inst{6}
\and
S.~Komossa \inst{8}
\and
J.M.~Mazzarella \inst{9}
\and
A.O. Petric \inst{6}
\and
J.A. Surace \inst{6}
\and
T. Vavilkin \inst{10}
\and
S. Veilleux \inst{3}
\and
N. Trentham \inst{11}
}

\institute{ICREA and Institut de Ci\`encies del Cosmos (ICC), Universitat de Barcelona (IEEC-UB), Mart\'i i Franqu\`es, 1, 08028 Barcelona, Spain
\and
Institute for Astronomy, 2680 Woodlawn Drive, Honolulu, Hawaii 96822, USA
\and
Department of Astronomy, University of Maryland, College Park, MD 20742, USA
\and
NASA Jenkins Predoctoral Fellow
\and
SAO Predoctoral Fellow, Harvard-Smithsonian Center for Astrophysics, Cambridge, MA  02138, USA
\and
Spitzer Science Center, California Institute of Technology, Pasadena, CA 91125, USA
\and
Department of Astronomy, University of Virginia, 530 McCormick Road, Charlottesville, VA 22904 and NRAO, 520 Edgemont Road, Charlottesville, VA 22903, USA
\and
Max Planck Institut f\"ur extraterrestrische Physik, Gie\ss enbachstra\ss e, 85748 Garching, Germany
\and
IPAC, California Institute of Technology, Pasadena, CA 91125, USA
\and
Department of Physics and Astronomy, State University of New York at Stony Brook, NY 11794, USA
\and
Institute of Astronomy, Madingley Road, Cambridge CB3 0HA, United Kingdom
}

\date{Accepted on 25 February 2011}

\abstract{

  We present X-ray data for a complete sample of 44 luminous infrared
  galaxies (LIRGs), obtained with the Chandra X-ray Observatory. These
  are the X-ray observations of the high luminosity portion of the
  Great Observatory All-sky LIRG Survey (GOALS), which includes the most
  luminous infrared selected galaxies, log~$(L_{\rm ir}/L_\odot) \geq
  11.73$, in the local universe, $z\leq 0.088$. X-rays were detected
  from 43 out of 44 objects, and their arcsec-resolution images,
  spectra, and radial brightness distributions are presented. With a 
  selection by hard X-ray colour and the 6.4 keV iron line, AGN
  are found in 37\% of the objects, with higher luminosity sources
  more likely to contain an AGN. These AGN also tend to be found in
  late-stage mergers. The AGN fraction would increase to 48\% if
  objects with [Ne V]$\lambda 14.3\mu $m detection are
  included. Double AGN are clearly detected only in NGC 6240 among 24
  double/triple systems. Other AGN are found either in single nucleus objects 
  or in one of the double nuclei at similar rates. Objects without
  conventional X-ray signatures of AGN appear to be hard X-ray quiet,
  relative to the X-ray to far-IR correlation for starburst galaxies,
  as discussed elsewhere. Most objects also show extended soft X-ray
  emission, which is likely related to an outflow from the nuclear
  region, with a metal abundance pattern suggesting enrichment by core
  collapse supernovae, as expected for a starburst.  }

\keywords{Infrared: galaxies - X-rays: galaxies - Galaxies: active - Galaxies: starburst}

\titlerunning{C-GOALS survey}
\authorrunning{K. Iwasawa et al.}
\maketitle

\section{Introduction}

The Great Observatory All-sky LIRGs Survey (GOALS, Armus et al. 2009)
is a multi-wavelength study of the
most luminous infrared galaxies in the local Universe, selected from
the 60~$\mu$m flux limited IRAS Revised Bright Galaxy Sample (RBGS: 
Sanders et al. 2003). As the brightest far-infrared selected galaxies
in the sky, these objects are the most amenable for study at all
wavelengths.

Luminous Infrared Galaxies (LIRGs: $L_{\rm ir} > 10^{11} L_{\odot}$
\footnote{$L_{\rm ir} \equiv L(8{-}1000\mu$m)}) have proven to be an
extremely important class of extragalactic objects. In the local
Universe they are more numerous than optically selected starburst and
Seyfert galaxies and quasi-stellar objects (QSOs) at comparable
bolometric luminosity.  Strong interactions and mergers of gas-rich
spirals appear to be the trigger for the most luminous infrared
objects, which are fueled by a mixture of intense starbursts and AGN,
with the latter becoming more dominant with increasing $L_{\rm
  ir}$. At the highest luminosities, ultraluminous infrared galaxies
(ULIRGs: $L_{\rm ir} > 10^{12} L_{\odot}$), may represent an important
stage in the formation of QSOs and powerful radio galaxies, and they
may also represent a primary stage in the formation of massive
ellipticals, the formation of globular clusters, and the metal
enrichment of the intergalactic medium (see Sanders \& Mirabel 1996,
for a more complete review and, e.g., Chapman et al 2005, Hopkins et al
2006, Veilleux et al 2009 for recent developments).
  
The X-ray survey data presented here (C-GOALS) is the X-ray component
of the GOALS multi-wavelength survey. This initial C-GOALS paper
presents data obtained by us and others with the Chandra X-ray
Observatory (Chandra, hereafter). Data obtained by us in Cycles $7+8$
have been combined with data from the Chandra Archive to produce the
first complete X-ray survey of the most luminous sources in the GOALS
sample. Previous X-ray investigations of LIRGs, either by Chandra, XMM-Newton
or Suzaku, have been presented in Ptak et al. (2003), Franceschini et
al (2003), Teng et al. (2005, 2009), and Grimes et al. (2005).

The C-GOALS sample and observations are described in \S2 and \S3, 
respectively, with the major results (X-ray images, flux density spectra, radial surface
brightness profiles) presented in \S4. Derived properties from the
X-ray spectra, and a discussion of trends of X-ray properties with
infrared luminosity are presented in \S5.  A Summary of our results is 
given in \S6.  Notes for each object are presented in the Appendix.  
[Note: A consistent set of detailed images for all targets can be
found in the electronic edition of this paper.]  The cosmology adopted
here is consistent with that  adopted by Armus et al. (2009). Cosmological distances
were computed by first correcting for the 3-attractor flow model of
Mould et al. (2000) and adopting $H_0 = 70$ km s$^{-1}$Mpc$^{-1}$,
$\Omega_{\rm V} = 0.72$, and $\Omega_{\rm M} = 0.28$ based on the
5-year WMAP results (Hinshaw et al. 2009), as provided by the
NASA/IPAC Extragalactic Database (NED).

\section{The Sample}

% Fig 1  Lir disribution

\begin{figure}
\begin{center}
  \centerline{\includegraphics[width=0.35\textwidth,angle=0]{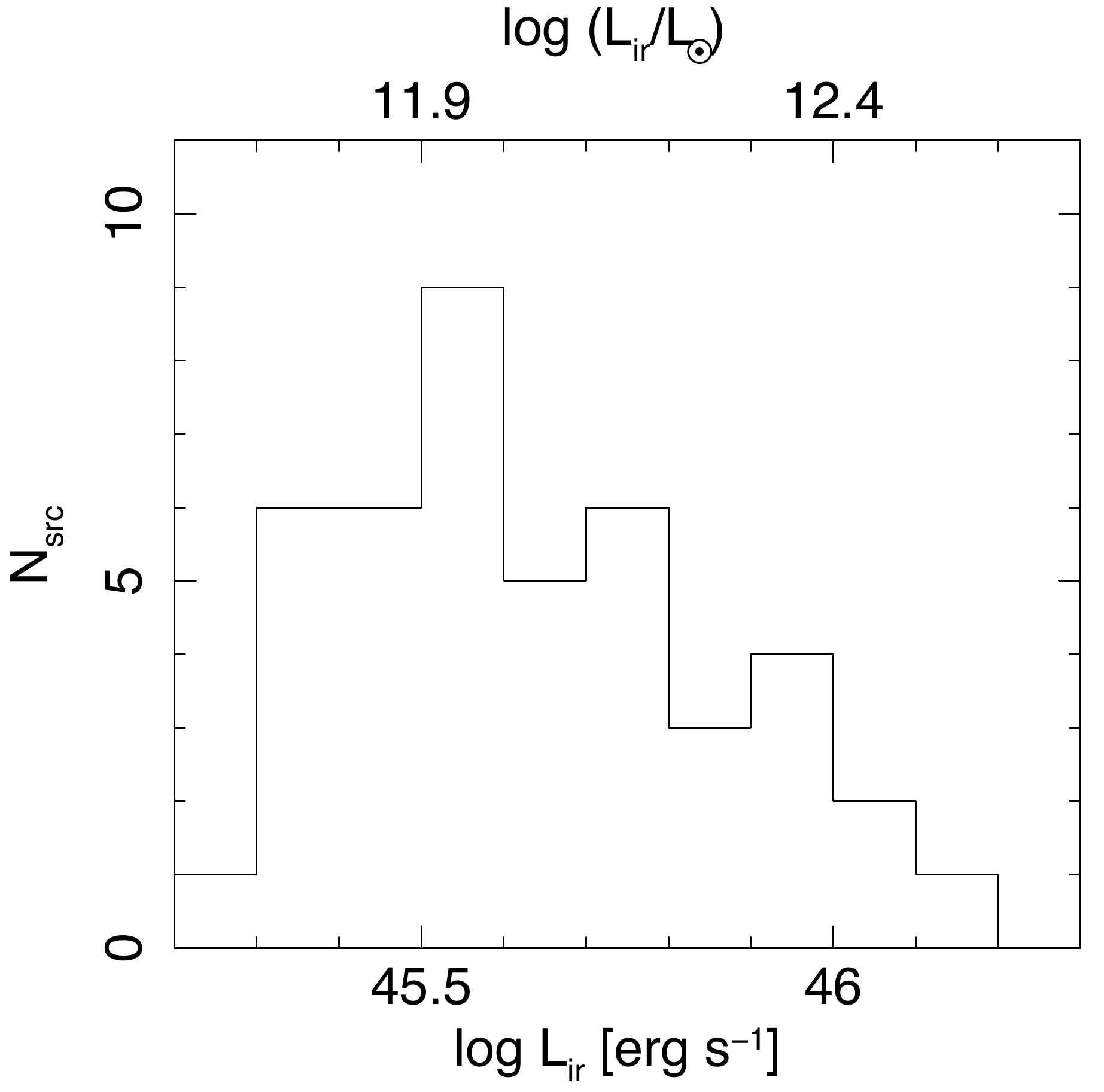}}
  \caption{The distribution of infrared luminosity, $L_{\rm ir}
    (8-1000 \mu$m) for the 44 objects in our C-GOALS sample with
    log~$(L_{\rm ir}/L_{\odot})=11.73-12.57$.  
    %%The median value is log~$(L_{\rm ir}/L_{\odot})=12.0$ corresponding 
    %% to log~$L_{\rm ir} = 45.6$ (erg s$^{-1}$). 
    For the infrared luminosity, we have chosen to use units of erg s$^{-1}$ (bottom axis) , 
    rather than the unit of $L_{\odot}$ (top axis)  normally used in infrared astronomy,  
    in order to be consistent with X-ray luminosity measurements.}
    %% Add log Lir (in solar units) to the top axis and make a statement as to why we are adopting 
    %% ergs/s rather than Lsun (i.e. to be consistent with X-ray measurements, etc. 
\end{center}
\end{figure}

The RBGS is a complete sample of 629 extragalactic objects with IRAS 
60$\mu$m flux density greater than 5.24 Jy, covering the full sky above 
Galactic latitude $\vert b\vert >5^{\circ }$. The GOALS sample contains 202 ``systems'' (77 of which
contain multiple galaxies), drawn from the RBGS with the luminosity
threshold $L_{\rm ir}\geq 10^{11}L_{\odot}$ (see Armus et al. 2009 for
further details).
The current C-GOALS sample represents the high luminosity part of the
GOALS sample, and is complete down to log~$(L_{\rm
  ir}/L_{\odot})=11.73$ .  The object with the largest IR luminosity, 
log~$(L_{\rm ir}/L_{\odot})= 12.57$,  is UGC 8058 (hereafter the more
common name, Mrk 231, is used in the text).  Table 1 gives basic data for all 44
galaxies in the C-GOALS sample.  The redshift range is $z = $0.010-0.088.  
The median IR luminosity of the sample log~$(L_{\rm ir}/L_{\odot})=11.99$ 
corresponding to log~$L_{\rm ir} = 45.58$ (erg s$^{-1}$.  The distribution of 
infrared luminosity is shown in Fig. 1.

\section{Observations and data reduction}

Twenty-six objects were observed with Chandra in Cycle-8 (PI:
D. Sanders) with a uniform 15~ks exposure on each target. All the
observations were carried out in imaging mode with the ACIS-S detector
operated in VFAINT mode. The Chandra data for the remaining 18 targets
were taken from the Archive, including four galaxies in the Cycle 7
mini survey of LIRGs with double nuclei (UGC 4881, VV 705,
F08572+3915, and F14348--1447) (PI: S Komossa). The exposure times for
these 4 objects are also 15~ks each. For other archival data, the
exposure times vary from 10~ks to 154~ks. The observation log for the
44 galaxies is shown in Table 2. Where multiple Chandra observations
are found, an imaging observation with the ACIS-S is chosen. For
Mrk~231, four ACIS-S observations exist in the archive, and we use all
four to construct and analyze the X-ray images, while only the first
observation was used for the analysis of the radial surface brightness
profiles and the X-ray colour, since the source is not variable and
the single observation provides good quality data for this purpose.
% above correction

In Table 2, the Galactic HI column density, estimated from the
Leiden/Argentine/Bonn (LAB) survey (Kalberla et al. 2005), is also
quoted.  Some of the Cycle-8 targets lie at low Galactic latitudes
%(the RBGS sources are selected at $\vert b\vert > 5^{\circ}$) 
and thus have been little studied because of high Galactic
extinction. The Galactic HI absorption column exceeds $10^{21}$ \psqcm\
for six objects.

The data reduction was performed using the Chandra data analysis
package CIAO version 3.4.1, and HEASARC's FTOOLS.

% Table 1 -- C-GOALS sample

\begin{table*}
\centering
\caption{The C-GOALS Sample.}
\begin{tabular}{rrlcccccl}
No. & IRAS Name & Optical ID & RA (NED) & Dec (NED) & $z$ & $D_{\rm L}$   & log~$(L_{\rm ir})$ & Other name\\
& {}              & {}               & (J2000)    & (J2000) & (km/s)    & (Mpc) & ($L_\odot$) & {}\\ 
{(1)}           & {(2)}            & {(3)}          & {(4)}       & {(5)}           & {(6)}       & {(7)} & {(8)} & {(9)}\\[5pt]
1 & F01364--1042 &	IRAS F01364--1042 &	01h38m52.92s&	-10d27m11.4s&	0.0483&	210.0&	11.85	& {}\\
2 & F04454--4838&	ESO 203-IG1&	04h46m49.50s&	-48d33m32.9s&	0.0529&	235.0&	11.86	& {}\\
3 & F05081+7936&	VII Zw 31&	05h16m46.44s&	+79d40m12.6s&	0.0537&	240.0&	11.99	&{} \\
4 & F05189--2524&	IRAS F05189--2524&	05h21m01.47s&	-25d21m45.4s&	0.0425&	187.0&	12.16	& {}\\
5 & F06259--4708&	ESO 255-IG7&	06h27m22.45s& 	-47d10m48.7s& 	0.0388& 	173.0& 	11.90	& {}\\
6 & 07251--0248&  	IRAS 07251--0248& 	07h27m37.55s& 	-02d54m54.1s& 	0.0875& 	400.0& 	12.39	& {}\\
7 & F08520--6850& 	ESO 60-IG16& 	08h52m31.29s& 	-69d01m57.0s& 	0.0463& 	210.0& 	11.82	& {}\\
8 & F08572+3915& 	IRAS F08572+3915& 	09h00m25.39s&	+39d03m54.4s&	0.0584&	264.0&	12.16	& {}\\
9 & 09022--3615 &	IRAS 09022--3615&	09h04m12.70s&	-36d27m01.1s&	0.0596&	271.0&	12.31	& {}\\
10 & F09111--1007&	IRAS F09111--1007&	09h13m37.61s&	-10d19m24.8s&	0.0541&	246.0&	12.06	& {}\\
11 & F09126+4432&	UGC 4881&	09h15m55.11s&	+44d19m54.1s&	0.0393&	178.0&	11.74 & Arp 55\\
12 & F09320+6134&	UGC 5101&	09h35m51.65s&	+61d21m11.3s&	0.0394&	177.0&	12.01	& {}\\
13 & F10038--3338&	ESO 374-IG 032$^{\mathrm{a}}$ &	10h06m04.80s&	-33d53m15.0s&	0.0341&	156.0&	11.78 &	\\
14 & F10173+0828&	IRAS F10173+0828&	10h20m00.21s&	+08d13m33.8s&	0.0491&	224.0&	11.86&	{}\\
15 & F10565+2448&	IRAS F10565+2448&	10h59m18.14s&	+24d32m34.3s&	0.0431&	197.0&	12.08	&{}\\
16 & F11257+5850&	NGC 3690&	11h28m32.25s&	+58d33m44.0s&	0.0104&	50.7&	11.93& Arp 299	\\
17 & F12112+0305&	IRAS F12112+0305&	12h13m46.00s&	+02d48m38.0s&	0.0733&	340.0&	12.36	& {}\\
18 & F12540+5708&	UGC 8058    &	12h56m14.23s&	+56d52m25.2s&	0.0422&	192.0&	12.57& Mrk 231	\\
19 & 13120--5453 &	IRAS 13120--5453&	13h15m06.35s&	-55d09m22.7s&	0.0308&	144.0&	12.32&	{}\\
20 & F13136+6223&	VV 250a&	13h15m35.06s&	+62d07m28.6s&	0.0311&	142.0&	11.81& Arp 238	\\
21 & F13182+3424&	UGC 8387&	13h20m35.34s&	+34d08m22.2s&	0.0233&	110.0&	11.73& IC 883, Arp 193 \\
22 & F13428+5608&	UGC 8696&	13h44m42.11s&	+55d53m12.7s&	0.0378&	173.0&	12.21& Mrk 273\\
23 & F14348--1447&	IRAS F14348--1447&	14h37m38.37s&	-15d00m22.8s&	0.0827&	387.0&	12.39 &	{}\\
24 & F14378--3651&	IRAS F14378--3651&	14h40m59.01s&	-37d04m32.0s&	0.0676&	315.0&	12.23&{}	\\
25 & F14547+2449&	VV 340a&	14h57m00.68s&	+24d37m02.7s&	0.0337&	157.0&	11.74& Arp 302\\
26 & F15163+4255&	VV 705&	15h18m06.28s&	+42d44m41.2s&	0.0402&	183.0&	11.92& I Zw 107	\\
27 & F15250+3608&	IRAS F15250+3608&	15h26m59.40s&	+35d58m37.5s&	0.0552&	254.0&	12.00	&{}\\
28 & F15327+2340&	UGC 9913&	15h34m57.12s&	+23d30m11.5s&	0.0182&	87.9&	12.28& Arp 220	\\
29 & F16330--6820&	ESO 69-IG6&	16h38m12.65s&	-68d26m42.6s&	0.0464&	212.0&	11.98&	{}\\
30 & F16504+0228&	NGC 6240&	16h52m58.89s&	+02d24m03.4s&	0.0245&	116.0&	11.93	& {}\\
31 & F17132+5313&	IRAS F17132+5313&	17h14m20.00s&	+53d10m30.0s&	0.0509&	232.0&	11.96	& {}\\
32 & F17207--0014&	IRAS F17207--0014&	17h23m21.96s&	-00d17m00.9s&	0.0428&	198.0&	12.46	&{}\\
33 & F18293--3413&	IRAS F18293--3413&	18h32m41.13s&	-34d11m27.5s&	0.0182&	86.0&	11.88&{}	\\
34 & F19115--2124&	ESO 593-IG8&	19h14m30.90s&	-21d19m07.0s&	0.0487&	222.0&	11.93&{}	\\
35 & F19297--0406&	IRAS F19297--0406&	19h32m21.25s&	-03d59m56.3s&	0.0857&	395.0&	12.45	&{}\\
36 & 19542+1110 &	IRAS 19542+1110&	19h56m35.44s&	+11d19m02.6s&	0.0650&	295.0&	12.12	&{}\\
37 & F20550+1655&	CGCG 448-020&	20h57m23.90s&	+17d07m39.0s&	0.0361&	161.0&	11.94& II Zw 96\\
38 & F20551--4250&	ESO 286-IG19&	20h58m26.79s&	-42d39m00.3s&	0.0430&	193.0&	12.06&	{}\\
39 & 21101+5810 &	IRAS 21101+5810&	21h11m30.40s&	+58d23m03.2s&	0.0390&	174.0&	11.81&	{}\\
40 & F22467--4906&	ESO 239-IG2&	22h49m39.87s&	-48d50m58.1s&	0.0430&	191.0&	11.84&{}	\\
41 & F22491--1808&	IRAS F22491--1808&	22h51m49.26s&	-17d52m23.5s&	0.0778&	351.0&	12.20&	{}\\
42 & F23128--5919&	ESO 148-IG2&	23h15m46.78s&	-59d03m15.6s&	0.0446&	199.0&	12.06&{}	\\
43 & F23180--6929&	ESO 77-IG14&	23h21m04.53s&	-69d12m54.2s&	0.0416&	186.0&	11.76&	{}\\
44 & F23365+3604&	IRAS F23365+3604&	23h39m01.27s&	+36d21m08.7s&	0.0645&	287.0&	12.20	&{}\\
\end{tabular}
\begin{list}{}{}
\item[$^{\mathrm{a}}$] When the IRAS Revised Bright Galaxy Sample (RBGS, Sanders et al. 2003)
was compiled, IRAS F10038--3338 was mistakenly cross-identified with the optical source 
IC 2545. The proper optical counterpart is ESO 374-IG 032. (See the Essential Notes in NED.)

%\item[$^{\mathrm{b}}$] When the RBGS was compiled, the best available heliocentric velocity for NGC 5010
%was $\rm 6400~km~s^{-1}$, adopted by NED from the RC3 (de Vaucouleurs et al. 1991). 
%The correct value of $V_{\rm Helio} = 2975 (\pm~27)~km~s^{-1}$ (Wegner et al. 2003, via NED),
%drops the luminosity below the definition of a LIRG. The source is listed here because it was
%included in our {\it Spitzer} imaging and spectroscopic observations.
\item[Column (1):] Through number of the object, which are also used in other tables and Fig. 2.
\item[Column (2):] Original IRAS source, where an ``F" prefix indicates the Faint Source Catalog and no prefix indicates the Point Source Catalog. 
\item[Column (3):] Optical cross-identification, where available from NED. 
%For many cases where the IRAS source corresponds to a pair of optically identified galaxies, we adopt the system name 
% instead of pair components. For example, IRAS F00163-1039 is identified in GOALS as Arp 256 rather than 
% ``MCG -02-01-051/2" as in Sanders et al. (2003).
\item[Column (4):] The best available source right ascension (J2000) in NED as of October 2008.
\item[Column (5):] The best available source declination (J2000) in NED as of October 2008.
\item[Column (6):] The best available heliocentric redshift, in NED as of October 2008.
\item[Column (7):] The luminosity distance in megaparsecs derived by correrecting the heliocentric velocity for the 3-attractor flow model
of Mould et al. (2000) and adopting cosmological parameters
$H_0 = 70$~km~s$^{-1}$~Mpc$^{-2}$, $\Omega_{\rm V} = 0.72$, and $\Omega_{\rm M} = 0.28$ based on the
5-year WMAP results (Hinshaw et al. 2009), as provided by NED.
\item[Column (8):] The total infrared luminosity in $\rm log_{10}$ Solar units computed using the flux densities reported 
in the RBGS and the lumiosity distances in column (7) using the formulae
$L_{\rm ir}/L_{\odot} = 4\pi (D_{\rm L [m]})^2~(F_{\rm ir}~[W~m^{-2}])/3.826\times 10^{26} [W m^{-2}]$,
where
$F_{\rm ir} = 1.8\times 10^{-14}\{13.48 f_{12\mu m}[Jy] + 5.16 f_{25\mu m}[Jy] + 2.58 f_{60\mu m}[Jy] + f_{100\mu m}[Jy]\} [W m^{-2}]$
(Sanders \& Mirabel 1996).
\item[Column (9):] Other conventionally used object name.
\end{list}
\end{table*}

% OBs log

\begin{table*}
\begin{center}
  \caption{Chandra observation log.$^{\mathrm{a}}$}
\begin{tabular}{rlcccccc}
No. & Galaxy & Obs ID & Date & Mode & Exp. Time & 0.4-7 keV$^{\mathrm{b}}$ & $N_{\rm H,Gal}$$^{\mathrm{c}}$ \\
& & & & (ks) & (cts) & ($10^{20}$ \psqcm) \\[5pt]
\multicolumn{7}{c}{\bf Cycle 8 targets} \\
1 & F01364--1042 & 7801 & 2007 Sep 10 & VFAINT & 14.57 & $46.0\pm 7.3$ & 2.0 \\
2 & ESO 203-IG1 & 7802 & 2008 Jan 17 & VFAINT & 14.85 & 0 ($<3$) & 1.4 \\
3 & VII Zw 31 & 7887 & 2007 May 27 & VFAINT & 14.98 & $173.8\pm 13.3$ & 7.4 \\
5 & ESO 255-IG7 & 7803 & 2007 May 27 & VFAINT & 14.57 & $341.8\pm 18.8$ & 3.8 \\
6 & 07251--0248 & 7804 & 2006 Dec 01 & VFAINT & 15.43 & $12.7\pm 3.6$ & 14.6 \\
%$221.7\pm 15.1$; $88.4\pm 9.5$; $31.7\pm 5.9$
7 & ESO 60-IG16 & 7888 & 2007 May 31 & VFAINT & 14.68 & $122.6\pm 11.1$ & 5.2 \\
9 & 09022--3615 & 7805 & 2007 Sep 04 & VFAINT & 14.85 & $265.3\pm 16.5$ & 26.6 \\
%$155.3\pm 127$ (N); $174.8\pm 13.2$ (S) 
%$53.9\pm 7.4$ (N); $30.2\pm 5.6$ (S) 
10 & F09111--1007 & 7806 & 2007 Mar 20 & VFAINT & 14.63 & $118.7\pm 11.1$ & 4.6 \\
13 & ESO 374-IG32 & 7807 & 2007 Mar 07 & VFAINT & 14.36 & $75.0\pm 9.1$ & 8.8 \\ 
%$92.2\pm 9.8$ (E); $26.5\pm 5.3$ (W) 
14 & F10173+0828 & 7808 & 2008 Jan 18 & VFAINT & 14.98 & $9.8\pm 3.2$ & 2.3 \\
19 & 13120--5453 & 7809 & 2006 Dec 01 & VFAINT & 14.67 & $300.7\pm 17.4$ & 21.3 \\
20 & VV 250 & 7810 & 2007 Aug 22 & VFAINT & 14.85 & $391.9\pm 20.1$ & 2.0 \\
21 & UGC 8387 & 7811 & 2007 Feb 19 & VFAINT & 14.07 & $251.4\pm 16.0$ & 1.0 \\
24 & F14378--3651 & 7889 & 2007 Jun 25 & VFAINT & 13.86 & $45.3\pm 6.8$ & 6.3 \\
25 & VV 340 & 7812 & 2006 Dec 17 & VFAINT & 14.86 & $331.4\pm 20.4$ & 3.3 \\
29 & ESO 69-IG6 & 7813 & 2007 Jun 21 & VFAINT & 14.54 & $330.1\pm 18.3$ & 9.1 \\
31 & F17132+5313 & 7814 & 2007 Apr 03 & VFAINT & 14.85 & $90.8\pm 10.0$ & 1.9 \\
33 & F18293-3413 & 7815 & 2007 Feb 25 & VFAINT & 14.04 & $444.6\pm 21.5$ & 9.7 \\
34 & ESO 593-IG8 & 7816 & 2007 Jun 09 & VFAINT & 14.97 & $158.1\pm 13.7$ & 8.1 \\
35 & F19297--0406 & 7980 & 2007 Jun 18 & VFAINT & 16.42 & $85.8\pm 9.7$ & 15.1 \\
36 & 19542+1110 & 7817 & 2007 Sep 10 & VFAINT & 14.98 & $324.3\pm 18.0$ & 14.0 \\
37 & CGCG 448-020 & 7818 & 2007 Sep 10 & VFAINT & 14.56 & $301.3\pm 18.2$ & 6.9 \\
39 & 21101+5810 & 7819 & 2007 Jul 01 & VFAINT & 14.85 & $21.7\pm 4.8$ & 37.2 \\
40 & ESO 239-IG2 & 7820 & 2007 Sep 10 & VFATIN & 14.57 & $151.5\pm 13.1$ & 0.9 \\
41 & F22491--1808 & 7821 & 2007 Jul 13 & VFAINT & 14.97 & $50.9\pm 7.3$ & 2.3 \\
43 & ESO 77-IG14 & 7822 & 2008 Jan 26 & VFAINT & 14.98 & $84.1\pm 9.3$ & 3.3 \\[5pt]
%$358.0\pm 19.1$ (E); $33.9\pm 6.1$ (W) 
%$284.8\pm 18.9$ (N); $46.6\pm 7.8$ (S) 
\multicolumn{7}{c}{\bf Archival data} \\
4 & F05189-2524 & 3432 & 2002 Jan 03 & FAINT & 14.86 & $2016.9\pm 45.0$ & 1.7 \\
8 & F08572+3915 & 6862 & 2006 Jan 26 & FAINT & 14.94 & $9.7\pm 3.2$ & 2.0 \\
11 & UGC 4881 & 6857 & 2006 Jan 12 & FAINT & 14.77 & $69.4\pm 8.4$ & 1.4 \\
12 & UGC 5101 & 2033 & 2001 May 28 & FAINT & 49.32 & $482.9\pm 22.3$ & 3.0 \\
15 & F10565+2448 & 4552 & 2003 Oct 23 & FAINT & 28.87 & $335.2\pm 18.8$ & 1.1 \\
16 & NGC 3690 & 6227 & 2005 Feb 14 & FAINT & 10.19 & $2526.0\pm 52.1$ & 0.9 \\
17 & F12112+0305 & 4110 & 2003 Apr 15 & VFAINT & 9.87 & $42.6\pm 6.6$ & 1.8 \\
18 & UGC 8058 & 1031$^{\mathrm{d}}$ & 2000 Oct 19 & FAINT & 39.25 & $2312.5\pm 66.2$ & 1.0 \\
%153.60 & $8642.4\pm 108.5$ & 1.0 \\
18 & UGC 8058 & 4028 & 2003 Feb 03 & VFAINT & 39.68 & $2205.6\pm 59.7$ & 1.0 \\
18 & UGC 8058 & 4029 & 2003 Feb 11 & VFAINT & 38.63 & $2070.5\pm 57.3$ & 1.0 \\
18 & UGC 8058 & 4030 & 2003 Feb 20 & VFAINT & 36.01 & $1876.6\pm 52.0$ & 1.0 \\
22 & UGC 8696 & 809 & 2000 Apr 19 & VFAINT & 44.19 & $2054.0\pm 45.8$ & 0.9 \\
23 & F14348-1447 & 6861 & 2006 Mar 12 & FAINT & 14.72 & $75.9\pm 8.8$ & 7.5 \\
26 & VV 705 & 6858 & 2006 Sep 11 & FAINT & 14.47 & $157.8\pm 12.6$ & 1.8 \\
27 & F15250+3608 & 4112 & 2003 Aug 27 & VFAINT & 9.84 & $26.6\pm 5.3$ & 1.5 \\
28 & UGC 9913 & 869 & 2000 Jun 24 & FAINT & 56.49 & $1555.1\pm 47.1$ & 3.9 \\
30 & NGC 6240 & 1590 & 2001 Jul 29 & FAINT & 36.69 & $10010.7\pm 103.5$ & 4.9 \\
32 & F17207-0014 & 2035 & 2001 Oct 24 & FAINT & 48.53 & $476.6\pm 23.1$ & 9.7 \\
38 & ESO 286-IG19 & 2036 & 2001 Oct 31 & FAINT & 44.87 & $767.8\pm 28.3$ & 3.3 \\
42 & ESO 148-IG2 & 2037 & 2001 Sep 30 & FAINT & 49.31 & $1052.2\pm 34.9$ & 1.6 \\
44 & F23365+3604 & 4115 & 2003 Feb 03 & VFAINT & 10.10 & $28.8\pm 5.4$ & 9.6 \\
\end{tabular}
\begin{list}{}{}
\item[$^{\mathrm{a}}$]  The 26 Cycle 8 Chandra observations (PI: D. Sanders) are listed first, followed
  by archival data for the additional 18 objects with log~($L_{\rm ir}/L_\odot) > 11.73$. All observations 
  were obtained in imaging mode with the ACIS-S, and the targets were placed at the
  nominal pointing position on the detector. 
  
 \item [$^{\mathrm{b}}$] The source counts are corrected for background and
  measured in the 0.4-7 keV band. The counts from separate components
  in a single objects are summed together. 
  
  \item [$^{\mathrm{c}}$] The Galactic absorption
  column density is taken from the LAB HI map
  by Kalberla et al. (2005). 

\item [$^{\mathrm{d}}$] This observation was used to make the radial surface-brightness profile.
\end{list}
\end{center}
\end{table*}

\section{Results}

% Table 3 --- Counts HR Fx Lx
\begin{table*}
\setlength{\tabcolsep}{0.03in}
\begin{center}
\caption{X-ray spectral properties for the C-GOALS sample} 
\begin{tabular}{rlccccccccc}
No. & Galaxy & $SX$ & $HX$ & {\sl HR} & $F_{\rm SX}$ & $F_{\rm HX}$ & $L_{\rm SX}$ & $L_{\rm HX}$ & {\sl SX/IR} & {\sl HX/IR} \\
& & (1) & (2) & (3) & (4) & (5) & (6) & (7) & (8) & (9) \\
\multicolumn{11}{c}{\bf Cycle 8 Data} \\
1 & F01364--1042 & $2.14\pm 0.41$ & $0.86\pm 0.30$ & $-0.43\pm 0.18$ & 4.5e-15 & 2.0e-14 & 2.4e40 & 1.5e41 & -5.05 & -4.26 \\
2 & ESO 203-IG1 & $0.00\pm 0.20$ & $0.00\pm 0.20$ & --- & $<$1.5e-15 & $<$3.0e-15 & $<$5.6e39 & $<$6.3e40 & $<$-5.69 &$<$-5.14 \\ 
3 & VII Zw 31 & $9.49 \pm 0.80$ & $2.01 \pm 0.38$ & $-0.65\pm 0.09$ & 3.0e-14 & 3.5e-14 & 2.0e41 & 2.7e41 & -4.27 & -4.14 \\
5 & ESO 255-IG7 N & $12.40 \pm 0.93$ & $2.77 \pm 0.46$ & $-0.64\pm 0.08$ & 3.5e-14 & 2.9e-14 & 1.2e41 & 1.6e41 & -4.41 & -4.28 \\
5 & ESO 255-IG7 C & $4.77 \pm 0.57$ & $1.35 \pm 0.32$ & $-0.56\pm 0.12$ & 1.5e-14 & 2.1e-14 & 5.3e40 & 1.2e41 & -4.76 & -4.41 \\
5 & ESO 255-IG7 S & $1.53 \pm 0.33$ & $0.63 \pm 0.23$ & $-0.42\pm 0.20$ & 4.5e-15 & 6.8e-15 & 1.5e40 & 3.8e40 & -5.31 & -4.90 \\
6 & 07251--0248 & $0.77 \pm 0.22$ & $0.05 \pm 0.06$ & $-0.87\pm 0.38$ & 3.3e-15 & $<$3e-15 & 6.2e40 & $< $5.8e40 & -5.18 & $<-5.21$ \\
7 & ESO 60-IG16 & $3.24 \pm 0.47$ & $5.09 \pm 0.59$ & $+0.22\pm 0.09$ & 8.1e-15 & 1.0e-13 & 5.5e40 & 7.3e41 & -4.66 & -3.54 \\
9 & 09022--3615 & $10.45 \pm 0.84$ & $7.38 \pm 0.72$ & $-0.17\pm 0.06$ & 2.9e-14 & 1.4e-13 & 4.4e41 & 2.0e42 & -4.25 & -3.59 \\
10 & F09111--1007E & $5.27\pm 0.60$ & $0.97 \pm 0.27$ & $-0.69\pm 0.13$ & 1.6e-14 & 1.2e-14 & 1.2e41 & 9.8e40 & -4.57 & -4.65 \\
10 & F09111--1007W & $1.20 \pm 0.29$ & $0.55 \pm 0.21$ & $-0.37\pm 0.22$ & 3.6e-15 & 4.5e-15 & 2.1e40 & 3.8e40 & -5.32 & -5.06 \\
13 & ESO 374-IG32 & $4.02 \pm 0.55$ & $1.22 \pm 0.32$ & $-0.62\pm 0.13$ & 1.3e-14 & 1.7e-14 & 3.8e40 & 6.3e40 & -4.78 & -4.56 \\
14 & F10173+0828 & $0.59 \pm 0.21$ & $0.06 \pm 0.06$ & $-0.82\pm 0.42$ & 2.1e-15 & 1.0e-15 & 1.1e40 & 6.6e39 & -5.40 & -5.62 \\
19 & 13120--5453 & $11.63 \pm 0.89$ & $9.00 \pm 0.79$ & $-0.13\pm 0.06$ & 3.7e-14 & 1.4e-13 & 1.1e41 & 4.5e41 & -4.86 & -4.25 \\
20 & VV 250 E & $16.90 \pm 1.08$ & $6.94 \pm 1.75$ & $-0.42\pm 0.06$ & 5.3e-14 & 9.5e-14 & 1.2e41 & 3.1e41 & -4.32 & -3.90 \\
20 & VV 250 W & $1.78 \pm 0.36$ & $0.54 \pm 0.22$ & $-0.53\pm 0.20 $ & 6.0e-15 & 1.0e-15 & 1.3e40 & 2.5e39 & -5.28 & -6.00 \\
21 & UGC 8387 & $13.77 \pm 0.99$ & $3.98 \pm 0.54$ & $-0.55\pm 0.07$ & 4.4e-14 & 4.3e-14 & 5.4e40 & 6.4e40 & -4.58 & -4.51 \\
24 & F14378--3651 & $1.92 \pm 0.38$ & $1.34 \pm 0.31$ & $-0.18\pm 0.15$ & 6.7e-15 & 2.1e-14 & 8.0e40 & 3.4e41 & -4.91 & -4.28 \\
25 & VV 340 N & $16.58 \pm 1.08$ & $2.55 \pm 0.55$ & $-0.74\pm 0.08 $ & 4.8e-14 & 7.3e-14 & 1.3e41 & 2.9e41 & -4.21 & -3.86 \\
25 & VV 340 S & $2.35 \pm 0.41$ & $0.92 \pm 0.33$ & $-0.44\pm 0.18 $ & 6.6e-15 & 3.6e-15 & 1.7e40 & 1.2e40 & -5.09 & -5.25 \\
29 & ESO 69-IG6 (N) & $8.88 \pm 0.79$ & $1.78 \pm 0.37$ & $-0.67\pm 0.10$ & 2.8e-14 & 1.8e-14 & 1.4e41 & 1.0e41 & -4.42 & -4.56 \\
29 & ESO 69-IG6 (S) & $2.19 \pm 0.39$ & $9.97 \pm 0.83$ & $+0.64\pm 0.09$ & 8.1e-15 & 1.7e-13 & 3.5e40 & 1.2e42 & -5.02 & -3.49 \\
31 & F17132+5313 & $5.13 \pm 0.60$ & $0.76 \pm 0.33$ & $-0.74\pm 0.14$ & 1.3e-14 & 1.1e-14 & 8.3e40 & 8.6e40 & -4.63 & -4.61 \\
33 & F18293--3413 & $22.92 \pm 1.29$ &$8.66 \pm 0.82$ & $-0.45\pm 0.05$ & 6.9e-14 & 1.3e-13 & 5.7e40 & 9.4e40 & -4.71 & -4.49 \\
34 & ESO 593-IG8 & $9.01 \pm 0.81$ & $1.65 \pm 0.44$ & $-0.69\pm 0.11$ & 2.1e-14 & 2.5e-14 & 1.2e41 & 1.9e41 & -4.44 & -4.24 \\
35 & F19297--0406 & $3.85 \pm 0.49$ &$1.21 \pm 0.30$ & $-0.52\pm 0.13$ & 1.2e-14 & 8.2e-15 & 3.0e41 & 1.8e41 & -4.56 & -4.78 \\
36 & 19542+1110 & $4.80 \pm 0.57$ & $17.12 \pm 1.07$ & $+0.56\pm 0.06$ & 1.8e-14 & 2.9e-13 & 2.2e41 & 4.1e42 & -4.36 & -3.09 \\
37 & CGCG 448-020 & $15.91 \pm 1.07$ & $4.41 \pm 0.64$ & $-0.57\pm 0.07 $ & 4.9e-14 & 6.7e-14 & 1.4e41 & 2.8e41 & -4.38 & -4.08 \\
39 & 21101+5810 & $1.17 \pm 0.29$ & $0.33 \pm 0.17$ & $-0.56\pm 0.25$ & 3.5e-15 & 4.0e-15 & 1.9e40 & 2.0e40 & -5.12 & -5.09 \\
40 & ESO 239-IG2 & $9.14 \pm 0.83$ & $1.31 \pm 0.36$ & $-0.75\pm 0.11$ & 3.0e-14 & 1.9e-14 & 1.4e41 & 9.7e40 & -4.28 & -4.44 \\
41 & F22491--1808 & $3.19 \pm 0.47$ & $0.14 \pm 0.14$ & $-0.91\pm 0.20$ & 8.3e-15 & 3.0e-15 & 1.3e41 & 0.6e41 & -4.67 & -5.01 \\
43 & ESO 77-IG14 N & $2.65 \pm 0.42$ & $0.94 \pm 0.26$ & $-0.47\pm 0.15$ & 7.8e-15 & 1.7e-14 & 9.6e40 & 6.8e40 & -4.36 & -4.51 \\
43 & ESO 77-IG14 S & $1.85 \pm 0.35$ & $0.22 \pm 0.13$ & $-0.78\pm 0.23$ & 6.0e-15 & 6.1e-15 & 2.4e40 & 2.5e40 & -4.96 & -4.95 \\[5pt]
\multicolumn{11}{c}{\bf Archival Data} \\
4 & F05189-2524 & $27.15 \pm 1.35$ & $107.00 \pm 2.69$ & $+0.60\pm 0.03$ & 8.2e-14 & 2.0e-12 & 3.4e41 & 1.3e43 & -4.21 & -2.63 \\
8 & F08572+3915 & $0.12\pm 0.09$ & $0.45\pm 0.18$ & $+0.57\pm 0.41$ & 9.0e-15 & 2.5e-14 & 8.0e40 & 2.0e41 & -4.83 & -4.44 \\
11 & UGC 4881 E & $2.09 \pm 0.38$ & $0.32 \pm 0.15$ & $-0.73\pm 0.21$ & 6.7e-15 & 7.8e-15 & 2.5e40 & 4.1e40 & -4.82 & -4.61 \\
11 & UGC 4881 W & $2.13 \pm 0.38$ & $0.13 \pm 0.12$ & $-0.88\pm 0.24$ & 7.1e-15 & 3.0e-15 & 2.7e40 & 1.3e40 & -4.19 & -4.51 \\
12 & UGC 5101 & $6.34 \pm 0.36$ & $3.49 \pm 0.27$ & $-0.29\pm 0.05$ & 1.9e-14 & 9.1e-14 & 7.0e40 & 4.7e41 & -4.74 & -3.92 \\
15 & F10565+2448 & $9.37 \pm 0.58$ & $2.21 \pm 0.30$ & $-0.62\pm 0.07$ & 2.8e-14 & 2.8e-14 & 1.2e41 & 1.6e41 & -4.58 & -4.46 \\
16 & NGC 3690 T & $204.84 \pm 4.64$ & $41.40 \pm 2.17$ & $-0.66\pm 0.03$ & 6.2e-13 & 6.8e-13 & 1.9e41 & 2.6e41 & -4.23  & -4.10 \\
16 & NGC 3690 E & $54.46 \pm 2.33$ & $15.43 \pm 1.24$ & $-0.56\pm 0.04$ & 1.5e-13 & 2.6e-13 & 7.7e40 & 9.6e40 & -4.49 & -4.40 \\
16 & NGC 3690 W & $58.71 \pm 2.42$ & $15.04 \pm 1.23$ & $-0.59\pm 0.04$ & 1.6e-13 & 2.6e-13 & 8.0e40 & 9.6e40 & -4.02 & -3.94 \\
17 & F12112+0305 & $3.22 \pm 0.57$ & $1.09 \pm 0.34$ & $-0.49\pm 0.17$ & 9.8e-15 & 1.9e-14 & 1.3e41 & 4.0e41 & -4.83 & -4.34 \\
18 & UGC 8058 & $34.23 \pm 1.04$ & $19.65 \pm 0.74$ & $-0.27\pm 0.03$ & 9.5e-14 & 4.0e-13 & 3.9e41 & 3.0e42 & -4.56 & -3.67 \\
22 & UGC 8696 core & $29.02 \pm 0.82$ & $16.20 \pm 0.62$ & $-0.28\pm 0.02$ & 7.4e-14 & 4.2e-13 & 2.5e41 & 2.5e42 & -4.39 & -3.39 \\
22 & UGC 8696 tail & $7.09 \pm 0.50$ & $0.00 \pm 0.31$ & $-1.00\pm 0.12$ & 1.7e-14 & 2.5e-15 & 5.6e40 & 1.1e40 & --- & --- \\
23 & F14348-1447 T & $3.80 \pm 0.51$ & $1.32 \pm 0.31$ & $-0.48\pm 0.13$ & 1.3e-14 & 2.4e-14 & 2.9e41 & 7.5e41 & -4.51 & -4.09 \\
23 & F14348-1447 S & $1.15 \pm 0.28$ & $0.80 \pm 0.24$ & $-0.18\pm 0.19$ & 4.1e-15 & 1.4e-14 & 9.0e40 & 4.5e41 & -4.90 & -4.20 \\
26 & VV 705 N & $8.34 \pm 0.76$ & $1.39 \pm 0.32$ & $-0.71\pm 0.10$ & 2.6e-14 & 1.6e-14 & 9.7e40 & 6.8e40 & -4.46 & -4.62 \\
26 & VV 705 S & $1.03 \pm 0.27$ & $0.06 \pm 0.07$ & $-0.90\pm 0.34$ & 3.5e-15 & 2.1e-15 & 1.3e40 & 8.6e39 & --- & --- \\
27 & F15250+3608 & $2.78 \pm 0.54$ & $0.00 \pm 0.18$ & $-1.0\pm 0.30$ & 8.7e-15 & $<$4.2e-15 & 6.2e40 & $<$4.5e40 & -4.87 & $<-$5.11 \\
28 & UGC 9913 & $21.90 \pm 0.70$ & $4.68 \pm 0.46$ & $-0.65\pm 0.04$ & 2.7e-14 & 6.6e-14 & 2.0e40 & 6.8e40 & -5.56 & -5.03 \\
30 & NGC 6240 & $209.55 \pm 2.45$ & $59.57 \pm 1.34$ & $-0.56\pm 0.01$ & 6.1e-13 & 1.2e-12 & 8.1e41 & 2.1e42 & -3.19 & -2.97 \\
32 & F17207-0014 & $7.13 \pm 0.40$ & $2.70 \pm 0.26$ & $-0.45\pm 0.05$ & 1.9e014 & 3.9e-14 & 9.9e40 & 2.2e41 & -5.04 & -4.70 \\
38 & ESO 286-IG19 & $14.83 \pm 0.58$ & $1.98 \pm 0.24$ & $-0.77\pm 0.05$ & 4.3e-14 & 3.3e-14 & 2.1e41 & 2.1e41 & -4.32 & -4.32 \\
42 & ESO 148-IG2 Tot & $14.46 \pm 0.56$ & $6.56 \pm 0.43$ & $-0.38\pm 0.04$ & 4.0e-14 & 1.1e-14 & 1.9e41 & 8.3e41 & -4.36 & -3.72 \\
42 & ESO 148-IG2 N &$2.78 \pm 0.24$ & $0.41 \pm 0.11$ & $-0.74\pm 0.10$ & 7.6e-15 & 4.2e-15 &3.7e40 & 2.4e40 & -4.47 & -4.66 \\
42 & ESO 148-IG2 S & $4.44 \pm 0.30$ & $5.48 \pm 0.34$ & $+0.11\pm 0.05$ & 1.3e-14 & 9.0e-14 & 5.7e40 & 6.6e41 & -4.75 & -3.69\\
44 & F23365+3604 & $1.67 \pm 0.41$ & $1.07 \pm 0.33$ & $-0.22\pm 0.20$ & 5.7e-15 & 1.1e-14 & 5.3e40 & 1.6e41 & -5.06 & -4.58 \\
\end{tabular}
\begin{list}{}{}
\item[Column (1):] Background corrected count rate in the
  0.5-2 keV band in unit of $10^{-3}$ ct s$^{-1}$
\item[Column (2):] Background
  corrected count rate in the 2-8 keV band in unit of $10^{-3}$ ct
  s$^{-1}$.
\item[Column (3):] X-ray colour, as defined by $HR = (H-S)/(H+S)$.
\item[Column (4):] Observed 0.5-2 keV band flux (\ergpspsqcm).
\item[Column (5):] Observed 2-7 keV band flux (\ergpspsqcm).
\item[Column (6):] 0.5-2 keV luminosity corrected for Galactic absorption (\ergps).
\item[Column (7):] 2-10 keV luminosity corrected for Galactic absorption (\ergps).
\item[Column (8):] Logarithmic luminosity ratio of the 0.5-2 keV and 8-1000 $\mu$m bands;
\item[Column (9):] Logarithmic luminosity ratio of the 2-10 keV and 8-1000 $\mu$m bands.
\end{list}
\end{center}
\end{table*}

Basic X-ray properties obtained from the Chandra data are presented in
Table 3. For each object, the ACIS-S count rates in the soft (0.5-2
keV) and hard (2-7 keV) bands, the X-ray colour, estimated X-ray
fluxes, X-ray luminosities in the two bands, and their logarithmic
ratio to the infrared luminosity, $L_{\rm ir} (8-1000\mu$m), are
listed. The count rates are computed for spatially distinctive
components within a single object when they are clearly resolved, as
opposed to the source counts in Table 2, which were collected from the
whole X-ray emitting region per object. Similarly, IR luminosity is
divided into respective components, estimated from IRAS HIRES processing
(Surace et al. 2004) and new estimates based on Spitzer MIPS
photometry (see Howell et al. 2010). The X-ray colour is defined as ${\sl HR} =
(H-S)/(H+S)$, often referred as the ``Hardness Ratio'', where $H$ and
$S$ are background-corrected counts in the 2-8 keV and 0.5-2 keV
bands, respectively.

We use the energy ranges, particularly for the hard X-ray band, that
are slightly different from one another, but optimized for different
purposes, which can sometimes be confusing. Here, we summarize the
choice of the energy ranges used in this paper. 1) 0.4-7 keV: This is
effectively the full ACIS-S bandpass, and used only for observed
counts; 2) 0.5-2 keV: The soft X-ray band is always this range; 3) 2-8
keV: The choice of this band is to match the conventional calculation
of hardness ratio, {\sl HR}, and used only for the hard band counts in
Table 3. The quoted counts are as observed, although the majority of
our sources have negligible counts above 7 keV where the background
would simply increases; 4) 2-7 keV: This is the nominal range for our
observed hard band flux, estimated from spectral fitting, and
optimized for the signal to noise ratio of the data for the majority
of our sources for the reason mentioned above; and 5) 2-10 keV: For
the hard band luminosity, we opted to use this extended range for a
purpose of comparison with results from other X-ray observatories,
since the 2-10 keV is the standard band for missions like XMM-Newton.

The images, spectra, colour and the selection of AGN based
on the colour, the correlation of X-ray and infrared luminosities and the 
surface brightness distribution of individual objects are presented in
the subsections below.

\subsection{Images}

% Fig. 2 -- CXO HST overlay

\begin{figure*}
\centerline{\includegraphics[width=0.95\textwidth,angle=0]{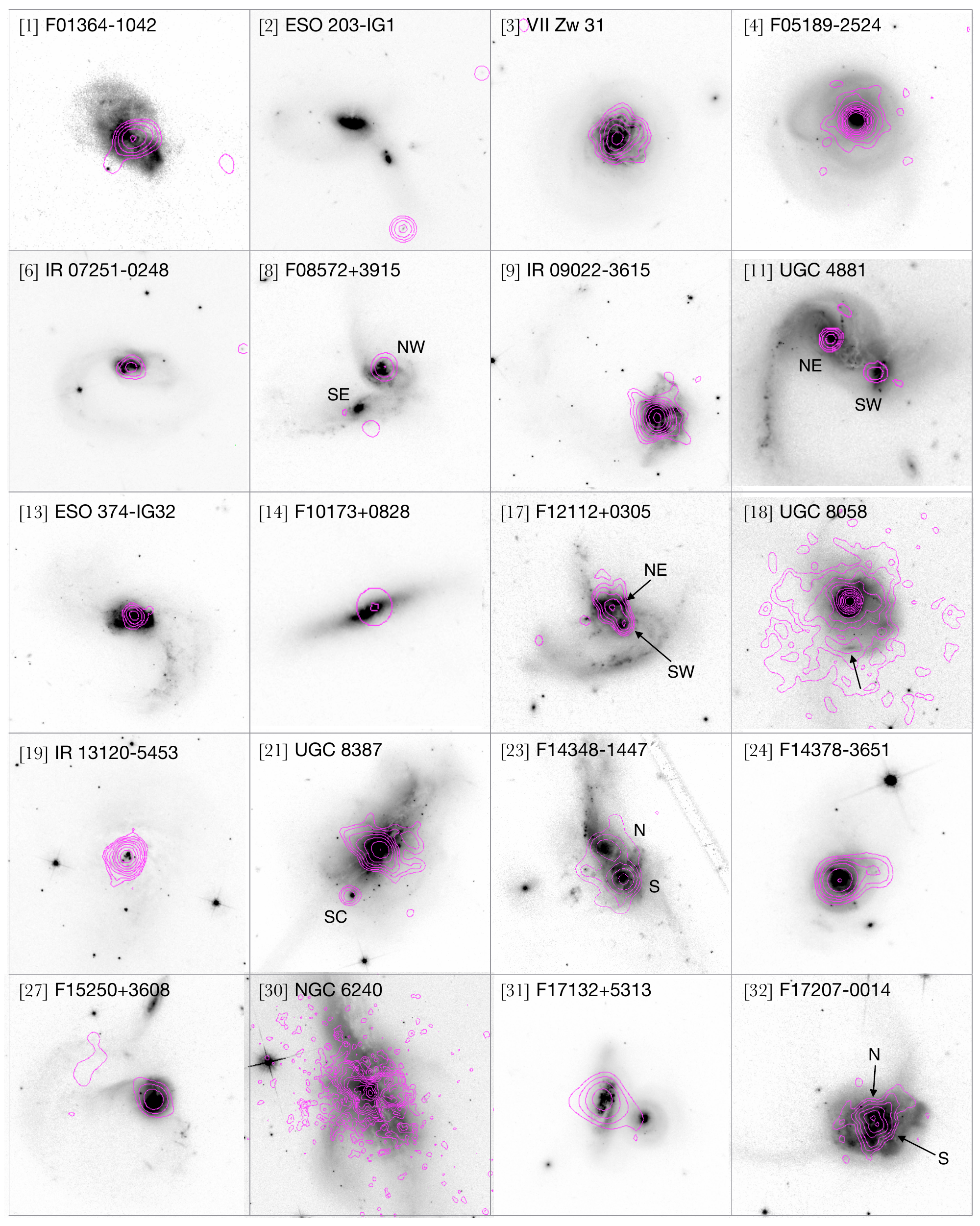}}
\end{figure*}

\begin{figure*}
\centerline{\includegraphics[width=0.95\textwidth,angle=0]{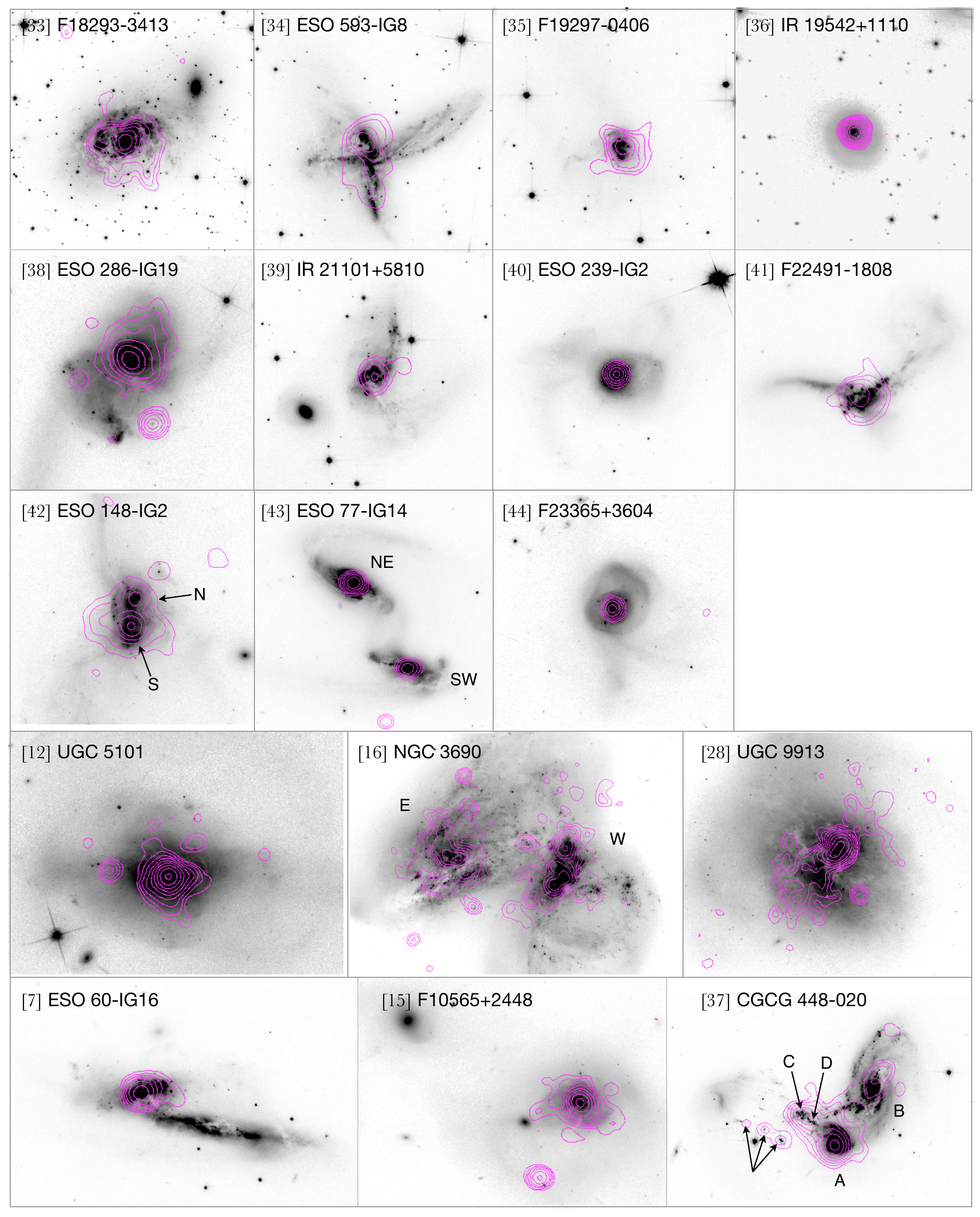}}
\end{figure*}

\begin{figure*}
\centerline{\includegraphics[width=0.95\textwidth,angle=0]{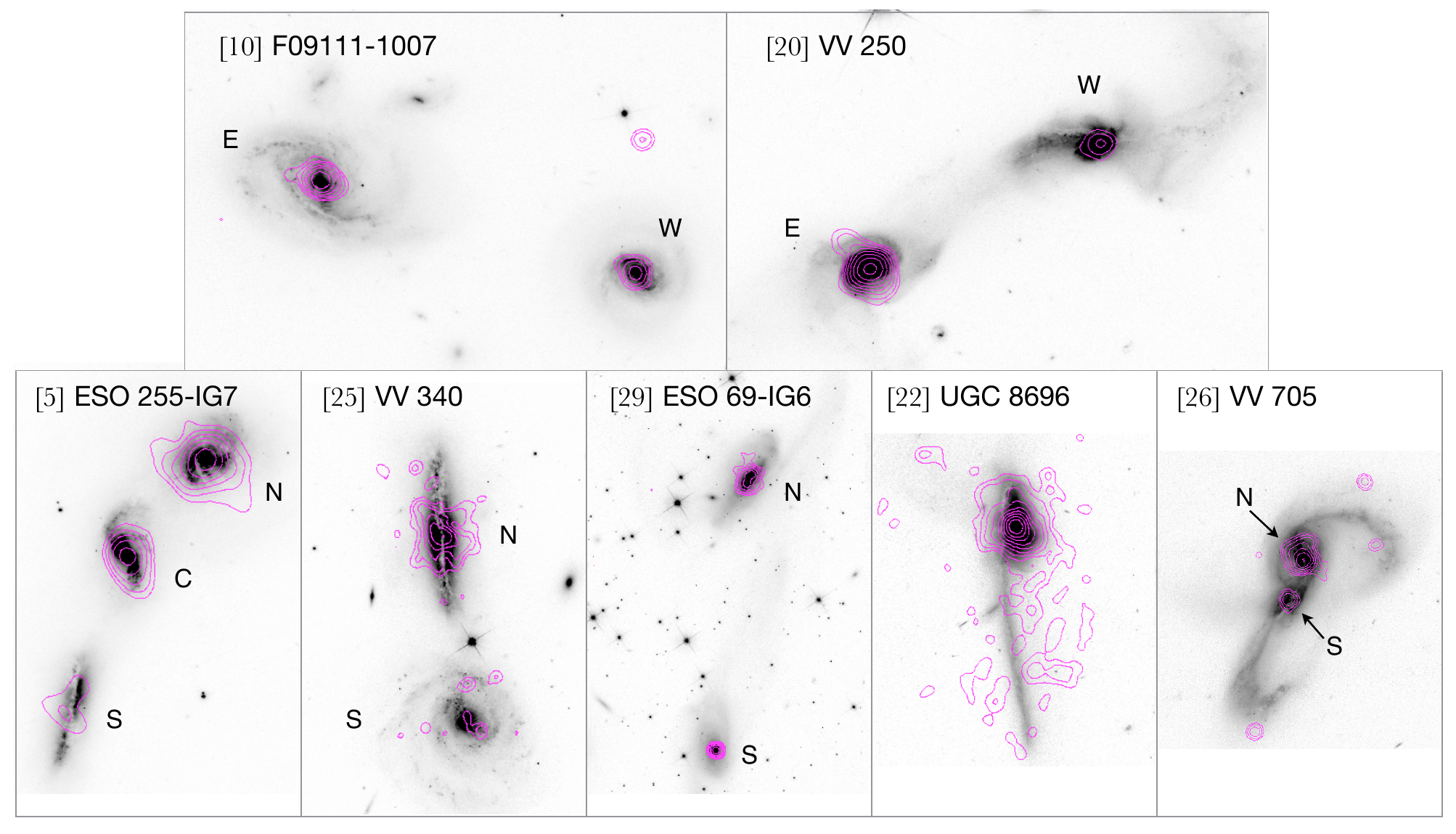}}
\caption{ X-ray and optical images of the 44 C-GOALS sample
  galaxies. X-ray (0.4-7 keV) brightness contours (magenta) are overlaid on the
  HST-ACS F814W image (grey scale) for each object. The orientation of
  all images is north up and east to the left. Eleven contour levels
  are defined in the fixed surface brightness range $4\times
  10^{-5} - 7\times 10^{-3}$ [counts s$^{-1}$
  arcsec$^{-2}$], which is divided into ten equal logarithmic
  intervals. For majority of objects, these contour levels were
  applied. For nine objects (F10173+0828, F17132+5313, F19297--0406,
  CGCG 448-020, F01364--1042, IRAS 21101+5810, F10565+2448, UGC 9913,
  and UGC 5101), a further lower interval was added to outline lower
  surface brightness features. For four bright objects, (UGC 8696, NGC
  3690, F05189--2524 and UGC 8058), custom contour levels were made for
  describing their X-ray morphology -- these are listed in Table 4.}
\end{figure*}

Figure 2 shows the X-ray brightness contours overlaid on the HST-ACS
F814W (I-band) image (Evans et al.,  in prep) for all 44 targets. The contours are made
from a 0.4-7 keV image, smoothed using a circular Gaussian filter
with typical dispersion of 1 arcsec. 
Note that the deep image of Mrk 231 reaches a significantly lower
brightness level, and thus the contour levels are drawn down to a
level which is a factor of 10 lower than for the other objects.

% Table 4 for objects with extra contours 

\begin{table}
\begin{center}
\caption{X-ray contours for bright objects}
\begin{tabular}{rlccc}
No. & Galaxy & Low & High & n \\
& & (1) & (2) & (3) \\[5pt]
4 & F05189--2524 & $3.0\times 10^{-5}$ & $4.2\times 10^{-3}$ & 10 \\
16 & NGC 3690 & $2.7\times 10^{-5}$ & $1.8\times 10^{-3}$ & 10 \\
18 & UGC 8058 & $8.5\times 10^{-7}$ & $3.7\times 10^{-4}$ & 11 \\ 
22 & UGC 8696 & $2.3\times 10^{-6}$ & $1.8\times 10^{-3}$ & 9 \\
\end{tabular}
\begin{list}{}{}
\item[Column (1)]The lowest contour (cts s$^{-1}$ arcsec$^{-2}$) in Fig. 2.
\item[Column (2)]The highest contour (cts s$^{-1}$ arcsec$^{-2}$) in Fig. 2.
\item[Coliumn (3)]The number of contour levels in Fig. 2.
\end{list}
\end{center}
\end{table}

% 6 panel figure ex IC883

The appearance of the X-ray images often differs dramatically between the
soft and hard bands. In addition to the full band (0.4-7 keV) image in
unsmoothed ($0.5^{\prime\prime}\times 0.5^{\prime\prime}$ pixel size)
and smoothed forms, images of the same region of the sky in the soft
(0.5-2 keV) and hard (2-7 keV) bands are shown. The same smoothing has
been applied to the 0.5-2 keV and 2-7 keV images. Linear scaling
is normally used for presentation, but for a few sources with 
strongly peaked emission, a logarithmic scale was used to show faint
extended features.  The scale bar represents an angular scale of 5
arcsec.  The soft (0.5-2 keV) and hard
(2-7 keV) band radial profiles, described in the following section,
are also shown.   An example for these figures is shown in Fig. 3 for UGC 8387.  
Similar style multi-panel figures were made for all 44 C-GOALS targets and 
these can be found in the online material. 

\begin{figure*}
\centerline{\includegraphics[width=0.8\textwidth,angle=0]{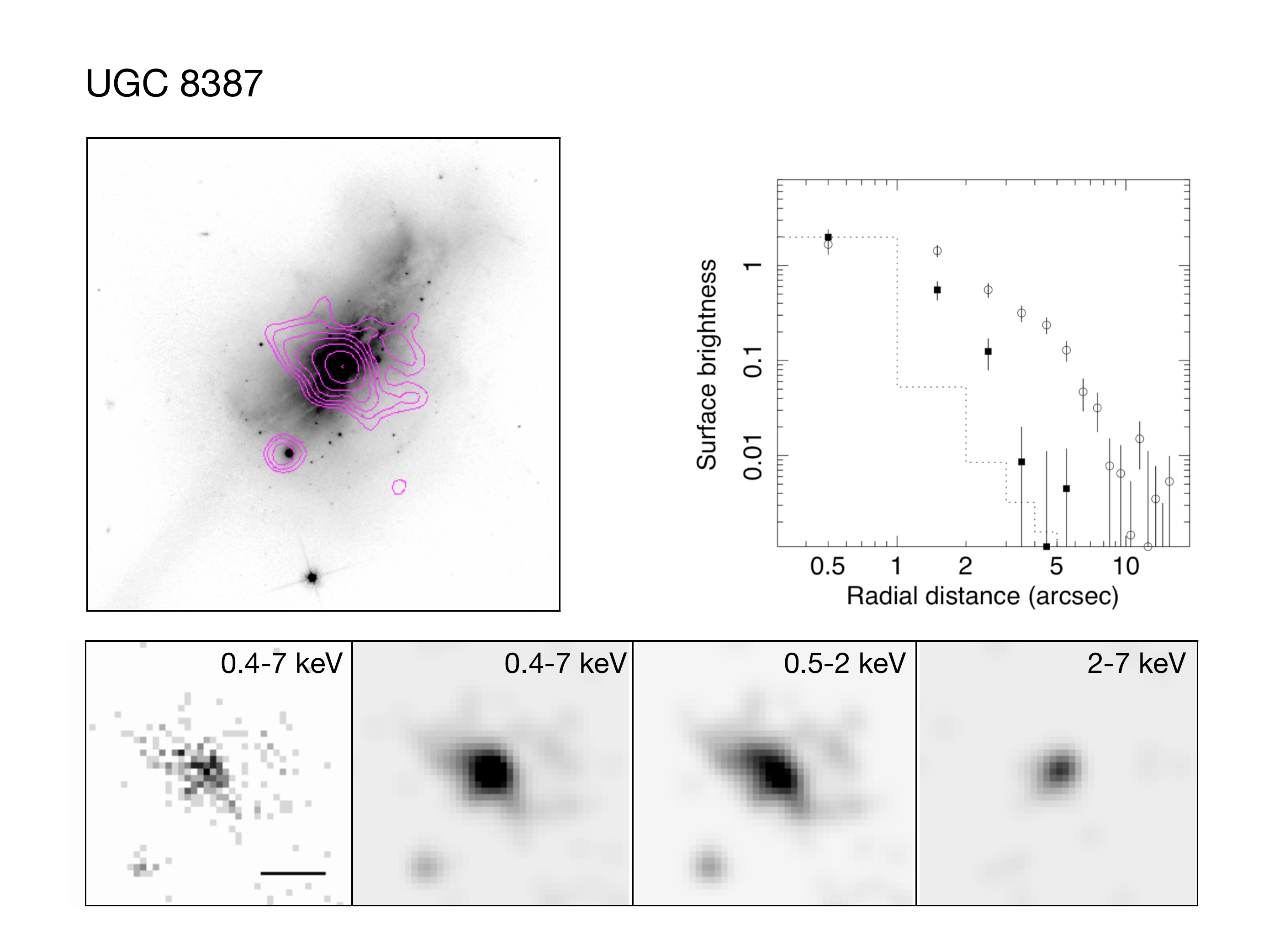}}
\caption{An example of X-ray images and the surface brightness
  distributions for individual objects is presented. The object shown
  here is UGC 8387 (= IC 883).  Similar multi-panel figures for the
  other 43 objects in our C-GOALS sample can be found in Appendix.
  {\it (Upper left)}: The X-ray (0.4-7 keV) brightness contours
  overlaid on the HST ACS I-band image.  {\it (Upper right)}: The
  radial surface brightness profiles in the 0.5-2 keV (open circles)
  and 2-7 keV (filled squares) bands are shown. These profiles are the
  azimuthal average and are measured from the peak of 2-7 keV emission
  at 1 arcsec resolution. The Point Spread Function (PSF) computed at
  3 keV is shown as a dotted histogram where the peak has been matched
  to the observed 2-7 keV peak for comparison. The bottom four-panel
  figure shows (left to right) the raw 0.4-7 keV image, the gaussian
  smoothed version of the same image, the 0.5-2 keV smoothed image,
  and the 2-7 keV smoothed image. The pixel size is $\approx
  0.5^{\prime\prime} \times 0.5^{\prime\prime}$. The scale bar
  indicates 5 arcsec.}
\end{figure*}

As seen clearly in UGC 8387, soft X-ray nebulae extending along the
minor axis of the galaxy, or in the direction displaced from the
optical light distribution, are often observed. These extended SX
nebulae are likely associated with a galactic-scale outflow from the
nuclear region, driven by either a starburst or AGN. Other examples
are F18293--3413, VV 340 N, UGC 9913 (hereafter Arp 220), NGC 6240 and
ESO 148-IG2.

\subsection{Flux density spectra}

X-ray spectra are traditionally shown as count rate spectra, i.e.,
data folded through the detector response. However, in this paper, in
the interest of comparing the X-ray spectra with other
multi-wavelength datasets in GOALS, the ACIS spectra have been
corrected for the detector response curve and further converted into
flux density units (Fig. 4). This correction may introduce some
uncertainty, particularly when a spectral bin is wide within which the
response varies rapidly, e.g., at the high energy end of the
bandpass. In spite of this caveat, we think that this presentation has
merit given that the spectral properties can be directly observed
without resorting to spectral fitting.  For this purpose, the flux
density range for all spectra was also kept to be 2 orders of
magnitude except where the dynamic range of the spectra exceeds 2
decades. The flux density is in units of [$10^{-14}$ erg cm$^{-2}$
s$^{-1}$ keV$^{-1}$], and can readily be converted into units of [W
m$^{-2}$ Hz$^{-1}$] by multiplying by $4.17\times 10^{-35}$. 

For some objects, data of spatially separate components are shown in Fig. 4, and when we
consider it appropriate, data for the total emission are shown to help
facilitate comparison with spectra taken from other X-ray
observatories at lower spatial resolution.  Finally, we note that
these are not ``unfolded spectra'', which are sometimes presented in
the literature. Instead, they have been corrected solely for the
detector effective area while preserving the energy resolution of the
detector, and thus are independent of any spectral model which might
be fitted. [Note: The spectra shown in Fig. 4 are for display purposes only;  all physical
quantities reported in this paper were obtained through conventional
spectral fitting to the count rate spectra with appropriate detector
responses.]

% Fig. 4 -- Fden spectra 

\begin{figure*}
\centerline{\includegraphics[width=\textwidth,angle=0]{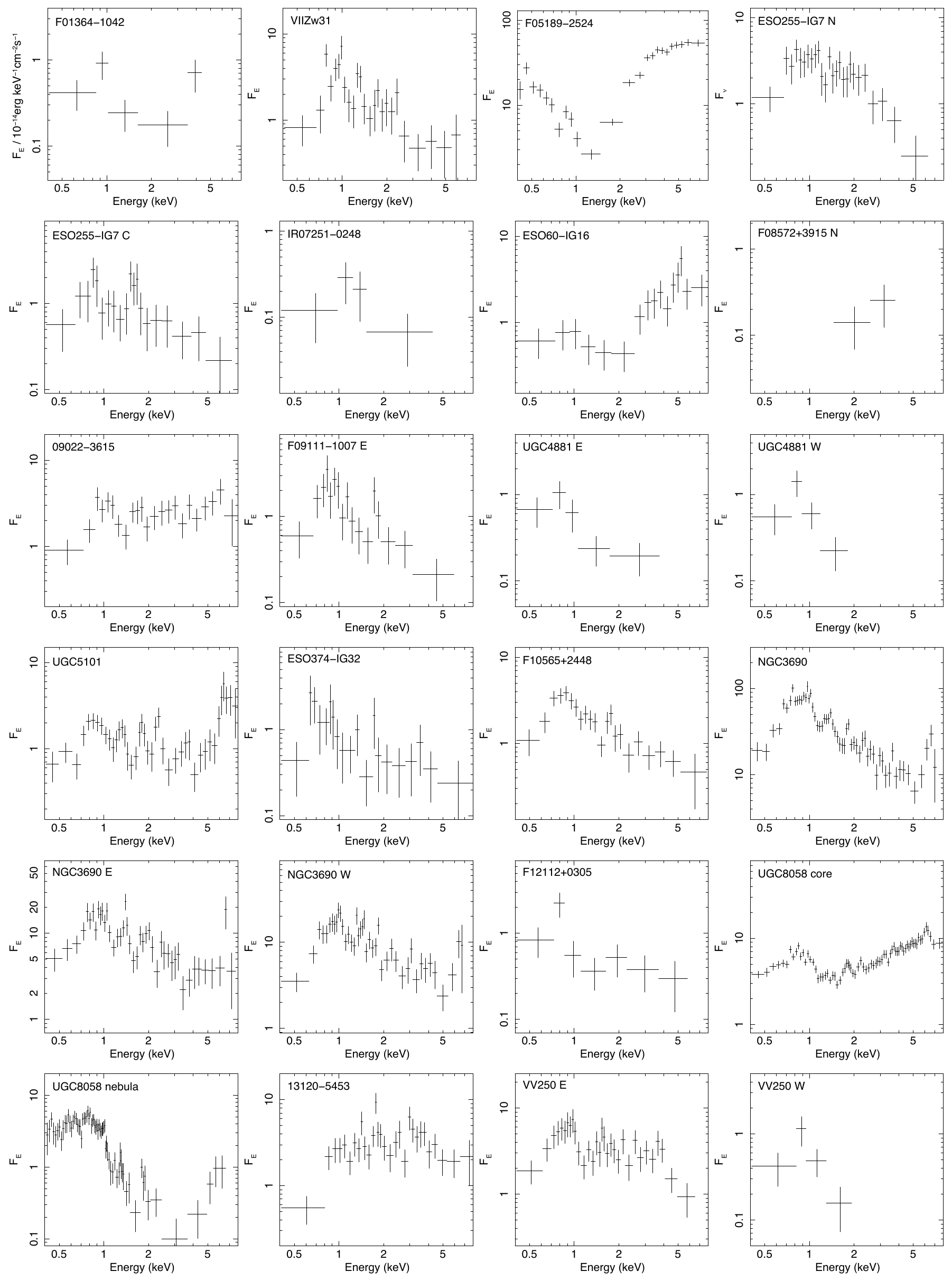}}
\end{figure*}
\begin{figure*}
\centerline{\includegraphics[width=\textwidth,angle=0]{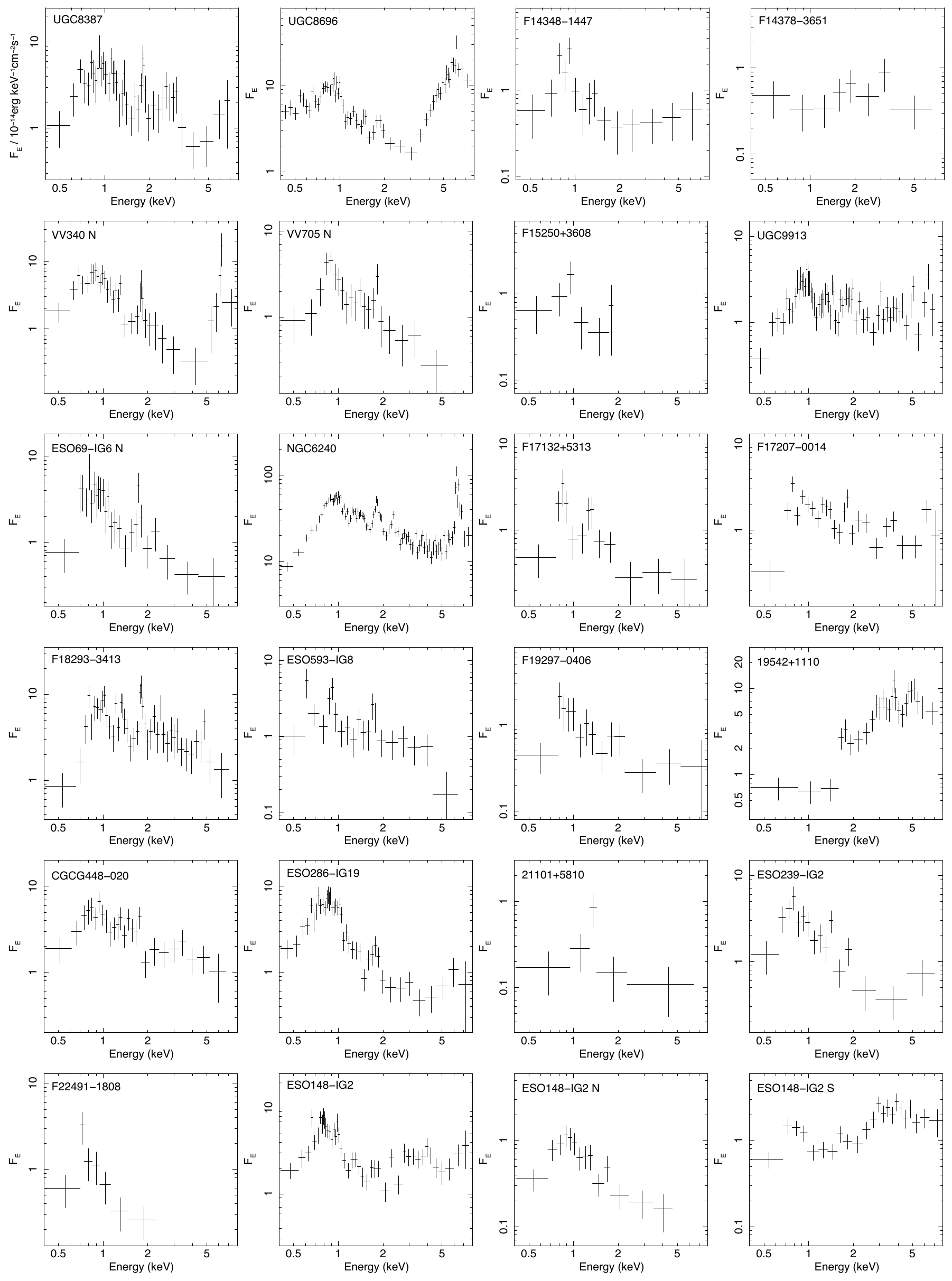}}
\end{figure*}
\begin{figure*}
%\centerline{\hbox{
%\includegraphics[width=0.21\textwidth,angle=270]{fig/eso77nfden2.ps}
%\includegraphics[width=0.21\textwidth,angle=270]{fig/eso77sfden2.ps}
%\includegraphics[width=0.21\textwidth,angle=270]{fig/f23365fden2.ps}
\centerline{\includegraphics[width=\textwidth,angle=0]{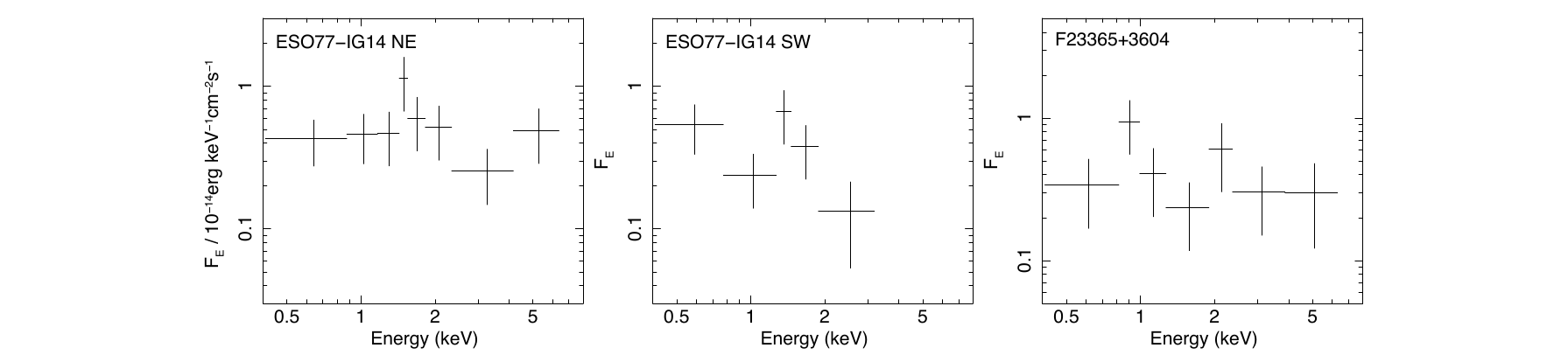}}
%}}
\caption{The X-ray flux density spectra of the C-GOALS sources, obtained from the Chandra ACIS. 
The energy scale is as observed. The flux density is in units of [$10^{-14}$ erg s$^{-1}$ cm$^{-2}$ keV$^{-1}$].
}
\end{figure*}

\subsection{X-ray colour and AGN selection}

% Fig 5 X-ray colour

\begin{figure}
\begin{center}
%\centerline{\includegraphics[width=0.3\textwidth,angle=270]{fig/irvshr.ps}}
%\centerline{\includegraphics[width=0.3\textwidth,angle=270]{fig/hrdist2.ps}}
%\centerline{\includegraphics[width=0.5\textwidth,angle=0]{fig/combohrdist.ps}}
\hbox{
\includegraphics[width=0.48\textwidth,angle=0]{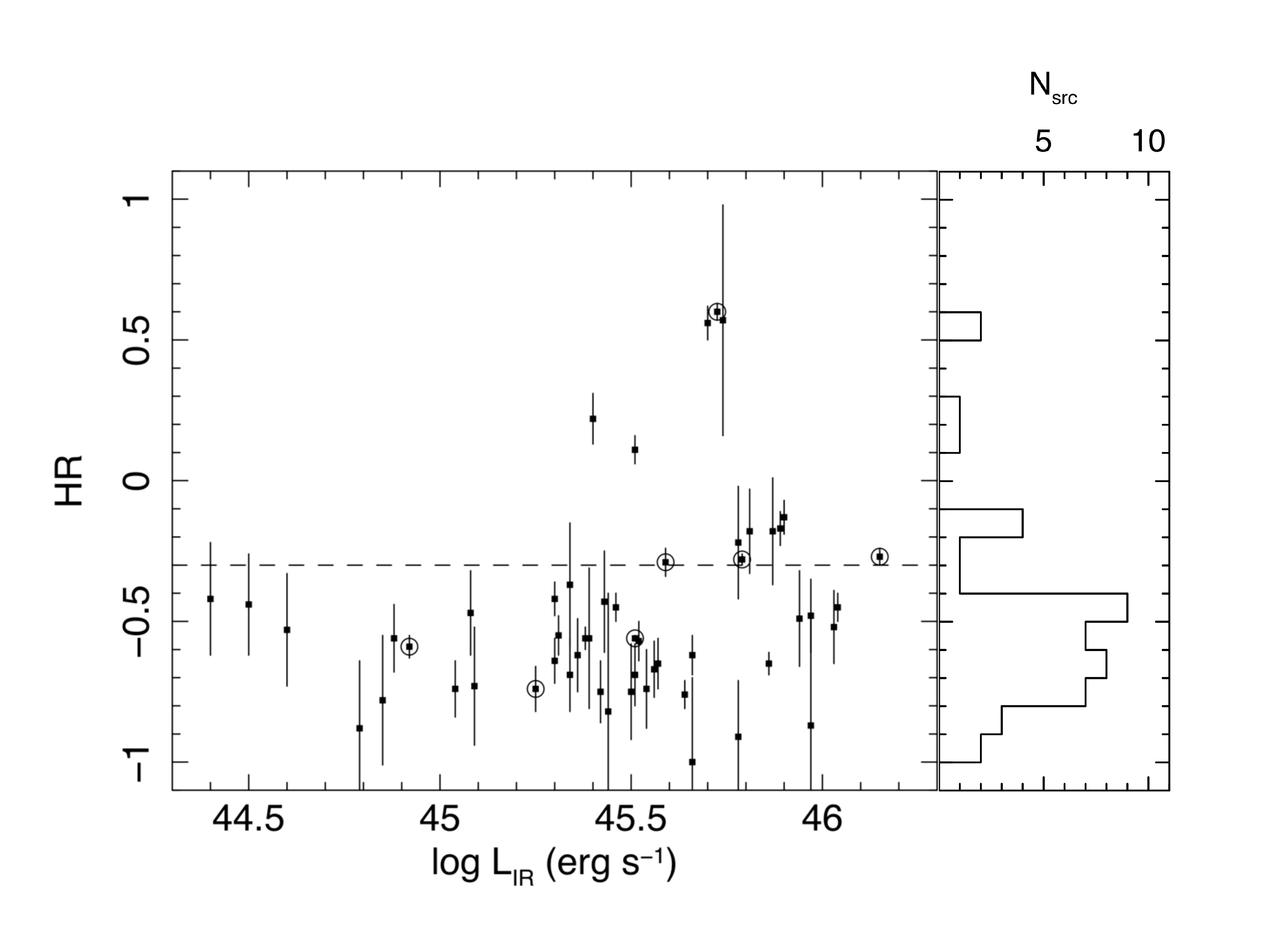}
}
\caption{The X-ray colour ({\sl HR}) as a function of infrared
  luminosity and its distribution. See the text for the definition of
  {\sl HR}. While the {\sl HR} is derived for individual components in
  a single object when the X-ray sources are well resolved into
  multiple components (i.e., the nuclear separation is $\sim
  4^{\prime\prime}$ or larger; see Table 3). The threshold for
  selecting AGN, based on the {\sl HR} $> -0.3$, is indicated by the
  dashed line. Objects in which a 6.4 keV Fe K$\alpha $ line detetion
  has been reported are marked by open circles. Three known Compton
  thick AGN with strong (6.4 keV) Fe K (NGC 3690 W, NGC 6240, VV340 N)
  lie below the {\sl HR} threshold. The {\sl HR} distribution
  histogram of {\sl HR} is attached on the right hand side. The median
  is {\sl HR} $= -0.56$. }
%Fitting a gaussian to the HR distribution gives a central
%  value {\sl HR}$ = -0.58$ and a dispersion of 0.20. }
\end{center}
\end{figure}

% Table 5 AGN table

\begin{table}
\setlength{\tabcolsep}{0.05in}
\begin{center}
  \caption{Classification of C-GOALS sources}
\begin{tabular}{rlccccl}
No. & Object & VO & YKS & $D_{\rm agn}$ & [Ne {\sc v}] & $X_{\rm agn}$\\%& MS \\
& & (1) & (2) & (3) & (4) & (5)\\[5pt]
\multicolumn{7}{c}{\bf X-ray selected AGN} \\
4 & F05189--2524 & S2 & S2 & 1 & Y & CL\\
7 & ESO 60-IG16 & -- & -- & -- & -- & C\\
8 & F08572+3915 NW & L & S2 & 0.6 & N & C\\
9 & 09022--3615 & -- & -- & -- & N & C\\
12 & UGC 5101 & L & S2 & 0.6 & Y & CL\\ 
16 & NGC 3690 W & H & -- & -- & -- & L\\
18 & UGC 8058 & S1 & S1 & 1 & N & CL\\
19 & 13120--5453 & -- & -- & -- & Y & C\\
22 & UGC 8696 & S2 & S2 & 0.8 & Y & CL\\
%F14348--1447 SW/NE & L/L & cp/cp & 0.7/0.6 & N \\%& b \\
23 & F14348--1447 SW & L & cp & 0.7 & N & C \\
24 & F14378--3651 & S2 & -- & -- & N & C\\
25 & VV 340 N & L & cp & 0.5 & Y & L\\  % armus 09
30 & NGC 6240 & L & L & 0.8 & Y & L\\
36 & 19542+1110 & -- & -- & -- & -- & C\\
42 & ESO 148-IG2 & -- & -- & -- & -- & C\\
44 & F23365+3604 & L & cp & 0.3 & N & C\\[5pt]%& d\\
\multicolumn{7}{c}{\bf Others (HXQ sample)} \\
1 & F01364--1042 & L & L & 0.8 & -- & \\%& c\\
2 & ESO 203-IG1  & -- & -- & -- & -- & \\% & a\\
3 & VII Zw 31 & H & -- & -- & Y & \\%& e\\
5 & ESO 255-IG7 & -- & H & 0 & -- & \\%& a\\
6 & 07251--0248 & -- & -- & -- & -- & \\%& c\\
10 & F09111--1007 & -- & -- & -- & Y & \\%& a\\
11 & UGC 4881 SW/NE & H/H & cp/cp & 0.3/0.2 & -- &\\%& b\\  
13 & ESO 374-IG32 & -- & cp & 0.3 & -- & \\%& b\\
14 & F10173+0828  & -- & -- & -- & -- & \\%& c\\
15 & F10565+2448 & H & cp & 0.2 & N & \\%& a\\
16 & NGC 3690 E & H & -- & -- & -- & \\%& b\\
17 & F12112+0305 & L & S2 & 0.7 & N & \\%& b\\
20 & VV 250 NW/SE & H/H & cp/cp & 0.4/0.3 & -- & \\%& a\\
21 & UGC 8387 & L & cp & 0.5 & Y & \\%& c\\    % aaron
26 & VV 705 S/N & L/H & cp/cp & 0.5/0.3 & -- &\\%& b\\ 
27 & F15250+3608 & L & cp & 0.4 & N & \\%& d\\
28 & UGC 9913 & L & L & 0.7 & N & \\%& b\\
29 & ESO 69-IG6 & -- & -- & -- & -- & \\%& a\\
31 & F17132+5313 & H & H & 0 & Y & \\%& b\\
32 & F17207--0014 & H & H & 0 & N & \\%& d\\
33 & F18293--3413 & -- & -- & -- & -- & \\%& b\\
34 & ESO 593-IG008 S/N & L/H & S2/cp & 0.4/0.2 & Y & \\%& b\\
35 & F19297--0406 & -- & -- & -- & N & \\%& b\\
37 & CGCG 448-020 & H & H & 0 & N & \\%& b\\     % Inami
38 & ESO 286-IG19 & H & H & 0 & -- & \\%& c\\
39 & 21101+5810  & -- & -- & -- & -- & \\%& a\\
40 & ESO 239-IG2 & -- & -- & -- & -- & \\%& c\\
41 & F22491--1808 & H & H & 0 & N & \\%& b\\ 
43 & ESO 77-IG14 & H & -- & -- & -- & \\%& a\\
\end{tabular}
\begin{list}{}{}
%\item[Note:] --- $\dagger $ denotes Compton thick AGN.
\item[Column (1):] Optical class based on Veilleux \& Osterbrock
  (1987) diagram from Veilleux et al. (1995, 1999). H: star-forming;
  S1: Seyfert 1; S2: Seyfert 2; L: LINER.
\item[Column (2):] The SDSS class from Yuan, Kewley \& Sanders
  (2010). The symbols are the same as in Column (1), plus an additional 
  class of  ``composite" objects;  cp: composite of starburst and AGN.
\item[Column (3):] The AGN fraction $D_{\rm agn}$ derived from
  [OI]/H$\alpha $ diagram (see Yuan et al. 2010 for details).
\item[Column (4):] Detection of [Ne V] 14.32 $\mu $m from Farrah et al.
  2007. New results on CGCG 448-020 (Inami et al. 2010), VV340 N (Armus
  et al. 2009), UGC 8387 (Modica et al. 2010), and other GOALS
  objects (Petric et al. 2010) are also included.
\item[Column (5):] AGN selection criteria: C: X-ray colour ($HR \geq 0.3$); L: 6.4 keV Fe K line.
%on CGCG 448-020 (Inami et al. 2010),
%  VV340 N (Armus et al. 2009), UGC 8387 (Modica et al., in prep) are
%  also included.
%\item[Column (5):] Merger stage adopted from Veilleux et al. (2002). 
%   a: wide binary; b: close binary; c: diffuse merger; d: compact merger; 
%   e: old merger.
\end{list}
\end{center}
\end{table}

The X-ray colour, or hardness ratio, {\sl HR}, gives the relative strength of the X-ray emission
above and below 2 keV (in counts). Since strong emission above 2 keV
is often associated with an absorbed X-ray source with a column density,
$N_{\rm H}$,  in the range of $10^{22}$-$10^{24}$\psqcm, which, in turn 
indicates the presence of an obscured AGN, it often 
serves as a crude probe of AGN. 

As described in Iwasawa et al. (2009), AGN are selected as follows: The
primary criterion is an hard X-ray spectrum as assessed by the X-ray
colour, {\sl HR}. The values of {\sl HR} as a function of $L_{\rm ir}$
are plotted in Fig 5. The median value of {\sl HR} is
$-0.56$.
%$0.58\pm 0.20$,
%derived from the HR distribution shown in Fig. 5 (lower panel).
Objects with {\sl HR} $> -0.3$ are classified as an AGN. This threshold is
chosen because ULIRGs known to host AGN (e.g. Mrk 231, UGC 8696 (= Mrk
273), UGC 5101) cluster just above this value. All of the optically
identified AGN, i.e., Seyfert 1 and Seyfert 2 galaxies, in our sample
are selected by this criterion.

Some Compton-thick AGN are missed by a {\sl HR} $> -0.3$ selection
because of their weakness in the hard band given that only reflected radiation 
is observed. The relative strength of the hard X-ray emission is
largely suppressed, giving a small value for {\sl HR}. Objects that show a
strong Fe~K line at 6.4~keV, a characteristic signature of a
Compton-thick AGN, are also classified as AGN (e.g. NGC 6240 --
Iwasawa \& Comastri 1998; Vignati et al. 1999; Ikebe et al. 2000;
Komossa et al. 2003; NGC 3690 West -- Della Ceca et al. 2002, Ballo et
al 2004; VV 340a -- Armus et al. 2009; UGC 5101 -- Imanishi et al.
2003). Note that high-ionization Fe K lines from FeXXV or FeXXVI at
6.7-7 keV that have been found in a few objects, (e.g., NGC 3690 East
-- Ballo et al. 2004; Arp 220 -- Iwasawa et al. 2005), are not
considered here as evidence for AGN since they can also originate from
hot gas produced by a starburst.

The selection by {\sl HR} and the 6.4 keV Fe K line finds 16 objects that 
contain AGN. The remaining objects  are characterized by relatively soft spectra,
defined by small {\sl HR} values or relative weakness of the hard
X-ray band emission (2-8 keV). The integral spectral properties of
this sample of 29 ``hard X-ray quiet'' (HXQ, as defined by small {\sl
  HR}) objects are reported in a separate paper (Iwasawa et al. 2009),
in which the detection and origin of a high-ionization Fe~K line in
the integrated spectrum is discussed. The X-ray classifications of the
sample are given in Table 5, along with optical and SDSS spectral
types, the mid-IR [Ne V] $\lambda 14.3 \mu $m detection, and the X-ray
AGN selection criteria, which are met. We
note that ESO 286-IG19 is classified as AGN in Francschini et al
(2003), based on the XMM-Newton data (see Appendix A for more detail),
while neither our X-ray or mid-IR criteria selected this object as an
AGN. Although ESO 286-IG19 remains as a viable AGN candidate, the
discussion below assume that this object is a HXQ.

In general, AGN being powerful X-ray emitters, an X-ray observation is
a sensitive probe of AGN unless they are hidden behind
Compton thick obscuration. However, it is not necessarily true that
AGN selected by the X-ray technique are energetically important for
the bolometric output of LIRGs. Here we have made an attempt to access the
importance of AGN contribution to the IR luminosity of the objects
hosting the X-ray selected AGN, using the observed X-ray properties.
However the uncertainty in these estimate should be quite large, as
explained below, and they can be considered as a guide only.

There are two steps for estimating the bolometric luminosity of an AGN
from the observed X-ray luminosity: 1) Absorption correction which
recovers the instrinsic X-ray luminosity by removing the flux
suppression effect due to absorption; 2) Bolometric correction which
converts the absorption-corrected X-ray luminosity to the bolometric
luminosity, assuming a typical spectral energy distribution (SED) of
an unobscured AGN (e.g., Elvis et al. 1994). In our rough estimates, the
absorption correction factors of 3 for a Compton thin AGN and 100 for
a Compton thick AGN (i.e., 1\% of 2-10 keV emission from a hidden AGN
is visible) are adopted for the 2-10 keV band. The bolometric
correction is assumed to be $L_{\rm bol}/L_{2-10}=30$, where
$L_{2-10}$ is absorption-corrected 2-10 keV luminosity, which is
adopted from Marconi et al. (2004) for objects with $L_{\rm bol}\sim
10^{12}L_{\odot}$. 

When applying the absorption correction, there is an ambiguity
inherent in the AGN selected by X-ray colour alone with a poor quality
spectrum. They show flat X-ray spectra (Fig. 4), as the {\sl HR}
analysis infers, but it is not trivial whether they are Compton thick
or thin objects without a good constraint on a Fe K line. As the
respective absorption correction factors differ significantly, the
classification of the obscuration type introduces a large uncertainty
in the AGN luminosity. Since only weak reflected light is observed
from a Compton thick AGN, we tentatively assume that the X-ray colour
selected AGN with log $(L_{\rm HX}/L_{\rm ir})<-4$ are Compton thick
AGN candidates\footnote{The hard X-ray to [OIII] $\lambda 5007$ ratio
  is often used as a diagnostic for a Compton thick AGN (e.g., Bassani
  et al 1999). However, this diagnostic may fail for dusty objects
  like LIRGs because emission lines from the narrow-line regions are
  likely suppressed as well.} and they are marked as such in Table
6. It is also noted that AGN with a reasonable quality spectrum
showing a low-energy cut-off due to moderate absorption, $N_{\rm
  H}\sim 10^{22}-10^{23}$ cm$^{-2}$, [F05189--2524, ESO 60-IG16, Mrk
273, 19542+1110] all have larger values between $-3.4$ and $-2.5$ in
log $(L_{\rm HX}/L_{\rm ir})$.

The absorption correction for a Compton thick AGN is also rather
uncertain, since it strongly depends on the geometry of the obscuring
matter, which is not well known. While a uniform correction factor of
$\times 100$ is used here, the final error in $F_{\rm agn}$ can be
easily larger than 0.5 dex, as demonstrated by $F_{\rm agn} > 1$ for
NGC 6240.  In any case, taking the face values, the median of log
$F_{\rm agn}$ is $-1$, and there are only two objects (Mrk 231 and NGC
6240) in which the AGN contribution exceeds 50\% of their infrared
luminosity. In summary, according to this simplistic estimate,
AGN-dominated objects would seem to be in the minority in our sample,
and the typical AGN contribution to the infrared luminosity would
appear to be $\sim 10$\%.

\begin{table}
\begin{center}
\caption{Estimates of AGN contribution to IR luminosity}
\begin{tabular}{rlccc}
No. & Galaxy & log (HX/IR) & CT & log $F_{\rm agn}$ \\
& & (1) & (2) & (3) \\[5pt]
4 & F05189--2524 & $-2.63$ & & $-0.65$ \\
7 & ESO 60-IG16 & $-2.54$ & & $-1.54$ \\
8 & F08572+3915 & $-4.44$ & $\ast $ & $-0.94$ \\
9 & 09022--3615 & $-3.59$ &  & $-1.59$ \\
12 & UGC 5101 &  $-3.92$ & $\circ $ & $-0.42$ \\
16 & NGC 3690 W & $-4.07$ & $\circ $ & $-0.57$ \\
18 & UGC 8058 & $-3.67$ & $\circ $ & $-0.17$ \\
19 & 13120--5453 &  $-4.25$ & $\ast $ & $-0.75$ \\
22 & UGC 8689 & $-3.39$ &  & $-1.39$ \\
23 & F14348--1447 S & $-4.90$ & $\ast $ & $-1.50$ \\
24 & F14378--3651 &  $-4.28$ & $\ast $ & $-0.78$ \\
25 & VV 340 N & $-3.86$ & $\circ $ & $-0.36$ \\
30 & NGC 6240 & $-2.97$ & $\circ $ & $+0.53$ \\
36 & 19542+1110 & $-3.05$ &  & $-1.09$ \\
42 & ESO 148-IG2 & $-3.72$ &  & $-1.72$ \\
44 & F23365+3604 & $-4.58$ & $\ast $ & $-1.08$ \\
\end{tabular}
\begin{list}{}{}
\item[(1)] Logarithmic luminosity ratio of the observed 2-10 keV and
  8-1000 $\mu$m bands, reproduced from Table 3.
\item[(2)] Compton thick AGN with Fe K detection are marked as $\circ
  $. Candidate Compton thick AGN (see text for detail) are marked as
  $\ast $. 
\item[(3)] Logarithmic fraction of the X-ray estimate for AGN contribution
  to the 8-1000$\mu$m luminosity.
\end{list}
\end{center}
\end{table}

The X-ray-selected AGN represent 37\% of the total sample
(16/44). This figure is comparable to that inferred from the mid-IR
diagnostics for the same IR luminosity range (Petric et al. 2010). When
the sample is divided into two luminosity bins, above and below the
median IR luminosity of log $(L_{\rm ir}/L_\odot) = 11.99$, 12 of these
16 AGN (75\%) are above the median, in agreement with previous studies
showing the fraction of AGN rising with increasing $L_{\rm ir}$ (e.g.,
Veilleux et al. 1995). If objects with [Ne V]$\lambda 14.3\mu $m
detection (Table 5), which is generally considered as evidence for
AGN, are included, the AGN fraction would increase up to 48\%.  In
terms of merger stage, X-ray-selected AGN tend to be found more in
mergers of later stages. For example, 50\% (9/18) of final stage
mergers (nuclear separation $< 1$ kpc) are X-ray-seleted AGN, while
only 26\% (7/27) of earlier stage mergers are X-ray selected AGN. We
refer to Evans et al. (in prep.) for details on the galaxy morphology
and merger-stage classifications based on our HST imaging.

NGC 6240 is the only object in the sample where two nuclei both show
evidence for harboring a powerful AGN \footnote{Mrk 266 (Mazzarella et
  al 2010) and Mrk 463 (Bianchi et al. 2008), two GOALS galaxies, but
  with $L_{\rm ir}$ below our current C-GOALS sample threshold, are other
  objects which are found to have X-ray evidence for AGN in both
  nuclei.} out of  24 objects that have double or tripple nuclei. It
should be noted that we do not consider NGC 3690 to have a double AGN,
as the Fe XXV line found in the eastern galaxy is not taken as an AGN
signature. Other AGN are found either in a single nucleus (8 objects)
or in one member of a double nucleus system  (7 objects).

\subsection{X-ray luminosities and correlation with $L_{\rm ir}$}

% Fig 6 Lx distribution 

\begin{figure}
\begin{center}
  \centerline{\includegraphics[width=0.4\textwidth,angle=0]{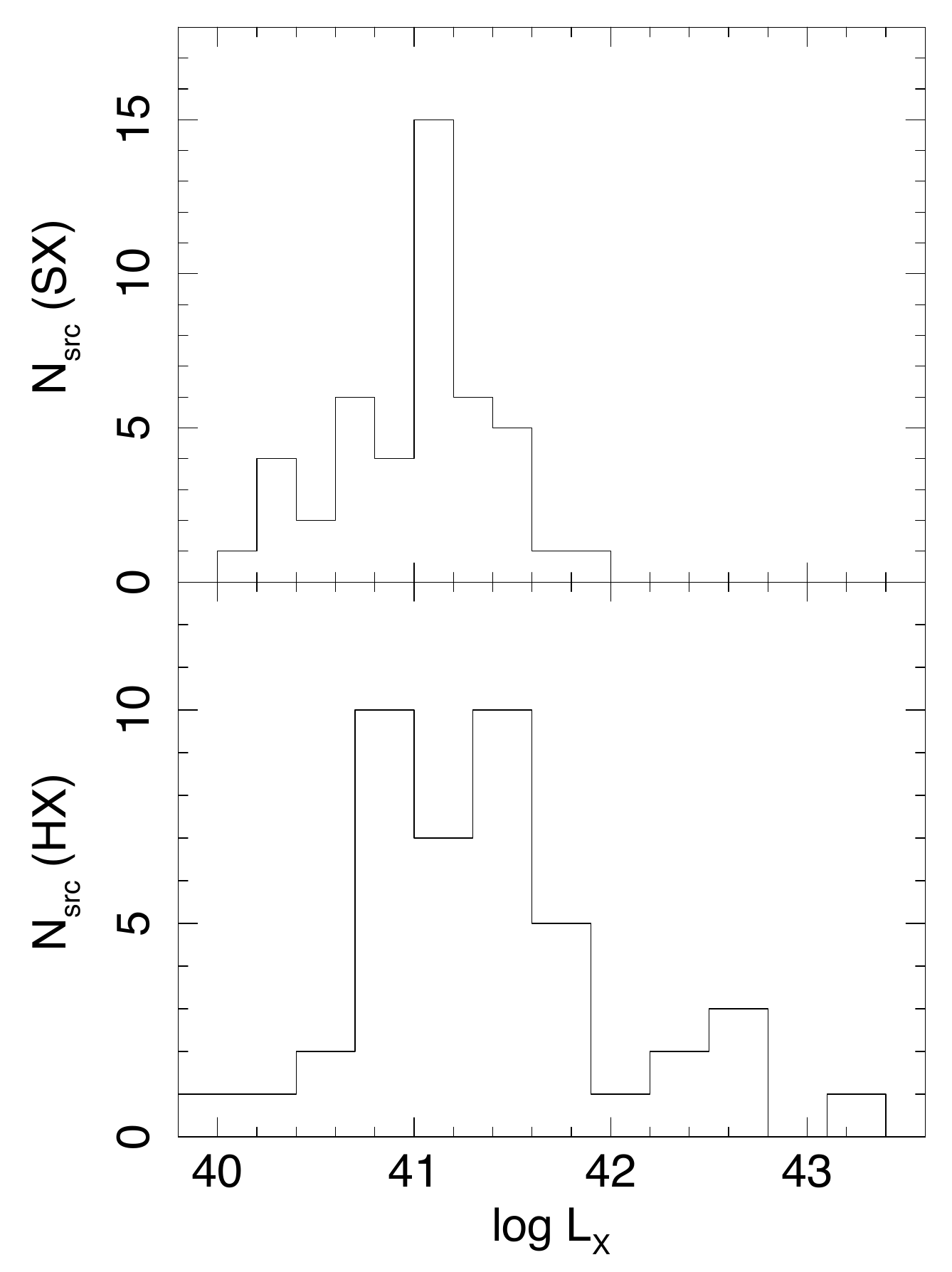}}
  \caption{X-ray luminosity distribution in the soft (0.5-2 keV; upper
    panel) and hard (2-10 keV; lower panel) bands.}
\end{center}
\end{figure}

% Fig 7 SX vs IR, HX vs IR 

\begin{figure}
\begin{center}
\centerline{\includegraphics[width=0.42\textwidth,angle=0]{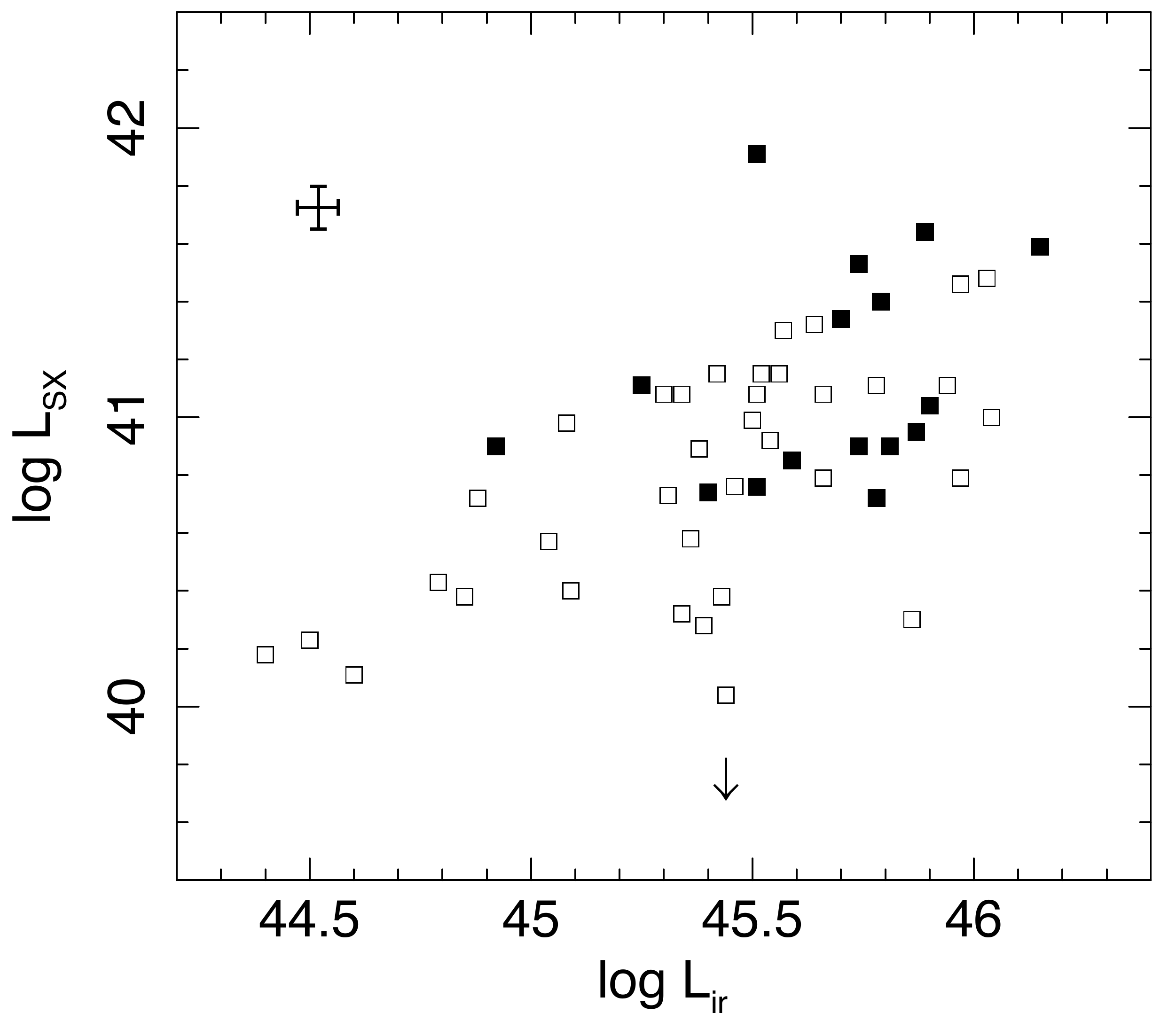}}
\centerline{\includegraphics[width=0.42\textwidth,angle=0]{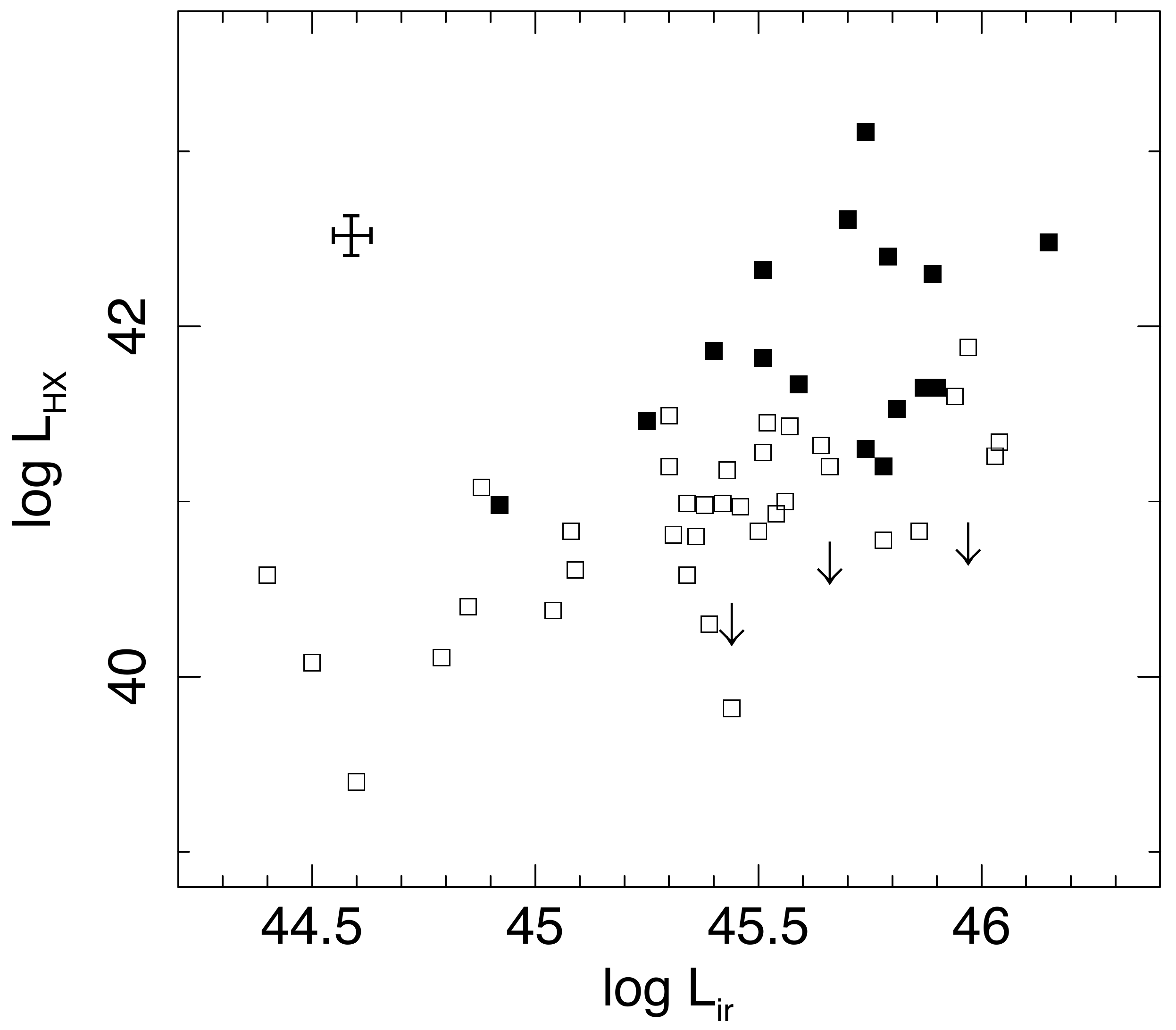}}
\caption{ Plots of X-ray (upper panel: 0.5-2 keV; lower panel: 2-10
  keV) versus infrared luminosity. The X-ray luminosity has been
  corrected only for Galactic absorption. The X-ray selected AGN are
  marked with filled squares. The typical error bars are indicated. For
  objects that are not detected, the 95 per cent upper limits are
  indicated by arrows. When multiple components are resolved in a
  single system, their luminosities are computed separately (Table
  3). The decomposition of the infrared luminosity in individual
  objects is described in Appendix A. There is a moderate correlation,
  with a correlation coefficient of $\sim 0.6$ for both plots. }
%The best-fit linear correlation
%    gives a logarithmic offiset log $(SX/IR) = -4.6\pm 0.1$ and log
%    $(HX/IR) = -4.3\pm 0.2$.
\end{center}
\end{figure}

% Fig .8 SX/IR and HX/IR distributions

\begin{figure}
\begin{center}
\centerline{\includegraphics[width=0.4\textwidth,angle=0]{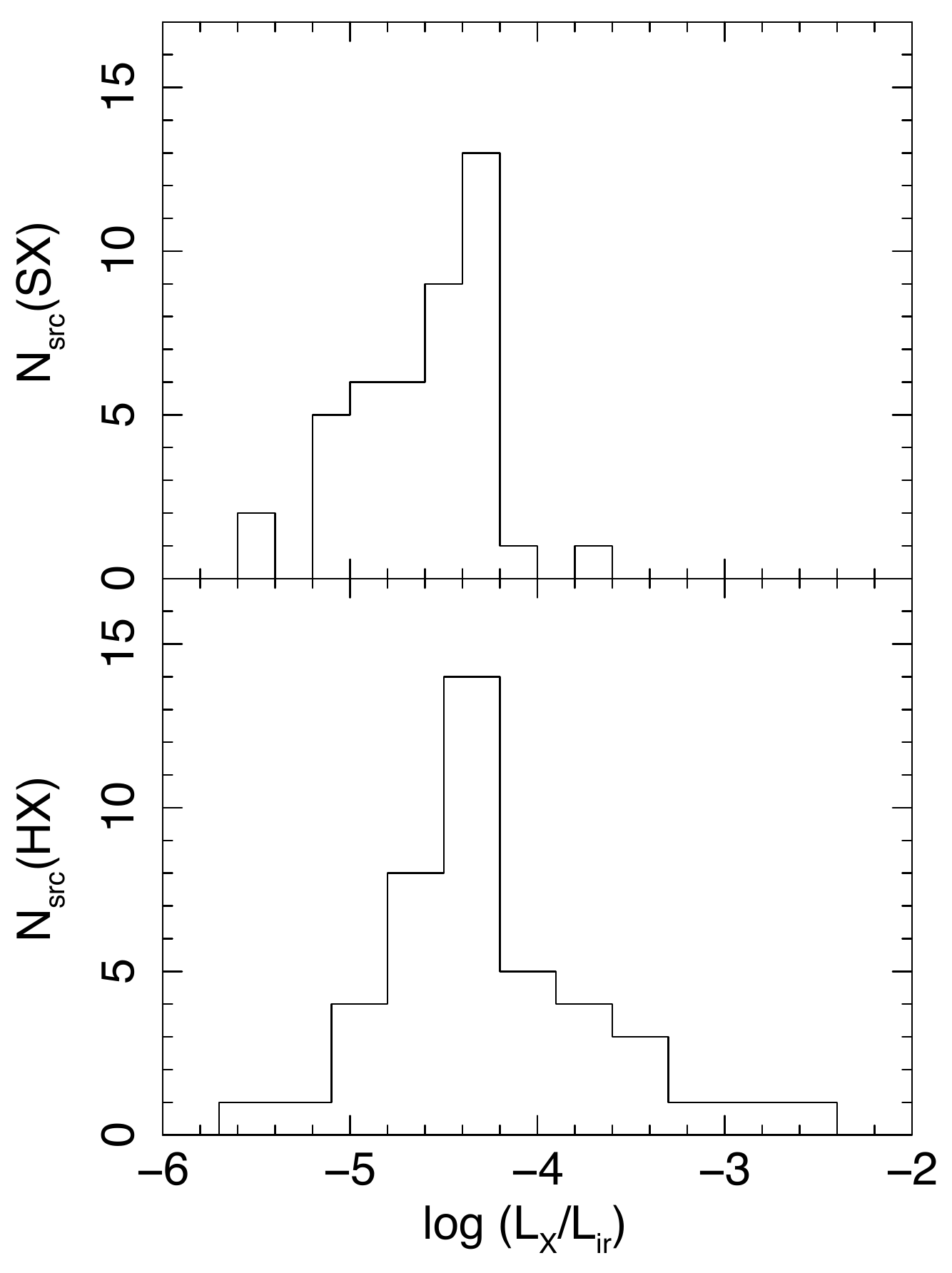}}
  \caption{X-ray to infrared luminosity ratio distribution. }
%Fitting a
%    Gaussian to the distribution of the $SX/IR$ gives a peak at $-4.53$
%    with dispersion of 0.34. For the $HX/IR$ distribution, which has
%    wider wings, a Lorentzian fits better than a Gaussian, and the peak
%    value is $-4.40$ with a dispersion of 0.63.}
\end{center}
\end{figure}

The X-ray fluxes were estimated based on a best-fitting spectral model
in the 0.5-2 keV and 2-7 keV bands (i.e., no correction of
absorption). For faint sources, the hard band data are often not good
enough to constrain the shape, in which case a power-law of $\Gamma =
2.1$ is assumed to derive the 2-7 keV flux (e.g., Ranalli et al.
2003). The X-ray luminosities are estimated by correcting only for
Galactic absorption.  [Note: Significant absorption intrinsic to the
sources is likely present. However, X-ray spectra of our objects are
complex with multiple components, while most of the spectra do not
have sufficient quality to uniquely decompose and estimate absorption
of each component. For this reason, we opt to present the luminosities
as observed.]  The hard band luminosity is for the 2-10 keV band,
which is estimated by extrapolating the 2-7 keV spectrum. For a few
objects, the luminosity values have been updated from those in Iwasawa
et al (2009). The distribution of the X-ray luminosity in the soft and
hard bands is shown in Fig. 6. The luminosity distribution in both
bands peaks at log ($L_{\rm X}) \approx 41.1$ \ergps, but the hard
band luminosity is spread over a wider range.
% Fitting a Lorentzian to the respective distribution gives a
% dispersion of 0.4 dex for the soft X-ray luminosity and 1.1 dex for
% the hard X-ray luminosity.
The median logarithmic values for the soft and hard band luminosities
are 41.1\ergps and 41.3 \ergps, respectively.

%Neither soft or hard band X-ray luminosities show a clear correlation
%with the $L_{\rm ir}$ among the objects in our sample (see
%Fig. 7). Removing obvious AGN candidates (see \S 4.3) does not improve
%the correlation, as discussed in Iwasawa et al. (2009)\footnote{In
%  Iwasawa et al. (2009), the IR luminosity ($8-100\mu$m) is relaced by
%  the FIR luminosity ($40-400\mu$m) to allow direct comparison with
%  the previous studies on the X-ray -- FIR correlation.}. 

A moderate correlation between the IR and X-ray luminosities can be
seen in Fig. 7, with a typical spread over an order of magnitude.  Note that
when multiple components are present in a single object, their
luminosities are plotted separately in Fig. 7. When integrated
luminosities in a single objects are used, as shown in Iwasawa et al.
(2009)\footnote{In Iwasawa et al. (2009), the IR luminosity
  ($8-1000\mu$m) is also replaced by the FIR luminosity ($40-400\mu$m)
  to allow direct comparison with the previous studies on the X-ray --
  FIR correlation.}, the correlation becomes less clear. In the soft X-ray
band, AGN are mixed in with the distribution of non-AGN (HXQ) objects,
while in the hard X-ray band, AGN tend to be the more luminous X-ray
sources, as expected. 

The X-ray to IR luminosity ratio distribution shown in Fig. 8 has a
significant spread, which is caused by the scatter around the
correlation between the luminosities (e.g. Fig. 7) rather than any
non-linear correlation. Typical values are log ($L_{\rm SX}/L_{\rm
  ir}) = -4.53\pm 0.34$ for the soft X-rays and log ($L_{\rm
  HX}/L_{\rm ir}) = -4.40\pm 0.63$ for the hard X-rays (the
uncertainties are the dispersion of the distributions).  When
X-ray-selected AGN are excluded, a linear relation between logarithmic
X-ray and IR luminosities (Fig. 7) is given by ${\rm log} L_{\rm SX} =
(-4.6\pm 0.1)+{\rm log} L_{\rm ir}$ for the soft X-ray band and ${\rm
  log} L_{\rm HX} = (-4.5\pm 0.1)+{\rm log} L_{\rm ir}$ for the hard
X-ray band. These values are significantly lower than those found for
local, star-forming galaxies with low star formation rates, e.g., log
$L_{\rm X}/L_{\rm FIR}\sim -3.7$ (Ranalli et al 2003). Note that a
direct comparison would need a correction for the different infrared
bandpasses (details of the direct comparison can be found in Iwasawa
et al 2009).

\subsection{Radial surface brightness profiles}

% Fig 6   Rmax vs Lsx
\begin{figure}
\begin{center}
%\centerline{\includegraphics[width=0.35\textwidth,angle=270]{fig/sxrmax.ps}}
\centerline{\includegraphics[width=0.42\textwidth,angle=0]{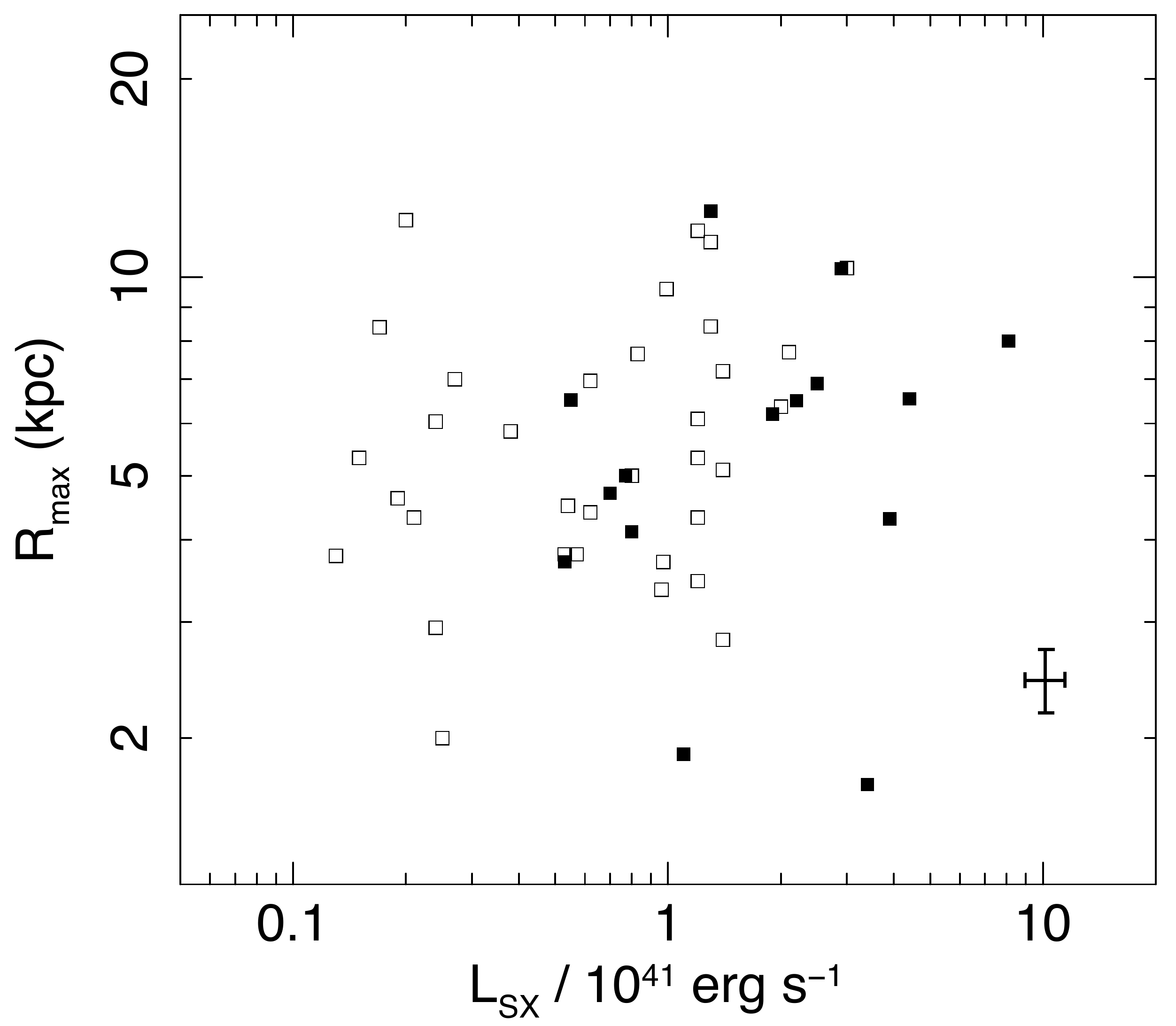}}
\caption{The soft X-ray emission radius, $R_{\rm max}$ (kpc) versus
  soft X-ray luminosity. Filled symbols are of X-ray-selected AGN and
  open symbols are others. The typical error bars are indicated. No
  trend can be seen. The median value of $R_{\rm max}$ is 5.3 kpc. }
% and fitting a Gaussian to the
%    distribution of $R_{\rm max}$ gives a peak value of 5.2 kpc with a 
%    dispersion of 2.4 kpc. }
\end{center}
\end{figure}

% Table SURBRI and compactness
\begin{table*}[H]
\begin{center}
  \caption{Properties derived from the X-ray radial surface brightness profiles.$^{\mathrm{a}}$}
\begin{tabular}{rlcccccc}
No. & Galaxy & ${\sl AS}$ & $r_{\rm max}$ & $R_{\rm max}$ & $r_{\rm hp}$ & $R_{\rm hp}$ & Morph \\
& & (kpc arcsec$^{-1}$) & (arcsec) & (kpc) & (arcsec) & (kpc) & HX \\[5pt]
 \multicolumn{8}{c}{\bf Cycle 8 Data} \\
1 & F01364--1042  & 0.93 & 6.5 & 6.04 & 2.0 & 1.86 &  P \\
3 & II Zw 31  & 1.06 & 6.0 & 6.36 & 2.0 & 2.12 &  R \\
5 & ESO 255-IG7 N  & 0.76 & 7.0 & 5.32 & 2.5 & 1.90 &  R \\
5 & ESO 255-IG7 C  & 0.76 & 5.0 & 3.80 & 2.0 & 1.52 &  A \\
5 & ESO 255-IG7 S  & 0.76 & 7.0 & 5.32 & 2.5 & 1.90 &  R \\
6 & 07251--0248  & 1.74 & 4.0 & 6.96 & 1.5 & 2.61 &  -- \\
7 & ESO 60-IG16  & 0.93 & 7.0 & 6.51 & 2.0 & 1.86 &  P \\
9 & 09022--2615  & 1.19 & 5.5 & 6.54 & 1.5 & 1.78 &  R \\
10 & F09111--1007 E  & 1.08 & 4.0 & 4.32 & 1.0 & 1.08 &  A \\
10 & F09111--1007 W  & 1.08 & 4.0 & 4.32 & 1.0 & 1.08 &  A \\
13 & ESO 374-IG32  & 0.73 & 8.0 & 5.84 & 2.5 & 1.82 &  R \\
14 & F10173+0828  & 0.99 & -- & -- & -- & -- &  A \\
19 & 13120--5453  & 0.63 & 3.0 & 1.89 & 1.0 & 0.63 &  P \\
20 & VV 250 E  & 0.63 & 5.5 & 3.46 & 1.5 & 0.95 &  R \\
20 & VV 250 W  & 0.63 & 6.0 & 3.78 & 2.5 & 1.57 &  A \\
21 & UGC 8387  & 0.50 & 9.0 & 4.50 & 3.0 & 1.50 &  R \\
24 & F14378--3651  & 1.37 & 3.0 & 4.11 & 1.0 & 1.37 &  P \\
25 & VV 340 N  & 0.70 & 18.0 & 12.60 & 5.5 & 3.85 &  R \\
25 & VV 340 S  & 0.70 & $>12.0$ & $>8.40$ & 7.5 & 5.25 &  -- \\
29 & ESO 69-IG6 (N)  & 0.94 & 3.0 & 2.82 & 1.0 & 0.94 &  R \\
31 & F17132+5313  & 1.02 & 7.5 & 7.65 & 2.5 & 2.55 &  R \\
33 & F18293--3413  & 0.38 & 10.0 & 3.80 & 3.0 & 1.14 &  R \\
34 & ESO 593-IG8  & 0.98 & 12.0 & 11.76 & 4.5 & 4.41 &  R \\
35 & F19297--0406  & 1.72 & 6.0 & 10.32 & 2.5 & 4.30 &  P \\
36 & 19542+1110  & 1.30 & 5.0 & 6.50 & 1.0 & 1.30 &  P \\
37 & CGCG 448-020  & 0.72 & 10.0 & 7.20 & 3.0 & 2.16 &  R \\
39 & 21101+5810  & 0.77 & 6.0 & 4.62 & 1.5 & 1.16 &  P \\
40 & ESO 239-IG2  & 0.85 & 6.0 & 5.10 & 1.5 & 1.27 &  P \\
41 & F22491--1808  & 1.53 & 5.5 & 8.42 & 2.0 & 3.06 &  P \\
43 & ESO 77-IG14 N  & 0.84 & 4.0 & 3.36 & 1.0 & 0.84 &  A \\
43 & ESO 77-IG14 S  & 0.84 & 3.5 & 2.94 & 2.0 & 1.68 &  P \\[5pt]
\multicolumn{8}{c}{\bf Archival Data} \\
4 & F05189--2524 & 0.83 & 2.0 & 1.7 & 1.0 & 0.83 & P \\
8 & F08572+3915 & 1.13 & -- & -- & -- & -- & A \\
11 & UGC 4881 E & 0.79 & 2.5 & 2.0 & 1.0 & 0.79 & P \\
11 & UGC 4881 W & 0.79 & 8.0 & 7.0 & 3.5 & 3.1 & -- \\
12 & UGC 5101 & 0.79 & 6.0 & 4.7 & 1.5 & 1.2 & P \\
15 & F10565+2448 & 0.87 & 7.0 & 6.1 & 2.0 & 1.7 & P \\
16 & NGC 3690 E & 0.24 & $>13$ & $>3.1$ & -- & -- & R \\
16 & NGC 3690 W & 0.24 & $>13$ & $>3.1$ & -- & -- & R \\
17 & F12112+0305 & 1.42 & 8.0 & 11.3 & 4.0 & 5.7 & P \\
18 & UGC 8058 & 0.85 & 5.0 & 4.3 & 1.0 & 0.85 & P \\
22 & UGC 8696 & 0.77 & 9.0 & 6.9 & 2.5 & 1.9 & R \\
23 & F14348--1447 & 1.58 & 6.5 & 10.3 & 3.0 & 4.8 & P \\
26 & VV 705 & 0.82 & 4.5 & 3.7 & 1.5 & 1.2 & A \\
27 & F15250+3608 & 1.10 & 4.0 & 4.4 & 1.5 & 1.6 & -- \\
28 & UGC 9913 & 0.41 & 30 & 12.2 & 8.5 & 3.5 & R \\
30 & NGC 6240 & 0.53 & 15 & 8.0 & 5.5 & 2.9 & R \\
32 & F17207--0014 & 0.87 & 11.0 & 9.6 & 3.0 & 2.6 & R \\
38 & ESO 286-IG19 & 0.85 & 9.0 & 7.7 & 3.5 & 3.0 & R \\
42 & ESO 148-IG2 & 0.88 & 7.0 & 6.2 & 3.5 & 3.1 & R \\
44 & F23365+3604 & 1.22 & 3.0 & 3.7 & 1.0 & 1.2 & A \\
\end{tabular}
\begin{list}{}{}
\item[$^{\mathrm{a}}$] The radius where the soft X-ray brightness goes below
  1\% of the peak brightness at $1^{\prime\prime}$ resolution is defined
  as $r_{\rm max}$ (arcsec). The corresponding physical radius, $R_{\rm max}$ (kpc), 
  is also listed. The half power
  radius, $r_{\rm hp}$ (arcsec), is where the cumulative soft X-ray counts exceed
  half of the integrated counts within $r_{\rm max}$ at $1^{\prime\prime}$ 
  resolution. The corresponding physical radius, $R_{\rm hp}$ (kpc),  is also
  given. Whether the 2-7 keV emission is resolved or not is
  indicated by ``P" (point-like), ``R" (resolved), and ``A" (ambiguous).
\end{list}
\end{center}
\end{table*}

\clearpage

Radial surface brightness profiles are derived in the soft (0.5-2 keV)
and hard (2-7 keV) bands at a resolution of  1 arcsecond.  In general,
the soft X-ray emission is spatially extended while the hard X-ray
emission is compact. A few exceptions are sources dominated by
absorbed hard X-ray emission as seen in objects like IRAS 19542+1110,
where both bands show equally compact emission.

With the $0.5^{\prime\prime}$ resolution of the Chandra ACIS, the
presence of extended soft X-ray emission is immediately clear by
visual inspection for all of the sources. To quantify the soft X-ray
extension, azimuthally-averaged surface brightness profiles are
produced in the soft (0.5-2 keV) and hard (2-7 keV) bands separately
for individual targets. An example of these surface brightness
profiles is shown in Fig. 3 for UGC 8387, and those for all the other
objects are presented in on-line material. The soft X-ray profiles
show a range of shapes -- from an exponetial profile to a more peaky,
power-law type profile. Since not all of the sources have sufficient
counts, the soft X-ray profiles are quantified by two characteristic
radii, rather than fitting a profile model. The maximum extension
radius, $r_{\rm max}$, and half power radius, $r_{\rm hp}$, are
defined as follows: The radius where the surface brightness falls to
1\% of the peak brightness (usually of the central bin of the surface
brightness profile) is defined as $r_{\rm max}$ in arcsec. This radius
gives a measure of the source extension, which has little dependence
on the depth of the image. The total source count, $C_{\rm max}$,
integrated within $r_{\rm max}$ is then derived . $C_{\rm max}$ can
differ from the total source count when the surface brightness, which
is lower than 1\% of the peak brightness, extends to large radii. The
radius where the accumulated source count reaches half of $C_{\rm
  max}$ is defined as the half power radius, $r_{\rm hp}$, in
arcsec. These quantities give a measure of the compactness/broadness
of the source extension. Since both $r_{\rm max}$ and $r_{\rm hp}$ are
determined as the innermost radial bin which satisfies the required
condition, their uncertainty is always 0.5 arcsec. For a compact
source (e.g., $r_{\rm hp}\sim 1^{\prime\prime}$), this half power
radius is dominated by the point spread function (PSF), and is likely
to be over-estimated. The radii in physical units (kpc) corresponding
to $r_{\rm max}$ and $r_{\rm hp}$ are also given in Table 7.

The hard X-ray sources are compact and often point-like. We note
however that there are some objects in which hard X-ray emission is
resolved, e.g., Arp 220, and UGC 8387 (=IC 883). As a
guide, the PSF simulated at an energy of 3
keV and normalized to the central peak of the hard X-ray emission is
plotted along side the observed surface brightness profiles. In Table 7,
the hard X-ray morphology is classified into three categories -- R/A/P.
When more than 2 data points deviate by more than
$2\sigma $ from the PSF  it is denoted as R (resolved).  When the data
points agree with the PSF within the error bars, it is denoted as P (point-like).  
Any other case is denoted as A (ambiguous).

Figure 9  shows a plot of  $R_{\rm max}$ versus the soft X-ray luminosity.  
No clear correlation is seen even if objects containing AGN are removed.
% However, if we remove AGN, which typically 
%have a compact core, and we remove low luminosity objects ($\leq
%3\times 10^{40}$ \ergps),  which generally show a flat X-ray
%brightness distribution most likely due to a collection of discrete X-ray
%sources rather than hot gas, then there might be a weak 
%correlation of  $R_{\rm max}$ versus the soft X-ray luminosity with a 
%slope of $\sim 0.3$.  If the soft X-rays are due
%to thermal emission and the total luminosity is dominated by
%bremsstrahlung continuum, the luminosity is proportional to $n^2V$. If
%the density, $n$, is constant between sources, the lumiosity would then be
%correlated with the projected size $V^{1/3}$, which is roughly
%consistent with the trend seen in Fig. 9.

The variety of shapes observed in the soft X-ray radial profiles 
could be related to the origin of the X-ray emission. Generally
speaking, a gravitationally bound, virialized system shows an
exponential profile while a power-law profile is expected from a
free-flowing outflow from a compact central source. However, given the
angular scale for our sample galaxies (0.24-1.74 kpc arcsec$^{-1}$), 
any compact nuclear starbursts are likely to be contained 
within the innermost $1^{\prime\prime}$ bin. There are
several objects showing a power-law type, peaky, soft X-ray profile, and
they are noted in the Notes on Individual Objects (Appendix A). Absorption can
modify the soft X-ray profile if it has a radial dependence, e.g.,
centrally concentrated absorption, which suppress the brightness of 
the compact nuclear component, e.g. Arp 220 and VV 340 N.

\section{Discussion}

In \S~4.3, we used relatively crude spectral information based on the X-ray
colour, (i.e. {\sl HR}),  to assess the presence of an AGN. Here we discuss further
characterizations of the ACIS spectra for sufficiently bright
objects. Before presenting the spectral analyses, a brief summary of the properties of 
X-ray sources in LIRGs and their spectra  are described
below.

The primary origin of the X-ray emission in LIRGs is considered to be
a starburst and/or AGN. Given the dusty nature of LIRGs, if an AGN is
present, it is likely to be an absorbed source. Such absorbed X-ray
sources can be selected by X-ray colour analysis, as shown in
\S~4.3. When the absorbing column density exceeds $10^{24}$ cm$^{-2}$
and the absorbed transmitted component moves out of the Chandra
bandpass, the identification of an AGN becomes difficult as it must
rely on the detection of the Fe~K line at 6.4 keV in faint reflected
emission (see the discussion in \S~4.3), and the data are often not of
sufficient depth to make a clear detection.  Such heavy nuclear
obscuration could, however, occur in many LIRGs, and sometimes the
hard X-ray band is the only available window to check for signatures
of an AGN, e.g. as for NGC 4945 (e.g., Iwasawa et al 1993; Spoon et al 2000).

In a starburst, there are various sources of X-ray emission (e.g.,
Persic \& Rephaeli 2002) whose origin can be traced back to
massive stars. Individual supernovae and their remnants, stellar-wind
heated ISM in star clusters, and X-ray binaries can all be X-ray
sources. In particular, high-mass X-ray binaries (HMXBs) are
considered to be the dominant source for the X-ray emission above 2
keV. A good correlation between the 2-10 keV luminosity and the star
formation rate have been found for nearby star forming galaxies
(Ranalli et al. 2003; Grimm et al. 2003; Gilfanov et al. 2004) as well as
for late type galaxies at higher redshift (Lehmer et al. 2008). The
galactic-scale emission nebulae, traced by H$\alpha $ and the soft X-ray
emission in local starburst galaxies like M~82, suggested that these
extended nebulae are produced by the shock heated interstellar medium, swept up by a
starburst-driven outflow (see e.g., Veilleux, Cecil \& Bland-Hawthorn
2005 for a review on galactic outflow phenomena, and Tomisaka \&
Ikeuchi 1988; Suchkov et al. 1996; Tenorio-Tagle \& Munoz-Yunon 1998;
Strickland \& Stevens 2000 for simulations). As hypothesized by
Chevallier \& Clegg (1985), hot ($T\sim 10^8$ K), high-pressure gas,
produced via the collective kinetic energy from supernovae and massive
stellar winds, is considered to drive the outflow. Given the
temperature, this hot gas could be a hard X-ray source, but it is also
expected to be rarefied, and  thus not very radiative. The presence of weak
emission from such gas has been suggested for M 82 (Griffiths et al. 2000;
Strickland \& Heckman 2007; Ranalli et al. 2008) and NGC 253 (Pietsch et
al 2001).

\subsection{X-ray spectra and derived properties}

Except for some sources which are clearly absorbed AGN, the 0.4-7 keV
Chandra spectra of the majority of the GOALS sources appear similar --
namely an emission-line dominated soft X-ray band (below 2 keV) and a
hard X-ray tail. The relative composition of these two components is
not widely different between objects, as the {\sl HR} distribution
(Fig. 5) shows.  Objects that have a spectrum of this type always show
totally different X-ray morphologies in the soft and hard X-ray bands
(see the images included as online material), suggesting distinct
origins for hard and soft X-rays.  Therefore, we will discuss the soft
and hard band spectra separately below\footnote{It is generally
  possible to reproduce a Chandra full-band spectrum for those objects
  whose spectra can be represented as the sum of a power-law and a
  thermal emission spectrum with sub-keV temperature and solar
  abundance. A good fit can often be found with the thermal spectrum
  accounting only for the emission-line bump around 1 keV, leaving not
  only the hard band ($\geq 2$ keV) emission but also the softest
  emission below $\sim 0.7$ keV to the power-law. This spectral
  decomposition would be invalid when the imaging data show a totally
  different spatial distribution between these two bands (as seen in
  our sample objects), because they cannot come from the same
  component represented by the power-law (see Appendix B).}.

As mentioned above, the hard X-ray emission is generally attributed to
HMXBs, {\it but this may not always be true for our objects}.  After
excluding objects with clear AGN signatures, the hard band spectra of
the remaining objects have relatively low signal to noise mainly due
to the relative weakness of the source, hence they have been referred
to (see \S 4.3) as HXQ galaxies  in Iwasawa et al. (2009).  Any spectral
feature like an Fe~K line is not evident in individual noisy
spectra. However, the stacked spectrum of the 29 HXQ galaxies in our
C-GOALS sample shows a strong high-ionization Fe K line (Fe XXV) at
6.7 keV with EW of $\simeq 0.9$ keV (Iwasawa et al. 2009). This means
that a non negligible fraction of the HXQ galaxies have strong Fe XXV
emission in their hard band spectra, but they are buried in the noise
in the individual spectra. Such a strong Fe XXV is not seen in the
spectrum of HMXBs in our Galaxy, except when they are in an eclipse
phase -- they normally show a much weaker Fe line at 6.4~keV
(e.g. White et al. 1983, Torrejon et al. 2010). This means that, {\it unlike
for local star-froming galaxies, HMXBs are not the primary source of
the hard X-ray band emission seen in the more luminous C-GOALS
objects}.  Also the 2-10 keV luminosity of the HXQ galaxies deviate
from the extrapolated correlation line on the X-ray quiet side ($\sim
0.7$ dex on average, Iwasawa et al. 2009). Most objects used to derive
the X-ray vs. FIR correlation (e.g., Ranalli et al. 2003) have $L_{\rm
  ir} < 10^{11}L_{\odot}$, while the lower bound of the IR luminosity
of our sample is $10^{11.73}L_{\odot}$.  {\it This suggests that a
transition in the nature of the dominant hard X-ray source in LIRGs
occurs somewhere in the range log $(L_{\rm ir}/L_\odot) = $11.0-11.73} .

%The soft X-ray emission below 2 keV is likely to be thermal gas which
%is spatially extended. 
In the soft X-ray band, at the spectral resolution of a CCD
spectrometer like the ACIS, heavy blending of emission lines makes it
difficult to distinguish between photoionized gas irradiated by an AGN
and thermal gas induced from a starburst. However, it is genrally
assumed that thermal emission is responsible for the emission-line
dominated spectra of LIRGs in the soft X-ray band, and this appears
even to be true for an object like NGC 6240 in which AGN signatures
appear only in the hard band (Netzer et al. 2005). Thus we also utilize
the thermal emission spectrum to compare with the soft X-ray
spectra. However, we still suspect that there might be a possible
contribution from photoionized gas to the soft X-ray spectra of some
objects, as will be discussed below.

In applying a thermal emission model, one problem specific to our
objects is the assumed solar abundance. The hot gas induced by a
starburst is expected to be polluted heavily by ejecta of core
collapse supernovae, which is rich in $\alpha $ elements relative to
iron. Therefore the abundance pattern should deviate significantly
from the solar pattern, and this has been found to be the case for
starburst knots in nearby galaxies, e.g., The Antennae (Fabbiano et al.
2004). When the data quality is good, a standard thermal emission
spectrum with a solar abundance pattern as computed by, e.g., MEKAL,
indeed does not agree with some of the observed emission line
strengths, and a modified abundance pattern, rich in $\alpha $
elements, gives a better description. This may not be evident for poor
quality spectra. Since the data quality between our spectra varies, we
first measure a spectral slope and then fit the thermal emission model
with the abundance pattern of core collapse SNe for all of the soft
X-ray data. These spectral fits were performed to the count rate
spectra by comparing the model folded through the detector response,
using XSPEC (ver. 11).

\subsubsection{Spectral slopes} 

% SX and HX slopes

As a first order characterization of the spectral shape, spectral
slopes in the soft (0.8-2 keV) and hard (3-7 keV) bands, derived from
power-law fitting are given in Table 8. The photon index, $\Gamma$, is the
slope parameter of a power-law model for describing a photon spectrum,
defined as $dN/dE\propto E^{-\Gamma }$ photons cm$^{-2}$ s$^{-1}$
keV$^{-1}$, and is related to the energy index $\alpha $ ($F_{\rm
  E}\propto E^{-\alpha}$) with $\Gamma = \alpha +1$. In order to
obtain meaningful constraints, the slopes were determined for those
spectral data which have more than 50~cts in the energy range of
interest. In fitting the power-law, Galactic absorption is assumed.

In most objects, the soft X-ray emission is dominated by (largely
unresolved) emission lines, and the spectrum turns over below 0.8 keV 
where the Fe~L complex becomes weak. The power-law fits for the soft X-ray
spectra are therefore mainly for the emission-line blend, not the
underlying continuum.

% Fig 7 Slope dist
\begin{figure}
\begin{center}
  \centerline{\includegraphics[width=0.4\textwidth,angle=0]{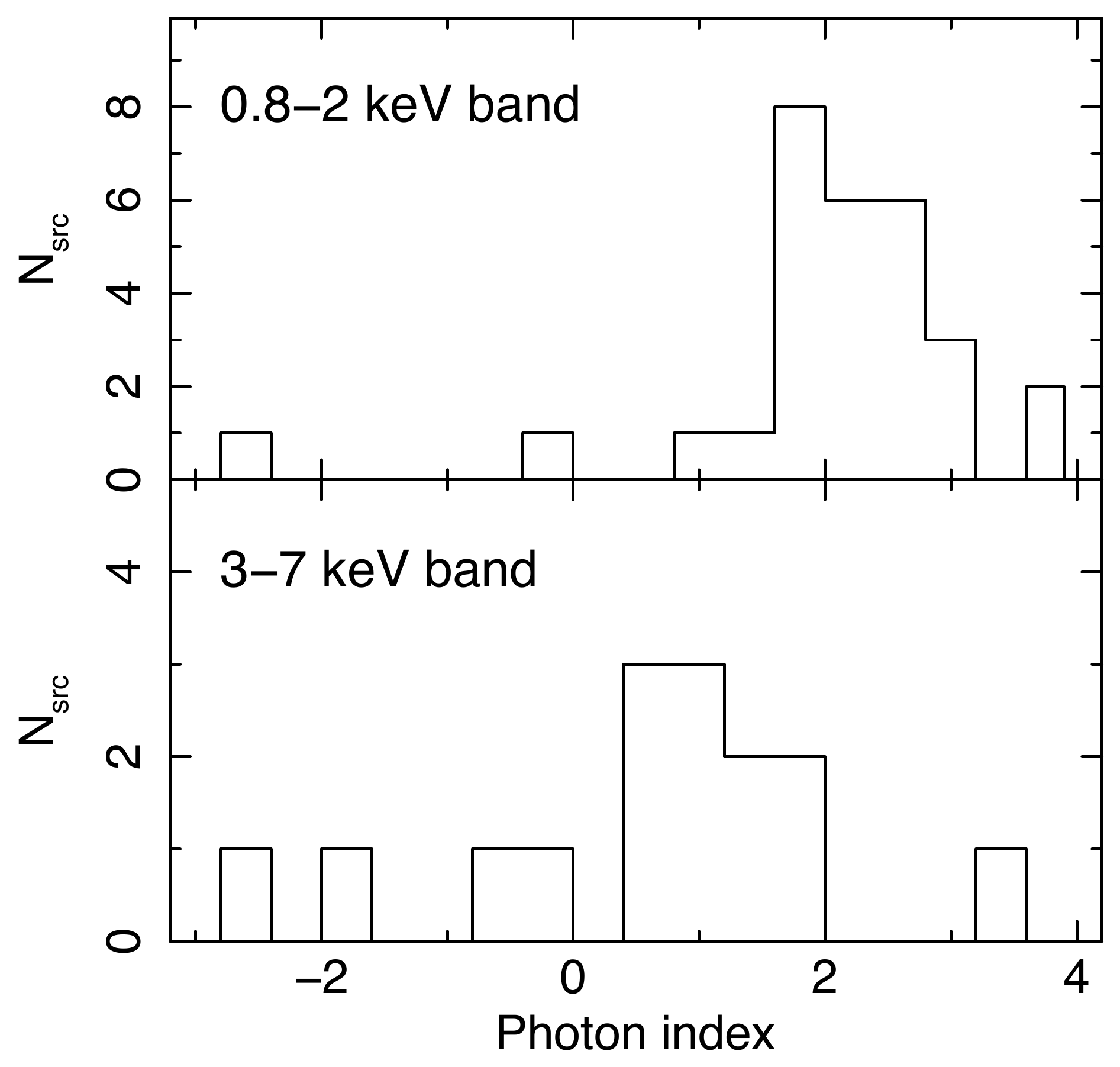}}
  \caption{Distribution of spectral slope (photon index) measured by a
    power-law fit in the 0.8-2 keV and 3-7 keV bands. A correction for Galactic
    absorption has been made.}
\end{center}
\end{figure}

\subsubsection{Prominent emission lines}

% Si Fe lines
The Fe~K line is a well known diagnostic of heavily obscured AGN, and
it is clearly seen in some spectra of our sample galaxies. Another
prominent spectral line we observe is Si K at $\sim 1.8$ keV. The
strongest line is usually from Si XIII at 1.85 keV. This line is
relatively isolated in the soft X-ray range so that the line
properties are easier to measure than is the case for other emission
lines. It could also provide a clue to the origin of the X-ray
emission.

When a line is detected at $> 2\sigma $ after fitting with a Gaussian,
its line flux and equivalent width with respect to the neighboring
continuum are measured.  Tables 9 and 10 lists measured values for the Si and Fe lines,
respectively.  Fe~K features are found at 6.4~keV (line emission from
cold Fe) and/or at higher energies (Fe XXV or Fe XXVI). The detection
of these Fe~K lines have been reported previously, based on
XMM-Newton, ASCA, and Chandra observations (Imanishi et al. 2003, Xia
et al. 2002, Armus et al. 2009, Komossa et al. 2003; Della ceca et al.
2002, Ballo et al. 2004; Clements et al. 2002, Iwasawa et al. 2005;
Iwasawa \& Comastri 1998; Turner \& Kraemer 2003, Braito et al.
2004). One addition is F17207--0014 for which a possible
high-ionization Fe~K line is detected ($2\sigma$) for the first time.

% SNII fits

% Fig 8 kT dist

\begin{figure}
\begin{center}
  \centerline{\includegraphics[width=0.4\textwidth,angle=0]{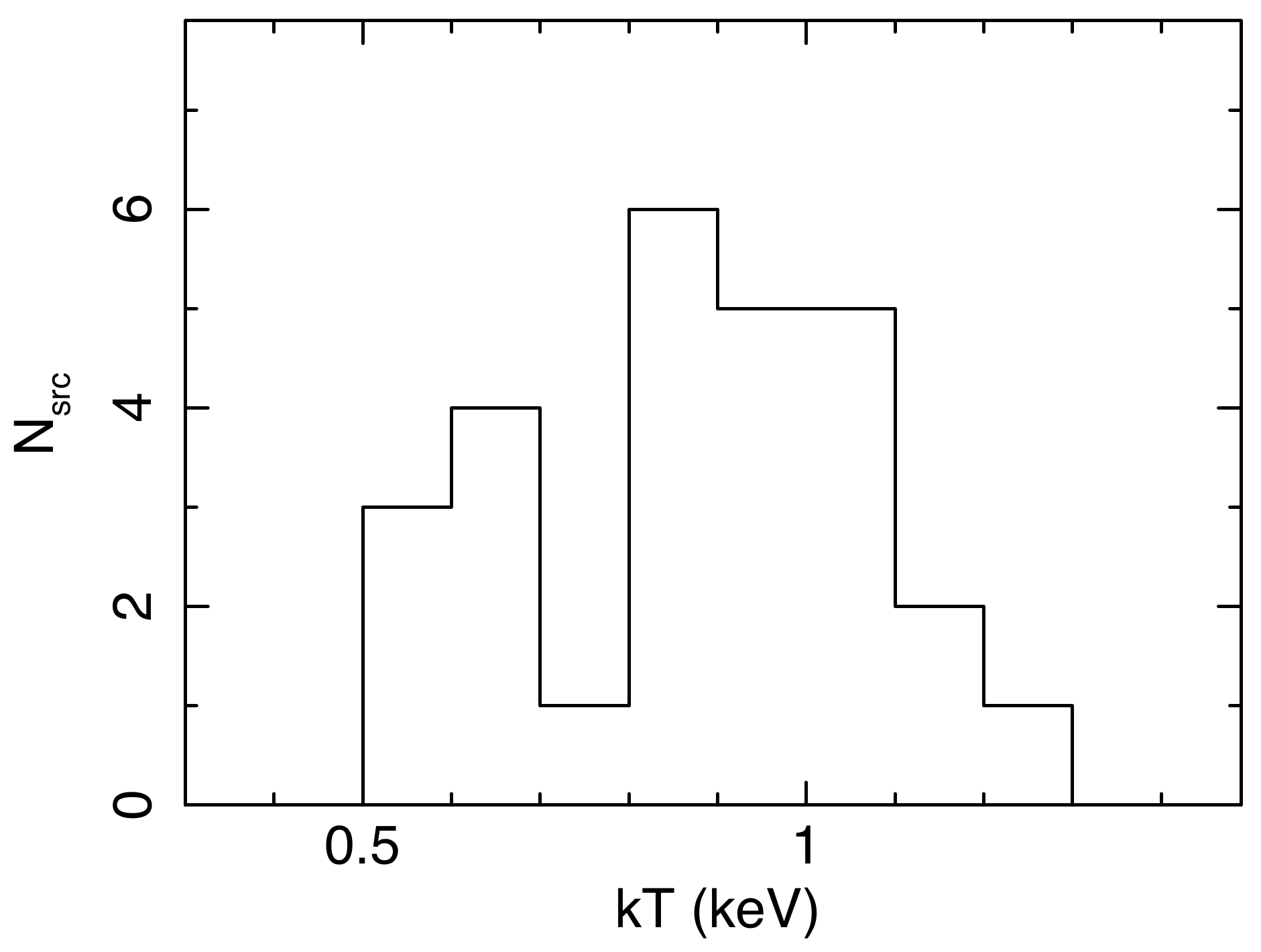}}
  \caption{The distribution of temperature, $kT$, when the thermal
    emission spectrum is fitted to the 0.4-2 keV data (see Table
    11). The median value is $kT = 0.88$ keV. }
%Fitting a Gaussian to
%    the $kT$ ditribution gives a peak value of $kT = 0.90$ keV with
%    dispersion of 0.22 keV. }
\end{center}
\end{figure}

\subsubsection{Metal abundance pattern}

As discussed above, the soft X-ray spectrum may be more complicated
than a standard thermal emission spectrum with a solar metal-abundance
pattern. For a more physically motivated characterization of the soft
X-ray emission than given by a simple power-law fit, we adopt a
thermal emission model with a Type II SN abundance pattern (e.g.,
Nomoto et al. 1997) to fit the 0.4-2 keV spectra. The nominal
metallicity ratio relative to Oxygen for various metals is assumed (as
given in Dupke \& Arnaud 2001) and varies together in the fit: (Mg,
Si)/O = 1, (Ne, S)/O = 0.67, (Ar, Ca, Ni)/O = 0.46, Fe/O = 0.27. We
note that this abandance pattern is one of those calculated for
core-collapse SNe and may not be very accurate, but it deviates
significantly from the solar pattern and will provide a good reference
against the observed data. The metallicity measurement is primarily
driven by the strength of the Fe~L complex around 1 keV at the
temperature assumed for the gas. The metallicity results are given
for the O abundance relative to solar ($Z_{\rm SNII}$ in Table 11) in
the interest of $\alpha $ elements, i.e., $Z_{\rm SNII}$ is an
``equivalent'' Oxygen abundance, which is mainly determined by the Fe
L feature and converted by the above Fe/O ratio, since given the
quality of the  data, the Oxygen line (e.g., OVIII at 0.65 keV) itself
cannot give a strong constraint. Here the solar abundance is that of
Anders and Grevess (1989).
Emission lines of He-like Mg and Si are located at
energies where the ACIS response is good, and in some qood quality
data their metallicity can be measured independently (when they are
strong). When this is the case, their metallicity measurements are
given separately. In this case, Si and Mg are assumed to have the same
abundance.

This prescription described above usually gives better agreement with
the data than that achieved with a solar abundance pattern, as
demonstrated for the spectrum of ESO 286-IG19 in Fig. 12. When a solar
abundance pattern is assumed, the metallicity is found to be only
$0.21^{+0.05}_{-0.03} Z_{\odot}$, and clearly underestimates
emission-lines such as Si XIII, which would be better described with
$\sim 2 Z_{\odot}$ (Table 11). The temperature measurement is also
affected by the choice of abundance pattern. In the example of ESO
286-IG19, the temperature derived from the solar abundance pattern is
$kT = 0.82\pm 0.03$, while the SNII pattern gives $kT = 0.70\pm 0.04$
keV.

The bright X-ray core of Mrk 231 is dominated by the unresolved
nuclear component even in the soft X-ray band. Given that a single thermal
emission model is not able to give a reasonable fit, the spectral
parameters are not presented in Table 11.

%caveat
The argument for a non-solar abundance pattern has so far been based
on a single-temperature fit. One caveat is that a multi-temperature
model could mimic the same effect (e.g., Buote \& Fabian 1998) in 
some cases, since it is difficult to distinguish between the two
interpretations with spectral data at the CCD resolution of the ACIS.
High-resolution spectroscopy that could identify individual lines could 
be used to determine the characteristic temperature of the gas. This would tell whether
the gas is in a multi-phase. At least for the very extended nebulae,
as seen in Mrk 231 and Mrk 273, which emit almost no emission above 2
keV but strong Mg XI and Si XIII, higher temperature gas is unlikely
to be present. We note $\alpha $-element rich gas would be produced
naturally if a starburst is responsible for the metal enrichment.

\subsubsection{Strong Si XIII emission lines and their origin}

% Diagnostics for AGN?

Even when assuming the $\alpha $-element rich abundance pattern of
core-collapse SNe, the observed Si metallicity sometimes requires an
even larger value (see Table 11). This may simply be due to more Si
enriched gas than theoretically predicted. However, the excess Si line
intensity could also be due to an additional source of the line
emission, e.g., photoionized gas illuminated by a hidden AGN. In fact,
among the objects listed in Table 11, all of the AGN that are identified
by spectral hardness and/or the Fe~K diagnostics show at least an
excess of $\times 2$ in the Si metallicity relative to $Z_{\rm
  SNII}$. Since the energies around Si XIII (1.84 keV) are where the
spectrum of the soft X-ray emitting thermal gas starts to decline
rapidly, an extra contribution from any AGN related component would stand
out relatively well.

Here, we examine the Si line strengths of starbursts and AGN in an
alternative way by using the Si XIII detected objects in Table 9. Since
the metallicity, $Z_{\rm SNII}$,  is driven by the Fe L emission (\S
5.1.3), the enhancement of $Z_{\rm Si}$ is basically the relative
strength of the Si XIII line to the Fe L bump around 0.8-1 keV. As the
heavily blended Fe L emission is not resolved at the CCD spectral
resolution, the mean 0.8-1 keV intensity is used as a crude measure of
Fe L emission, assuming that Fe L dominates in this band. Thus, the
mean Si XIII / Fe L ratio can be assessed between starburst and
AGN. 

The starburst sample consists of 7 objects, F10565+2448, NGC 3690 E,
Arp 220, F17207--0014, F18293--3413, CGCG 448-020 and ESO
286-IG19. Note that objects with mid-IR [Ne V] detection are not included (see
below). The AGN sample consists of 5 known Compton-thick AGN, UGC 5101, Mrk
231, 13120--5453, VV 340 N and NGC 6240.

When  making a mean spectrum of each sample, a redshift correction 
was made due to the fact that the energy-scale shifts between objects 
are not negligible. The spectral data of individual objects are chosen
so that the rest-frame 0.50-2.18 keV range is covered and then they
are binned at (rest-frame) 21 eV intervals. These spectra are
corrected for the effective area in the same way as those spectra in
Fig. 4 (\S 4.2), which is necessary when the redshift correction for
the energy scale is made. Since the soft X-ray band is sensitive even
to variations in Galactic column ($N_{\rm H}$ ranges between $9\times
10^{19}$ cm$^{-2}$ and $2.1\times 10^{21}$ cm$^{-2}$ for the relevant
objects here), an absorption correction for the Galactic column is also
made.  Individual spectra are then normalised to the 0.8-1 keV
intensity before computing an average from the individual members of each
sample.

The mean 0.5-2.2 keV spectra of the starburst and AGN samples are
shown in Fig. 13a and 13b, where the 0.8-1 keV intensity of both spectra 
has been set to the same level.  The Si line strength relative
to the Fe L emission can be readily compared. The Si line of the mean
AGN spectrum appears to be stronger than that of the starburst by a
factor of $2.0\pm 0.6$. Albeit that this is a relatively crude measure, the result is
consistent with an enhanced Si line in AGN spectra, as suggested by the
metallicity fitting (Table 11).

Furthermore, an additional mean spectrum was constructed using the same
procedure, for five objects which are not selected as AGN by the
X-ray criteria, but which show a detectable mid-IR [Ne V] line in their
Spitzer IRS spectra (Petric et al. 2010): VII Zw 31, F09111--1007, UGC
8387, F17132+5313 and ESO 593-IG8 (see Table 5). The mean spectrum of
these Ne V detected objects also suggests an enhancement of Si XIII;  
although the data are still noisy, the enhancement relative to the
starburst sample is a factor of $1.8\pm 0.9$.

While it is premature to propose such enhanced Si XIII as an
alternative to the Fe K diagnostic for a hidden AGN, it is interesting
to note that both X-ray selected Compton thick AGN and mid-IR selected
AGN candidates appear to show strong Si XIII on average, even when the data
quality is not good in the Fe K band. A contribution from an  AGN
photoionized component in addition to a thermal component can be a
feasible explanation, unless the photoionized gas region is
affected by internal absorption.

% ESO 286 spectra with MEKAL and SNII 

\begin{figure*}
\begin{center}
  \hbox{\centerline{\includegraphics[width=0.38\textwidth,angle=0]{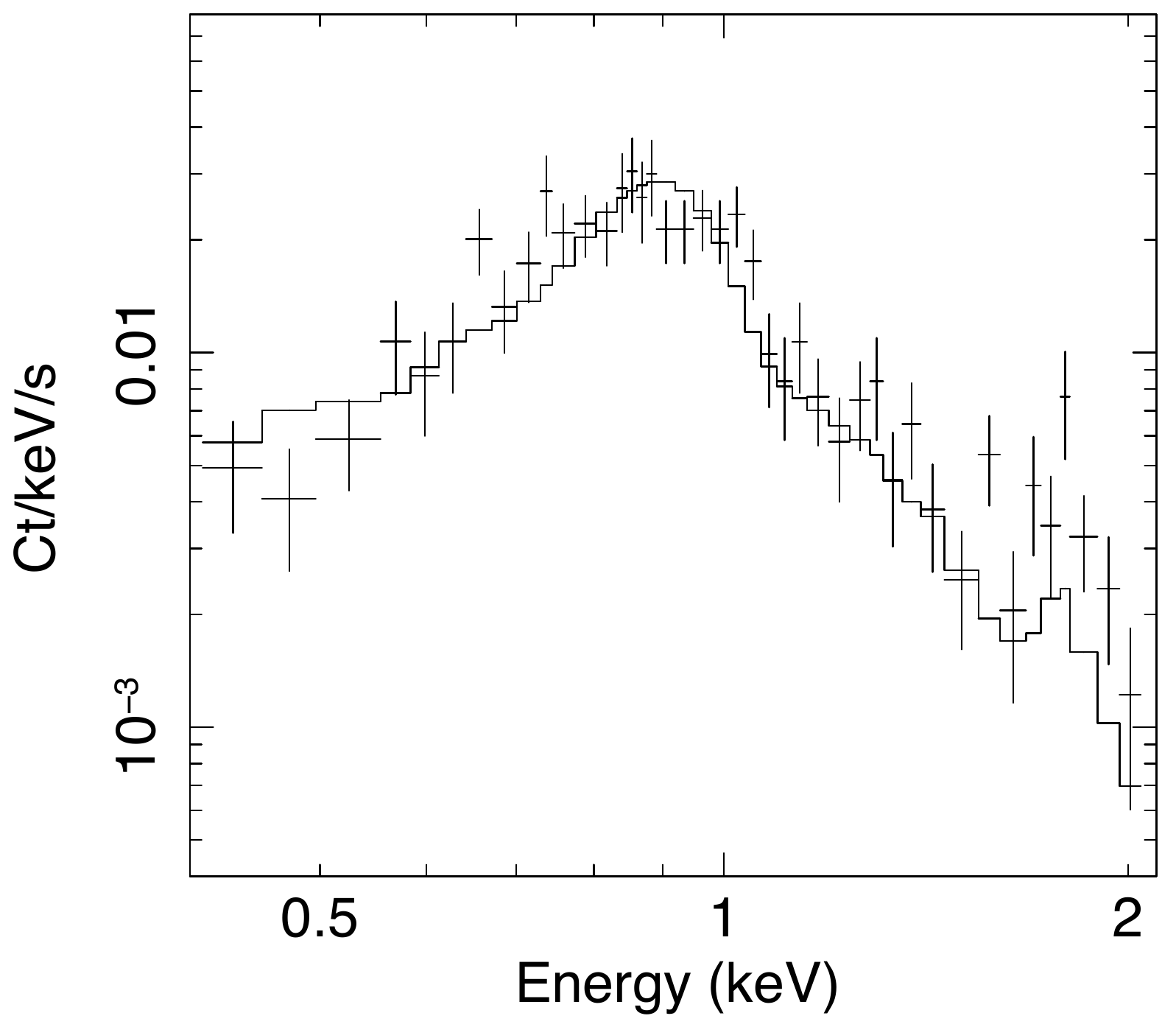}
\includegraphics[width=0.38\textwidth,angle=0]{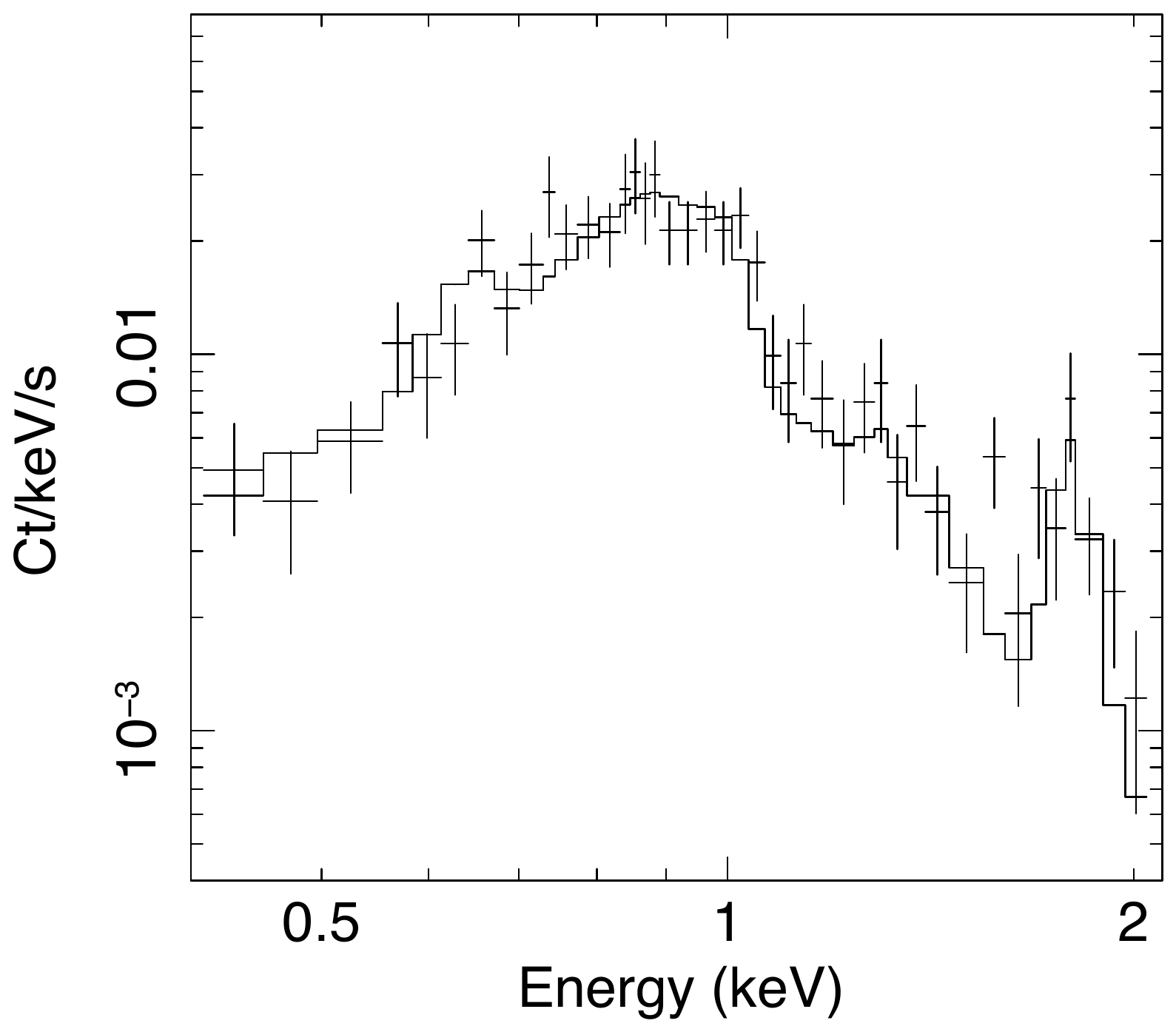}}}
\caption{The Chandra ACIS 0.4-2 keV spectrum of ESO 286-IG19, fitted
  with a thermal emission spectrum (MEKAL) with a solar abundance
  pattern (left) and that of the core-collapse SNe (right), in which
  OVIII (0.65 keV), NeIX (1.0 keV), MgXI (1.36 keV) and SiXIII (1.85
  keV) are more pronounced. The $\chi^2$ values for the two fits are
  53 and 37 for 41 and 40 degrees of freedom, respectively. }
\end{center}
\end{figure*}

% Table 4  -- spectral slopes

\begin{table}
\begin{center}
  \caption{X-ray spectral slopes.$^{\mathrm{a}}$}
% VV340 hx 28 ct
\begin{tabular}{rlcc}
No. & Galaxy & $\Gamma_{\rm S}$ & $\Gamma_{\rm H}$ \\[5pt]
\multicolumn{4}{c}{\bf Cycle 8 targets} \\
3 & VII Zw 31 & $2.7^{+0.4}_{-0.4}$ & --- \\
5 & ESO 255-IG7 N & $2.0^{+0.7}_{-0.6}$ & --- \\
5 & ESO 255-IG7 C & $2.0^{+0.5}_{-0.4}$ & --- \\
7 & ESO 60-IG16 & $2.1^{+1.4}_{-1.3}$ & $0.7^{+0.3}_{-0.8}$ \\
9 & 09022-3615 & $2.3^{+0.6}_{-0.6}$ & $-0.1^{+0.8}_{-0.3}$ \\
10 & F09111-1007 E & $3.1^{+0.9}_{-0.8}$ & --- \\
19 & 13120-5453 & $1.0^{+0.4}_{-0.4}$ & $2.6^{+1.5}_{-0.9}$ \\
20 & VV 250 E & $1.7^{+0.4}_{-0.4}$ & $3.6^{+0.9}_{-0.8}$ \\
21 & UGC 8387 & $2.5^{+0.4}_{-0.4}$ & --- \\
25 & VV 340 N & $3.8^{+0.4}_{-0.4}$ & $-2.5^{+1.7}_{-0.5}$ \\
29 & ESO 69-IG6 N & $3.0^{+0.4}_{-0.4}$ & --- \\
31 & F17132+5313 & $2.4^{+1.4}_{-1.1}$ & --- \\
33 & F18293-3413 & $2.7^{+0.4}_{-0.3}$ & $2.0^{+0.4}_{-1.0}$ \\
34 & ESO 593-IG8 & $2.2^{+0.6}_{-0.6}$ & --- \\
35 & F19297-0406 & $3.6^{+1.5}_{-1.4}$ & --- \\
36 & 19542+1110 & $\sim -2.8$ & $1.3^{+0.2}_{-0.5}$ \\
37 & CGCG 448-020 & $2.3^{+0.3}_{-0.3}$ & $1.7^{+0.5}_{-0.4}$ \\
40 & ESO 239-IG2 & $2.7^{+0.4}_{-0.4}$ & --- \\[5pt]
\multicolumn{4}{c}{\bf Archival data} \\
4 & F05189-2524 & $\sim 0$ & $0.48^{+0.20}_{-0.08}$ \\
12 & UGC 5101 & $2.0^{+0.3}_{-0.3}$ & $\sim -0.7$ \\
15 & F10565+2448 & $2.5^{+0.3}_{-0.3}$ & --- \\
16 & NGC 3690 E & $1.9^{+0.2}_{-0.2}$ & $1.0^{+0.7}_{-0.7}$ \\
16 & NGC 3690 W & $1.9^{+0.2}_{-0.2}$ & $1.1^{+0.5}_{-0.5}$ \\
18 & UGC 8058 core & $1.9^{+0.2}_{-0.1}$ & $0.71^{+0.15}_{-0.06}$ \\
18 & UGC 8058 nebula & $5.2^{+0.3}_{-0.3}$ & --- \\
22 & UGC 8696 & $2.8^{+0.2}_{-0.2}$ & $-1.8^{+0.1}_{-0.3}$ \\
26 & VV 705 N & $2.4^{+0.6}_{-0.5}$ & --- \\
28 & UGC 9913 & $1.8^{+0.2}_{-0.2}$ & $0.8^{+0.5}_{-0.3}$ \\
32 & F17207--0014 & $2.0^{+0.3}_{-0.3}$ & $0.7^{+0.9}_{-0.4}$ \\
38 & ESO 286-IG19 & $3.8^{+0.2}_{-0.2}$ & $0.7^{+1.0}_{-2.0}$ \\
42 & ESO 148-IG2 N & $3.0^{+0.5}_{-0.4}$ & --- \\
42 & ESO 148-IG2 S & $1.3^{+0.3}_{-0.3}$ & $1.4^{+0.3}_{-0.4}$ \\
\end{tabular}
\begin{list}{}{}
\item[$^{\mathrm{a}}$] Photon indices derived by fitting a power-law to the 0.8-2 keV
  ($\Gamma_{\rm S}$)  and 3-7 keV ($\Gamma_{\rm H}$) bands after correcting for Galactic
  absorption.   Measurements are given only when the
  detected source counts exceed 50~cts in the respective bands in order
  for the spectral slope to have meaningful constraints.
\end{list}
\end{center}
\end{table}

% Table 8 -- Si line

\begin{table}
\begin{center}
  \caption{Si {\sc xiii} (1.84 keV) measurements. $^{\mathrm{a}}$}
\begin{tabular}{rlccc}
No. & Galaxy & $I_{\rm Si}$ & $EW_{\rm Si}$ & Note\\
& & $10^{-6}$ ph\thinspace cm$^{-2}$\thinspace s$^{-1}$ & keV & \\[5pt]
10 & F09111--1007 E & $0.74^{+4.5}_{-3.7}$ & 0.44 &\\
12 & UGC 5101 & $0.66^{+0.22}_{-0.30}$ & 0.23 &\\
15 & F10565+2448 & $6.1^{+3.4}_{-2.8}$ & 0.17 &\\
16 & NGC 3690 E & $2.3^{+1.1}_{-1.1}$ & 0.15 &\\
18 & UGC 8058 core & $0.60^{+0.26}_{-0.27}$ & 0.05 &\\
18 & UGC 8058 nebula & $0.30^{+0.15}_{-0.14}$ & 0.39 &\\
19 & 13120--5453  & $2.1^{+0.8}_{-0.8}$ & 0.20 &\\
21 & UGC 8387 & $1.5^{+0.7}_{-0.7}$ & 0.26 &\\
25 & VV 340 N & $1.1^{+0.4}_{-0.5}$ & 0.26 &\\
28 & UGC 9913 & $1.3^{+1.5}_{-0.5}$ & 0.23 & ${\mathrm b}$\\
29 & ESO 69-IG6 N & $1.3^{+0.6}_{-0.6}$ & 0.34 &\\
30 & NGC 6240 & $20.4^{+3.5}_{-4.0}$ & 0.30 & ${\mathrm b}$\\
32 & F17207--0014 & $0.50^{+0.42}_{-0.20}$ & 0.14 &\\
33 & F18293--3413 & $4.0^{+0.8}_{-1.0}$ & 0.42 &\\
34 & ESO 593-IG8 & $0.85^{+0.54}_{-0.43}$ & 0.26 &\\
37 & CGCG 448-020 & $1.1^{+0.7}_{-0.5}$ & 0.16 & ${\mathrm c}$\\
38 & ESO 286-IG19 & $0.59^{+0.28}_{-0.26}$ & 0.16 &\\
\end{tabular}
\end{center}
\begin{list}{}{}
\item[$^{\mathrm a}$] Detections with $2\sigma$ or higher significance are listed. 
\item[$^{\mathrm b}$] This line intensity is of a blend of Si~{\sc xiii} and Si~{\sc xiv} 
  and is measured by fitting a broad Gaussian.
\item[$^{\mathrm c}$] This emission-line is not found at the expected
  energy of Si {\sc xiii},  but at $1.60\pm 0.06$ keV.
\end{list}
\end{table}

% Table 9 -- Fe line

\begin{table}
\begin{center}
  \caption{Fe K line features.$^{\mathrm{a}}$}
\begin{tabular}{rlccc}
No. & Galaxy & $E$ & $I_{\rm FeK}$ & $EW_{\rm FeK}$ \\
& & keV & $10^{-6}$ ph s$^{-1}$ cm$^{-2}$ & keV \\[5pt]
\multicolumn{5}{c}{\bf Cold line} \\
4 & F05189--2524 & 6.4 & $6.0^{+4.0}_{-2.8}$ & 0.12 \\
12 & UGC 5101 & 6.4 & $0.68^{+0.55}_{-0.31}$ & 0.30 \\
18 & UGC 8058 & 6.4 & $1.0^{+0.3}_{-0.4}$ & 0.10 \\
22 & UGC 8696 & 6.4 & $6.7^{+2.7}_{-2.4}$ & 0.24 \\
25 & VV 340 N & 6.4 & $1.8^{+1.1}_{-0.9}$ & 1.2 \\
30 & NGC 6240 & 6.4 & $15.4^{+2.0}_{2.0}$ & 0.40 \\[5pt]
\multicolumn{5}{c}{\bf High-ionzation line} \\
16 & NGC 3690 E & $6.65^{+0.03}_{-0.02}$ & $7.4^{+2.1}_{-2.5}$ & --- \\
28 & UGC 9913 & $6.64^{+0.03}_{-0.03}$ & $0.92^{+0.51}_{-0.32}$ & 0.93 \\
30 & NGC 6240 & $6.65^{+0.02}_{-0.02}$ & $8.0^{+2.0}_{-1.6}$ & 0.16 \\
32 & F17207--0014 & $6.90^{+0.05}_{-0.05}$ & $0.82^{+0.73}_{-0.41}$ & 0.82 \\
\end{tabular}
\begin{list}{}{}
\item[$^{\mathrm a}$] Line detections with $2\sigma$ or higher significance are
  listed. The cold Fe~K$\alpha$ at 6.4~keV and the high-ionization 
  Fe~K$\alpha$, Fe ~{\sc xxv} or Fe~{\sc xxvi} are listed separately. The
  line centroid energy is measured in the rest frame in keV. 
\end{list}
\end{center}
\end{table}

% Temperature from snii fit

\begin{table}
\setlength{\tabcolsep}{0.05in}
\begin{center}
  \caption{Temperature, absorption and metallicity.$^{\mathrm{a}}$}
\begin{tabular}{rlcccc}
No. & Galaxy & $kT$ & $N_{\rm H}$ & $Z_{\rm SNII}$ & $Z_{\rm Si}$ \\
& & keV & $10^{21} cm^{-2}$ & $Z_{\odot}$ & $Z_{\odot}$ \\[5pt] 
\multicolumn{6}{c}{\bf Cycle 8 targets} \\
3 & VII Zw 31 & $1.0^{+0.1}_{-0.1}$ & --- & $4_{-2}$ & $8^{+20}_{-5}$ \\
5 & ESO 255-IG7 N & $1.8^{+0.8}_{-0.5}$ & --- & $0.4^{+1.6}_{-0.4}$ & --- \\
9 & 09022-3615 & $1.1^{+0.5}_{-0.2}$ & --- & $0.05^{+0.1}_{-0.05}$ & $0.4^{+0.2}_{-0.2}$ \\
10 & F09111-1007 E & $0.90^{+0.18}_{-0.13}$ & --- & $1.0^{+1.2}_{-0.5}$ & $4^{+7}_{-2}$ \\
13 & ESO 374-IG32 & $0.68^{+0.15}_{-0.12}$ & --- & $0.6^{+3}_{-0.4}$ & $1.7^{+3.7}_{-1.2}$ \\
19 & 13120-5453 & $0.82^{+0.26}_{-0.14}$ & --- & $0.7^{+0.8}_{-0.3}$ & --- \\
20 & VV 250 & $1.0^{+0.1}_{-0.1}$ & --- & $0.7^{+0.5}_{-0.2}$ & $2.6^{+1.9}_{-0.7}$ \\
21 & UGC 8387 & $1.0^{+0.1}_{-0.1}$ & --- & $2.2^{+2.3}_{-0.7}$ & $7^{+9}_{-3}$ \\
25 & VV 340 N & $0.83^{+0.08}_{-0.07}$ & --- & $1.3^{+0.8}_{-0.4}$ & $3.0^{+2.6}_{-1.1}$ \\
29 & ESO 69-IG6 N & $0.88^{+0.10}_{-0.10}$ & --- & $4.5^{+20}_{-1.9}$ & $10^{+18}_{-6}$ \\
31 & F17132+5313 & $0.59^{+0.10}_{-0.10}$ & --- & $0.7_{-0.4}$ & --- \\
33 & F18293-3413 & $0.61^{+0.08}_{-0.07}$ & $7.2^{+0.8}_{-1.0}$ & $1.4^{+4.9}_{-0.7}$ & --- \\
34 & ESO 593-IG8 & $0.95^{+0.16}_{-0.11}$ & --- & $0.53^{+0.62}_{-0.29}$ & $2.1^{+1.5}_{1.0}$ \\
35 & F19297-0406 & $1.1^{+0.3}_{-0.2}$ & ---  & $5_{-4}$ & --- \\
37 & CGCG 448-020 & $0.66^{+0.23}_{-0.09}$ & $2.9^{+0.7}_{-1.0}$ & $0.09^{+2.3}_{-0.09}$ &--- \\
40 & ESO 239-IG2 & $0.76^{+0.07}_{-0.06}$ & --- & $1.2^{+5}_{-0.6}$ & $3.5^{+10}_{-1.8}$ \\[5pt]
\multicolumn{6}{c}{\bf Archival data} \\
%ESO 148-IG2 all & $0.73^{+0.04}_{-0.04}$ & --- & $1.1^{+0.3}_{-0.2}$ & $2.7^{+1.2}_{-0.6}$ \\
12 & UGC 5101 & $0.89^{+0.08}_{-0.07}$ & --- & $0.9^{+0.2}_{-0.2}$ & $4.7^{+1.5}_{-1.4}$ \\
15 & F10565+2448 & $1.0^{+0.1}_{-0.1}$ & ---  & $1.1^{+0.5}_{-0.3}$ & $3.0^{+1.6}_{-1.0}$ \\
16 & NGC 3690  & $0.66^{+0.04}_{-0.04}$ & $2.4^{+0.3}_{-0.3}$ & $0.30^{+0.05}_{-0.05}$ & $0.43^{+0.10}_{-0.10}$ \\
18 & UGC 8058 neb & $0.50^{+0.03}_{-0.03}$ & --- & $0.31^{+0.06}_{-0.06}$ & $0.96^{+0.43}_{-0.33}$ \\
22 & UGC 8696 & $0.88^{+0.04}_{-0.04}$ & --- & $0.52^{+0.08}_{-0.07}$ & $1.1^{+0.2}_{0.2}$ \\
22 & UGC 8696 neb & $0.58^{+0.04}_{-0.04}$ & --- & $1.3^{+0.8}_{-0.3}$ & $1.7^{+2.2}_{0.8}$ \\
23 & F14348--1447 & $0.92^{+0.19}_{-0.15}$ & --- & $1.2^{+10}_{-0.8}$ & --- \\
26 & VV 705 N & $0.94^{+0.12}_{-0.09}$ & --- & $2.8^{+3.7}_{-1.7}$ & $8^{+9}_{-4}$ \\
28 & UGC 9913 & $0.90^{+0.07}_{-0.06}$ & $2.9^{+0.4}_{-0.3}$ & $0.37^{+0.12}_{-0.10}$ & $0.70^{+0.28}_{-0.21}$ \\
30 & NGC 6240 & $0.91^{+0.03}_{-0.03}$ & $3.0^{+0.x}_{-0.x}$ & $0.24^{+0.03}_{-0.03}$ & $0.54^{+0.06}_{-0.05}$ \\
32 & F17207--0014 & $0.59^{+0.12}_{-0.10}$ & $5.2^{+1.0}_{-1.0}$ & $0.30^{+0.32}_{-0.15}$ & $0.59^{+0.25}_{-0.23}$ \\
38 & ESO 286-IG19 & $0.70^{+0.04}_{-0.04}$ & --- & $1.0^{+0.3}_{-0.3}$ & $2.0^{+0.6}_{-0.5}$ \\
42 & ESO 148-IG2 N & $0.84^{+0.13}_{-0.10}$ & --- & $2.7^{+10}_{-1.3}$ & $6^{+8}_{-3}$ \\
42 & ESO 148-IG2 S & $1.01^{+0.18}_{-0.18}$ & --- & $0.73^{+0.42}_{-0.27}$ & $3.2^{+1.8}_{-1.0}$ \\
\end{tabular}
\begin{list}{}{}
\item[$^{\mathrm{a}}$] The 0.4-2 keV data are fitted with a thermal
  emission spectrum assuming the metal abundance pattern of
  core-collapse supernovae (see text for details). When excess
  absorption above the Galactic value is required, the best-fit column
  density ($N_{\rm H}$) is listed. $Z_{\rm SNII}$ is the metallicity
  for Oxygen, which is tied to the other elements with the assumed
  abundance pattern of Type II SNe. This value is mainly driven by the
  Fe metallicity, which is assumed to be 0.27 times that of the Oxygen
  metallicity, and which fits the Fe~L hump around 1~keV. When excess
  metallicity for Si (and Mg) is required, the Silicon metallicity is
  fitted independently ($Z_{\rm Si}$).
\end{list}
\end{center}
\end{table}

\begin{figure}
\begin{center}
\includegraphics[width=0.37\textwidth,angle=0]{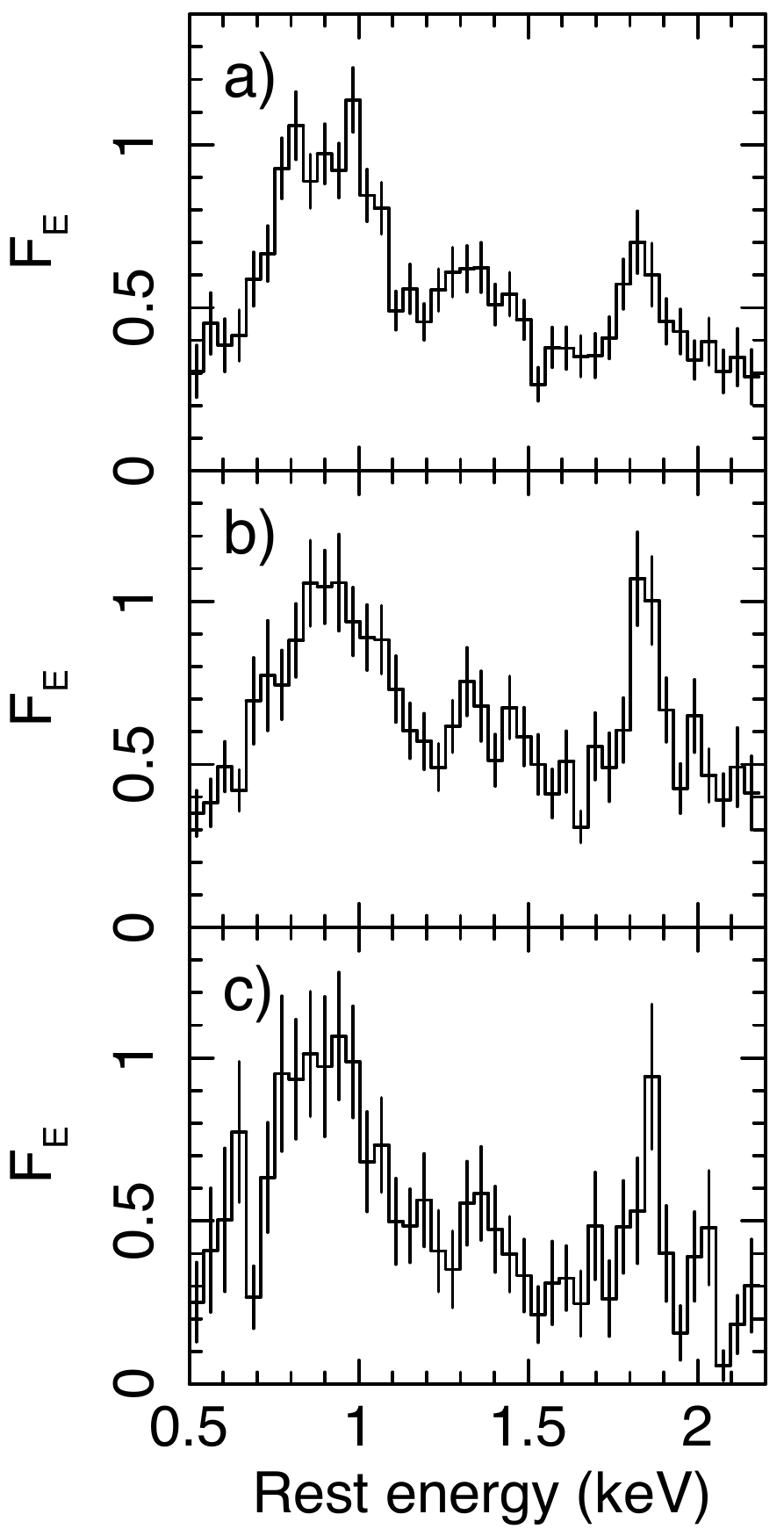}
\caption{The mean 0.5-2.2 keV spectra of a) seven starbursts; b) five
  Compton-thick AGN; and c) five non X-ray AGN but with [Ne V]$\lambda
  14.3\thinspace \mu$m detection. The energy scale is in the rest
  frame. The individual spectra have been corrected for Galactic
  absorption and the detector effective area and then normalized to
  the 0.8-1 keV intensity before averaging within each sample. These
  mean spectra are constructed in order to facilitate comparison of the strength of Si
  XIII at 1.84 keV relative to the 0.8-1 keV band where Fe L emission
  dominates between the different classes of objects (see text for
  detail).}
\end{center}
\end{figure}

\subsection{Discrete X-ray emission from star clusters}

There are a few objects that show discrete X-ray sources separate from
the main body of the galaxy emission. Some of these discrete X-ray
sources have clear optical counterparts identified in our high
resolution HST-ACS images. The optical counterparts are star clusters
or dwarf galaxies in the tidal tails. In most cases, only a few X-ray
counts are detected in the soft band. Inferred X-ray luminosities are
on the order of $10^{39}$ \ergps. They are much more powerful than the
super star clusters in our Galaxy, for which the primary X-ray
production mechanism is considered to be shock heating by stellar
winds. If the X-ray emission in the C-GOALS objects comes from a
single source in a star cluster, then this could be an example of an ultra-luminous X-ray
source (ULX). Prime examples are the southern source in UGC 8387,
and the eastern triple source in CGCG 448-020 (see
Fig. 2).  

\section{Summary}

\begin{list}{}{}
\item[1.] We present Chanda-ACIS data for a complete sample of 44
  objects which represent the high luminosity portion of the GOALS
  sample. The objects in this C-GOALS sample have 
  log ($L_{\rm ir}/L_\odot) = 11.73-12.57$ with  $z = 0.010-0.088$, and
  represent the most luminous IR selected galaxies in the local
  universe. X-rays are detected from 43 out of the 44 objects, and their
  arcsec-resolution images, spectra, and radial brightness profiles
  are presented.

\item[2.] Objects with a clear AGN X-ray signature represent 37\%
  (16/44) of the total sample, and 75\% (12/16) of these AGN are found
  in the higher IR luminosity half of the sample at log $(L_{\rm
    ir}/L_\odot) > 12.0)$.  The AGN fraction would increase to 48\% if
  objects with [Ne V]$\lambda 14.3\mu $m detection are included. These
  AGN, however, appear to be a relatively minor ($\sim 10$\%)
  component in the total energy output of the host galaxies apart from
  a few exceptions.

\item[3.] While most objects show evidence for strong
  interactions/mergers, NGC 6240 remains the only object to clearly
  show a double AGN in our X-ray spectra. Eight AGN are found in
  single nucleus objects and an additional seven have X-ray AGN in one of
  the double nuclei.

\item[4.] For objects without obvious X-ray AGN signatures, X-ray
  luminosities are found to be lower than expeted from their IR
  luminosities based on the correlation found for nearby star-forming
  galaxies with lower star formation rates. The hard X-ray emission of
  these objects does not always appear to be due to HMXBs, given that
  the stacked spectrum shows strong Fe XXV, a signature of hot gas.

\item[5.] The extended soft X-ray emission found in many objects shows
  a spectrum consistent with a thermal emission spectrum with
  an abundance pattern with enhanced $\alpha $ elements relative to
  iron, as expected for an interstellar medium enriched by core collapse
  SNe produced in a starburst.

\item[6.] A comparison between the soft X-ray spectra of starburst
  galaxies and galaxies containing Compton thick AGN shows that, on
  average, the latter show stronger Si XIII emission at 1.85 keV which
  may be due to a contribution of AGN photoionized gas besides thermal
  emission from a starburst.

\end{list}

\begin{acknowledgements}
  This research has made use of software packages CIAO provided by the
  Chandra X-ray Center (CXC) and FTOOLS provided by NASAS's HEASARC.
  This research has also made use of the NASA/IPAC Extragalactic
  Database (NED), which is operated by the Jet Propulsion Laboratory,
  California Institute of Technology, under contract with the National
  Aeronautics and Space Administration. The Chandra data archive is
  maintained by the Chandra X-ray Center at the Smithsonian
  Astrophysical Observatory. DBS acknowledges support from Chandra
  grants GO6-7098X and GO7-8108A. Vivian U also acknowledges support
  from a NASA Jenkins Predoctoral Fellowship and a SAO
  Harvard-Smithsonian Predoctoral Fellowship. We thank the referee for
  helpful comments.
\end{acknowledgements}

%\end{document}

\appendix
\section{Notes on individual objects}

{\bf IRAS F01364-1042 :}\  The optical emission-line spectrum is consistent with a 
shock-heated, LINER spectral type, similar to the extended optical 
emission in Arp 220 and NGC 6240.
% (V95, $D_{\rm agn}=0.8$ in YKS10). 
The detected X-ray emission is too faint
to infer its origin. A point-like hard X-ray source is located at
the nucleus of the galaxy. The soft X-ray emission is elongated along
the SE-NW direction, possibly due to two blobs straddling the nuclear
source.

\smallskip

\noindent {\bf ESO 203-IG1 :}\ 
This is the only object in our sample from which no X-ray counts are
detected. There are two faint X-ray sources to the SSW ($15^{\prime\prime}$) and
NWW ($20^{\prime\prime}$), which have optical counter parts in the HST-ACS image
that are much fainter than ESO 203-IG1 and unrelated to the IRAS
galaxy.

\smallskip

\noindent {\bf VII Zw 31 :}\ The soft X-ray image shows a relatively
symmetric source with a slight elongation along the NW-SE
direction. There are two faint spurs extending to the NW. This is a
``typical" soft X-ray source showing an exponential radial brightness
profile.  A strong Mg X line at 1.3 keV is visible in the spectrum.
The hard X-ray emission is resolved (a compact but not point-like
source) with a weaker source at $3^{\prime\prime}$ to the West of the
compact nucleus. While this object is not met the X-ray AGN selection
criteria, the mid-IR [Ne V] line is detected (Petric et al 2010). 

\smallskip

\noindent {\bf IRAS F05189--2524 :}\  There are two Chandra observations of this
galaxy (ObsID 2034, 3432) with 20 ks and 15 ks exposures,
respectively. The Chandra results have been published in Ptak et al.
(2003) and Grimes et al. (2005). Since this is a bright X-ray source,
we only use the ObsID 3432 data, which has a similar exposure time to that of
our Cycle 8 targets.

The soft X-ray image shows a faint extensions to the NW and SW, but
is dominated by a compact source at the nucleus. The hard X-ray source is
point-like. The low surface-brightness envelope is consistent with the
wing of the PSF, which is broader than that in the soft X-rays.

The spectrum has a hard X-ray bump, typical of the absorbed emission from 
an AGN. When a photon index of 1.8 is assumed, the absorbing
column density is found to be \nH $=9\pm 1\times 10^{22}$ \psqcm, by
fitting the 3-8 keV spectrum. The absorption correction factor  
for the 2-10 keV luminosity is 1.7.   A 6.4~keV Fe~K line is detected at
the $2 \sigma $ level. 

There is a steep rise of the spectrum below 1.5 keV. 
In analogy to Seyfert 2 galaxies with similar absorption, this is
likely due to emission line blends of photoionised gas, unresolved at
the CCD resolution.

\smallskip

\noindent {\bf ESO 255-IG7 :}\ This is a triple system of linearly
aligned galaxies. All three galaxies are detected with MIPS but the
northern galaxy is much brighter than the other two (Mazzarella et al.
2011). The X-ray brightness shows a similar pattern. We tentatively
assume that the total IR luminosity is distributed among the three galaxies in the
ratio of N:C:S = 8:3:1. The northern galaxy is the brightest and has a
softer spectrum than the others. The X-ray spectrum is strangely
featureless in the soft X-ray band.
%The SDSS class is pure HII (YKS10). 
The central galaxy, in contrast, shows
stronger rest energy emission-lines at 0.85 keV and 1.7 keV. 
%The SDSS class is composite with $D_{\rm agn}=0.1$ (YSK10). 
Resolved hard X-ray emission is detected from all the three
galaxies. The X-ray colour varies gradually from the northern to
southern galaxies, with {\sl HR}(N:C:S) $= -0.64, -0.56, -0.42$.

\smallskip

\noindent {\bf IRAS 07251--0248 :}\ 
This galaxy is the highest redshift object in our sample.
%and located close to the Galactic plane ($b = +6.7$) with high Galactic extinction. 
The X-ray source is very faint, detected only in the soft band.

\smallskip

\noindent {\bf ESO 60-IG16 :}\ 
A compact X-ray source is located at the nucleus of the NE galaxy of
this galaxy pair, as is the mid/far-IR emission seen in
the MIPS image (Mazzarella et al. 2011). The soft X-ray image shows a faint tail 
to the SW of length $7^{\prime\prime}$, while the hard X-ray emission is point-like. The X-ray
spectrum is typical of Seyfert 2 galaxies with an excess in
the hard band due to transmitted emission from an AGN 
absorbed by \nH $\sim 1\times 10^{23}$ \psqcm. The
absorption-corrected 2-10 keV luminosity, $10^{42}$ \ergps, is in the
range of Seyfert nuclei. The weakness of the soft X-ray emission is
partly due to the large Galactic extinction (see Table 2).

\smallskip

\noindent {\bf IRAS F08572+3915 :}\ The northwest nucleus of this
double system is suspected to have an heavily obscured AGN, as the
Spitzer IRS spectrum shows a very deep silicate absorption and little
PAH emission (e.g., Spoon et al. 2007). The northwest nucleus is
detected only in the hard band, (2-4 keV), with a small number of
counts ($\sim 10$). There is a faint, soft X-ray source near the
southeast nucleus but its physical association is unclear. The hard
X-ray colour (${\sl HR} = +0.57$) of the northwest nucleus strongly
suggests the presence of an obscured AGN. However, the very small hard
X-ray to IR luminosity ratio, log $(HX/IR) = -4.4$, points to a
Compton-thick AGN, although the limited X-ray counts do not allow us
to confirm it with the Fe K line detection.
% similar to NGC 4418? Maiolino et al
The Chandra result is also reported in Teng et al. (2009). 
%The optical class is LINER for both northern and southern galaxies
%(V95). The SDSS class is composite with $D_{\rm agn}=0.3$ for
%the northern galaxy and 0.2 for the southern galaxy.

\smallskip

\noindent {\bf IRAS 09022--3615 :}\ 
%Another galaxy located at low Galactic latitude ($b = 7.0$). 
The X-ray spectrum is flat ($HR = -0.17$) and the 2-10 keV luminosity
is $10^{42}$ \ergps, which qualifies this galaxy as an AGN.  However, the hard
X-ray source is marginally resolved. The [NeV] line is not
detected (Farrah et al. 2007).

\smallskip

\noindent {\bf IRAS F09111--1007 :}\  
Two galaxies with projected separation of $40^{\prime\prime}$ are both MIPS
sources; however, the western galaxy, which is less luminous in
the optical, is a stronger 24~$\mu$m source.  HIRES processing of the
IRAS image (Surace et al. 2004) suggests that the two galaxies each make 
comparable contributions to the IRAS flux, contrary to some previous
suggestions (Murphy et al. 1996, Duc, Mirabel \& Maza 1997, Goncalves,
Veron-Cetty \& Veron 1999).   In the X-ray, the eastern galaxy is brighter, 
but the western galaxy has a harder spectrum although the HR does not
meet our criterion for AGN selection. The spectrum of the eastern
galaxy shows strong Si XIII (see Table 9).

\smallskip

\noindent {\bf UGC 4881 :}\ This system is a pair of galaxies separated by
$11^{\prime\prime}$.  The eastern component is brighter in the IR and
radio than the western component. Howell et al. (2010) estimated the
IRAS luminosity ratio of the E-W pair is $\sim$2:1, based on the 
Spitzer-MIPS data. 
%The optical class (V95) is HII for the SW galaxy and LINER
%for the NE galaxy. The SDSS class is composite for both with $D_{\rm
%  agn}=0.3$ and 0.2, respectively. 
SN1999 GW, reported at the position
(RA, Dec) = (09h15m54.7s, +44d19m55s)$_{\rm J2000}$, is located near
the western nucleus. Three X-ray counts are detected at the position
of the supernova in our 15 ks exposure, but the association of the
X-rays with the SN is not clear given the limited counts.

X-ray emission is detected both from the eastern and western
nuclei. The eastern componet is brighter and relatively peaky at the
nucleus while the X-ray source for the western galaxy is diffuse and
has a rather flat brightness distribution.  The X-ray spectrum is soft
with $HR = -0.73, -0.88$.

\smallskip

\noindent {\bf UGC 5101 :}\ 
%z= 0.0394
The hard X-ray spectrum ($HR = -0.29\pm 0.05$) and the detection of a
cold Fe~K line with XMM-Newton demostrate that this galaxy contains an
obscured AGN (Imanishi et al. 2003). A high-ionization Fe~K line might
be present, but its detection is below $2 \sigma$ in the Chandra
data. Strong Si XIII (Table 9) as well as S XIV needs a large
enhancement in abundance of these metals in order to be explained by a
thermal emission spectrum alone. A significant contribution of
photoionized gas from an AGN would better explain these spectral
features. The same Chandra data have been published in Ptak et al.
(2003) and Grimes et al. (2005).

%Grimes 05 Ptak 03 

\smallskip

\noindent {\bf ESO 374-IG32 :}\  This galaxy pair hosts an OH megamaser (Baan \&
Kl\"ockner 2006). 
%The SDSS class is of the composite type with $D_{\rm agn}=0.3$ (YKS10). 
The X-ray source is compact but
resolved. The soft X-ray emission is elongated in the NE-SW direction
while the hard X-ray source has a faint spur to the south.

\smallskip

\noindent {\bf IRAS F10173+0828 :}\  This edge-on galaxy hosts a mega-maser source  
(Mirabel \& Sanders 1987), and is expected to have an heavily obscured
source that powers the large IR luminosity (Goldader et al. 1997). A very faint X-ray source (10 cts) is found at the nucleus
only in the soft band.

\smallskip

\noindent {\bf IRAS F10565+2448 :}\ 
The western galaxy is the IRAS source as it is the only source in the
MIPS 24 $\mu$m image (Mazzarella et al. 2011), and the X-ray emission also comes only from the
western galaxy. 
%The SDSS class is HII. 
Much of the X-ray emission arises from the nucleus as
indicated by a peaky X-ray distribution. The hard X-ray emission is point-like
and the soft X-ray emission has an extension up to $7^{\prime\prime}$ ($\sim r_{\rm
  max}$). There is a weaker point-like X-ray source $12^{\prime\prime}$ to the
south (10h59m18.48s, 24d32m23.3s, J2000), for which the counterpart is
not visible in the HST-ACS I-band image, but it is possibly detected in the Spitzer-IRAC
bands. It is probably a background AGN.

\smallskip

\noindent {\bf NGC 3690 (= Arp 299) : }\
This nearby pair of galaxies are often referred to as Arp 299. Although
the easten galaxy is referred to as NGC 3690 and the westen galaxy as IC
694 by several authors, here we follow the RBGS and NED conventions and refer to the 
two galaxies as NGC 3690 East (E) and NGC 3690 West (W). Both galaxies
are strong infrared sources. The IRAS flux distribution between the two
galaxies estimated by Surace et al. (2004) using the HIRES
algorithm is E:W = 3:1, which is different from that assumed in
Iwasawa et al. (2009) based on the 38 $\mu$m result by Charmandaris,
Stacy \& Gull (2002). The X-ray to IR luminosity ratios for the
eastern and western components are derived using the 3:1 IRAS image
ratio.

A detailed study based on a 25 ks Chandra ACIS-I observation (ObSID
1641) has been published by Zezas, Ward \& Murray (2003).  In this
paper, we used the 10~ks ACIS-S observation. As mentioned in Zezas et
al (2003), there are many discrete sources distributed over the galaxy pair,
as well as surrounding diffuse emission. With the overlay on the
high-resolution HST-ACS image, association of the discrete X-ray
sources with optical counterparts becomes clear in many cases.

Strong Fe~K line emission was first detected with BeppoSAX (Della
Ceca et al. 2002). A subsequent XMM-Newton observation resolved the
two nuclei and found that the spectra of both nuclei show Fe lines but
at different energies -- at 6.4 keV in NGC3690 W and at 6.7 keV in NGC
3690 E (Ballo et al. 2004). The Chandra data alone provide detection of
these lines at the $2\sigma $ level. From the hard band image, the 6.4
keV line should originate from the B1 knot of the eastern galaxy and
serves as strong evidence for this region to contain a Compton-thick
AGN. We do not consider, contrary to Ballo et al. (2004), that the 6.7
keV immediately implies an AGN (see also Neff et al. 2004 for radio
data), but we do assume that it should have the same origin as that in
Arp 220. There are two hard X-ray knots in the eastern galaxy, for
which eastern knot has a more absorbed spectrum than the western knot.

The surface brightness profiles are measured separately for the
eastern and western galaxies. Because of the proximity of the two
galaxies, the measurements are limited within $13^{\prime\prime}$
radius. The soft band profile does not reach the 1\% level to define
$r_{\rm max}$, so we do not give the compactness in Table 7.

\smallskip

\noindent {\bf IRAS F12112+0305 :}\ 
% z = 0.0733
Two X-ray sources aligned in the NE-SW direction with a separation of
$\sim 2.5^{\prime\prime}$ are resolved in the Chandra image. They are
coincident with the positions of the two optical nuclei of this double system. Both nuclei
are hard X-ray sources, with the SW nucleus being slightly brighter
and harder in X-ray colour. The quality of the spectrum from this
short exposure observation is poor. A faint soft X-ray arm, similar to
that seen in Mrk 266 (Mazzarella et al. 2010 and references therein),
emanates from the NE nucleus. The same Chandra data are reported in
Teng et al. (2005) and the XMM-Newton results are reported in
Franceschini et al. (2003) where this object is considered to be
starburst dominated.

%XMM Fra03 SB
% teng05

\smallskip

\noindent {\bf UGC 8058 (= Mrk 231):}\ Mrk 231 is the only optically
classified Seyfert 1 galaxy in our C-GOALS sample and it is also a
low-BAL object. The infrared SED is considered to be powered primarily
by the AGN (e.g. van der Werf et al. 2010), but note that mid-IR [NeV] is not detected (Farrah et
al 2007). There are four separate Chandra-ACIS exposures giving a
total exposure time of 160~ks, making this the longest exposure time
by far among the observations of our sample. Results on part or all of
these data have been reported in Ptak et al. (2003), Grimes et al.
(2004), and Gallagher et al. (2002, 2005).

Almost all of the emission above 2 keV comes from the unresolved
nucleus. Despite the Seyfert 1 optical classification, the X-ray spectrum
does not at all resemble that which is typically observed in Seyfert 1
galaxies, i.e., a steep ($\Gamma\sim 2$) power-law with or without
weak absorption by partially ionized gas (the ``warm'' absorber). The
Mrk 231 spectrum above 3~keV is flat ($\Gamma \simeq 0.7$). A Fe~K feature is
visible at 6.4 keV ($EW \sim 100$ eV), in the 160~ks exposure spectrum, 
similar to that previously seen in the XMM-Newton data
(Franceschini et al. 2003; Turner \& Kraemer 2003; Braito et al.
2004).  Given the detection of an hard X-ray excess with the BeppoSAX PDS
(Braito et al. 2004), the nucleus is considered to host a
heavily obscured AGN with an absorbing column of $N_{\rm
  H}\sim 2\times 10^{24}$ \psqcm.  The flat 3-7 keV continuum
is attributed to the reflected AGN continuum. The constant flux measured
in the four separate observations (Gallagher et al. 2005) is compatible with
this interpretation. However, the observed EW of the Fe~K line is much
smaller than the expected nominal value.

The soft X-ray emission also peaks strongly at the nucleus. This
causes $r_{\rm max}$ to be only $5^{\prime\prime}$, because of how 
this parameter is defined (the radius where the brightness falls to 1\% of the
peak brightness).

A large, low surface brightness extension of more than
$30^{\prime\prime}$ ($= 25$ kpc) down to the $2\times 10^{-17}$ erg
s$^{-1}$ cm$^{-2}$ arcsec$^{-2}$ level is detected in the deep X-ray 
image. Much of the extended emission beyond $5^{\prime\prime}$ from
the nucleus is emitted below 1~keV.  Between 1~keV and 2~keV, the
emission is due predominantly to emission lines of Mg XI and Si XIII
(see Fig. 13 in \S~5.1.4).

The deep X-ray exposure reveals a horse-shoe structure $15^{\prime\prime}$
south of the nucleus (indicated by an arrow in Fig. 2), where a faint
optical counterpart is visible in the HST image. Note this is not the
UV horse-shoe structure with star forming knots, located closer ($\sim
4^{\prime\prime}$) to the nuclues, imaged in the optical/UV
(Mu\~noz-Marin et al. 2007).

With the long exposure time, a reasonably good quality spectrum of the
extended, low surface-brightness emission was obtained. The
temperature implied from the fit to this very extended low surface
brightness emission ($kT \simeq 0.5$ keV) is lower than the spatially
averaged values for other galaxies and little emission is detected
above 2 keV. The spectrum fitted with the model above is also shown
for comparison (Fig. A.1). The diameter of the nebula is $\sim 50$kpc.

% Mrk231 nebula soft X spectrum
\begin{figure}
\begin{center}
  \centerline{\includegraphics[width=0.4\textwidth,angle=0]{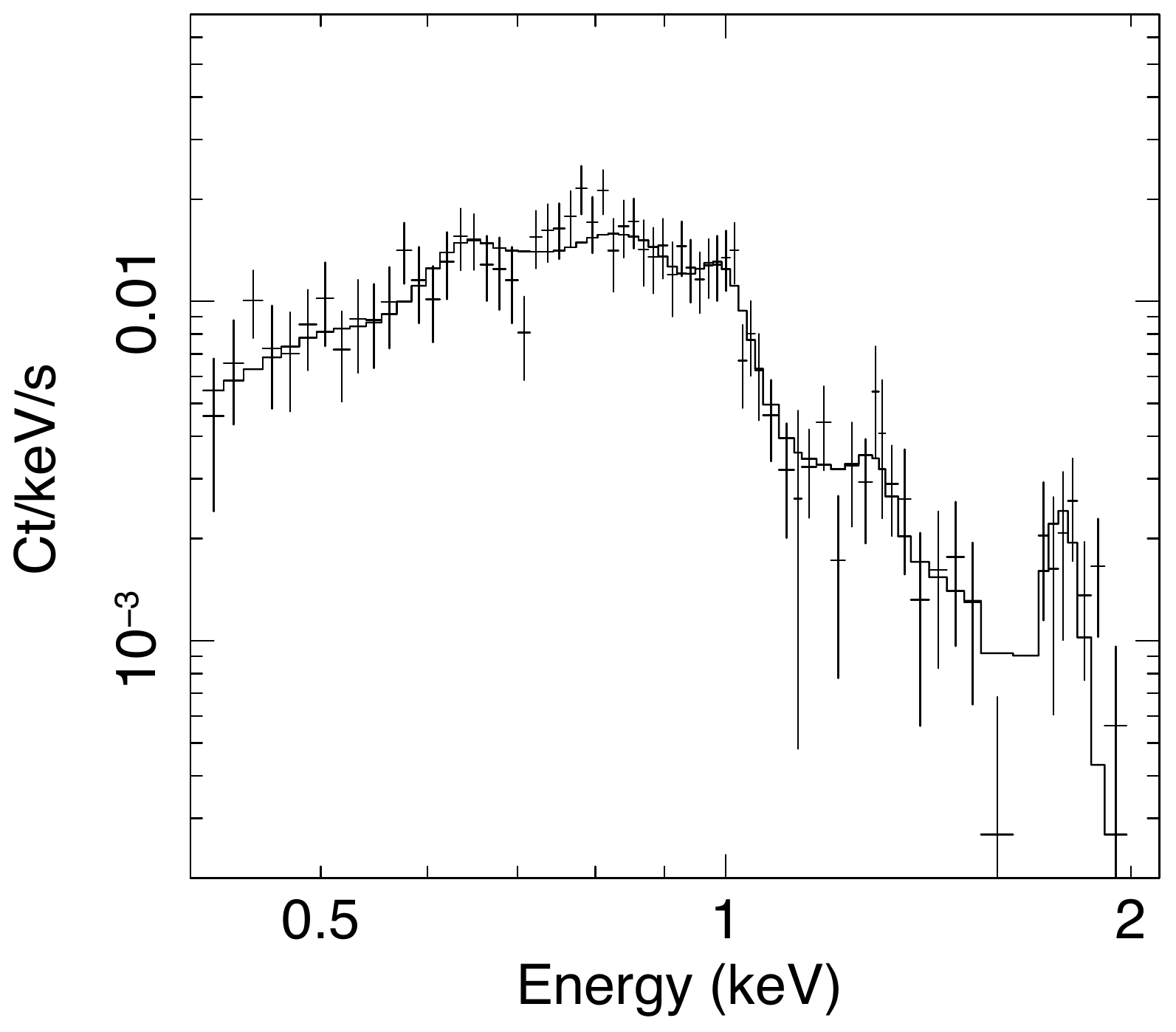}}
  \caption{The Chandra ACIS spectrum of the extended, low surface
    brightness emission around Mrk 231, fitted with a thermal
    emission model.}
\end{center}
\end{figure}

\smallskip

\noindent {\bf IRAS 13120--5453 :}\ 
%Another galaxy located at low Galactic latitude ($b = 7.6$) with
%a large Galactic extinction. 
The X-ray source is compact both in the soft and hard bands. The
radial brightness profile is slightly broader than the PSF but the
extension is limited below $1^{\prime\prime}$. The hard X-ray colour
(${\sl HR} = -0.13$) classifies this object as an AGN. The large
Galactic extinction (Table 2) is part of the reason for the hard X-ray
colour.  The small hard X-ray to IR ratio, log (HX/IR) $< -4$, places
this object as a Compton thick AGN candidate (Table 6).
%The [NeV] line is detected (F07).
%The 4-panel image is in logarithmic scale. 

\smallskip

\noindent {\bf VV 250 :}\  This is a widely separated galaxy pair with a clear
tidal structure conneting the two galaxies. The eastern galaxy
is brighter in X-rays than the western galaxy, with a count ratio of
$\sim $10:1. The  IRAS HIRES estimate gives a   
flux ratio of E:W = 0.85:0.15 (Surace et al. 2004). 
%The optical class in V95 is HII or LINER for the western galaxy and HII for the eastern galaxy (V95). The SDSS class is composite for both nuclei with $D_{\rm
%  agn}=0.4$ and 0.3, respectively (YKS10). 
The hard X-ray core is marginally
resolved. A broad Fe~L bump is seen below 1~keV in the spectrum, but
the other emission lines, e.g.,  Si XIII, do not stand out. The X-ray
emission associated with the Western galaxy is patchy and distributed
roughly towards the direction of the Eastern galaxy.

%The 4-panel X-ray image for the Eastern galaxy is in logarithmic scale.

\smallskip

\noindent {\bf UGC 8387 (= IC 883) : }\  The soft X-ray emission is elongated along
the minor axis of the galaxy's main body, with faint filaments forming
a loop on the NW side. There is a relatively sharp boundary in the
soft X-ray emission on the SE side. A compact, but resolved hard X-ray
source is located at the nuclear position, which produces a faint tail
in the energy spectrum above 4 keV. There is a compact soft X-ray
source about $9^{\prime\prime}$ to the SE of the galaxy nucleus, which has an
optical counterpart in the HST-ACS image, most likely a star
cluster. The total counts detected from this object is $15.8\pm 4.0$
cts in the 0.4--7 keV band. The X-ray color, $HR = -0.64\pm 0.30$,
suggests that the X-ray spectrum is softer than the source associated
with the galaxy main body. The details of this galaxy together with
the other GOALS datasets are discussed in a separate paper (Modica et
al 2010). Very faint [NeV] is detected in the Spitzer IRS high
resolution spectrum, which indicates that a weak AGN might be present
in this galaxy. 
%The optical class of V95 is LINER and the SDSS class
%is of a composite with $D_{\rm agn}=0.5$ (YKS10). 
The X-ray spectrum of this object is one of
those exhibiting enhanced Si XIII emission (Table 9).

\smallskip

\noindent {\bf UGC 8696 (= Mrk 273) :}\ 
 %z = 0.0378
A detailed analysis of the Chandra data for this ULIRG, which has an
absorbed AGN, has been published in Xia et al. (2002).  Analyses of
the same data set are also given in Ptak et al. (2003), Satyapal et
al.  (2004), Grimes et al. (2005), and Gonz\'alez-Mart\'in et
al. (2006). The XMM-Newton results have been reported by Balestra et
al. (2005). The absorbing column density towards the X-ray source is
\nH $\sim 4\times 10^{23}$ \psqcm, and the absorption corrected 2-10
keV luminosity is $0.9 \times 10^{43}$ \ergps. A cold Fe K line at 6.4
keV is detected with ${\sl EW}\sim 0.24$ keV. Our analysis of the hard
X-ray emission seems to indicate that the absorbed AGN may be located
at the southern nucleus. The soft X-ray tail, which emits little above
2 keV and has luminosity of $5.5\times 10^{40}$ \ergps, extends to the
south. This X-ray emission is, however, not associated with the tidal
tail imaged in the optical, but rather is located behind it. The
details concerning the nuclear position, the nature of the double
nuclei, and the soft X-ray tail can be found in Iwasawa et al. (2011).

\smallskip

\noindent {\bf IRAS F14348--1447 :}\ 
%z=0.0827
This is a double system with two nuclei separated by $\sim
4^{\prime\prime}$.  In the IRAC bands, the southern nucleus is
brighter (Mazarella et al. 2010b), while the MIPS image is
unresolved. We assumed the northern and southern nuclei to have IR
flux ratio of N:S = 1:3. The southern nucleus is also a brighter X-ray
source and shows point-like hard X-ray emission. While the HR value
for the total emission is $-0.48$, when selecting only the southern
nuclear region, $HR = -0.18\pm 0.19$, indicating the possible presence
of an AGN, which is possibly Compton thick, based on the small HX/IR
ratio (Table 6). Imanishi et al. (2007) present Spitzer IRS spectra
and also suggest the presence of an AGN based on the L-band
spectrum. The northern nucleus is enveloped by soft X-ray
emission. Based on the XMM-Newton data, Franceschini et al. (2003)
classified this galaxy as a starburst dominated source.

%XMM Fra03 SB

\smallskip

\noindent {\bf IRAS F14378--3651 :}\ A faint, point-like source is
detected at the nucleus of this galaxy in the hard band while there is
an additional fainter blob (or an extension) at $3^{\prime\prime}$ to
the west, which is also detected in the soft X-ray band. The hard
X-ray colour ($HR = -0.18$) classifies this object as an AGN with a
relatively low 2-10 keV luminosity of $\sim 3\times 10^{41}$ \ergps,
which agrees with the Seyfert 2 classification derived from the
optical spectroscopy (Kim et al. 1998). The small HX/IR ratio means
this object is a Compton thick AGN candidate (Table 6). There is a
faint structure in the HST-ACS image, which might be a star cluster
within the NW extended structure noted by Bushouse et al. (2002).
%No [Ne V] is detected in the IRS spectrum (F07).

\smallskip

\noindent {\bf VV 340 :}\  This galaxy pair consists of an edge-on galaxy to the
north and a face-on galaxy to the south. 
%The optical class for the norhern galaxy is LINER/HII (V95), while the SDSS class is composite with $D_{\rm agn}=0.5$ (YKS10). 
The IRAS HIRES estimate yields a flux ratio N:S = 0.85:0.15 (Surace et
al 2004). The northern galaxy is brighter in the 24~$\mu$m image and
the detection of [NeV] in the IRS spectrum plus the Chandra spectrum
with a possible 6.4~keV line (see below) suggest the presence of a
heavily obscured AGN [see Armus et al. (2009) for the GOALS
multi-wavelength datasets on this object]. The northern galaxy is
brighter in X-rays than the southern galaxy. The soft X-ray emission
of the northern galaxy shows a complex morphology and has a flat
brightness profile. It extends along the minor axis of the edge-on
galaxy, suggesting a relation with a nuclear outflow.
%(fitting the $\beta$ model yields the core radius of 7 arcsec with
%$\beta = 1.1$). 
The soft X-ray peak is displaced by $2^{\prime\prime}$ to NE from the nuclear
position, where the hard X-ray emission peaks.  
This is reflected in the surface brightness profile, with the
innermost bin being depressed. Absorption at the nucleus is a
probable cause.

While the hardness ratio, $HR = -0.74\pm 0.08$, indicates a soft
spectrum, there is a small bump above 4 keV which peaks at the
rest-energy of 6.4 keV. The quality of the data is insufficient to
warrant a significant detection of the Fe K line at 6.4 keV ($\sim
2\sigma$), but the spectral shape is strongly suggetive of a heavily
obscured, possibly Compton-thick AGN. The absorbing column density has
to be close to $10^{24}$ \psqcm. The hard X-ray source is resolved
into a nuclear source and small blobs located both North and South
up to $7^{\prime\prime}$ away, which are in alignment with the radio morphology
(Condon et al. 1990). A strong Si XIII ($EW \sim 0.3$ keV) at 1.87 keV is noted in
the soft band spectrum, which could be partly due to extended gas
photoionied by the AGN in addition to thermal emission from a
starburst. The Chandra data has been published as a part of the water
maser sample of Kondratko et al. (2006).

The X-ray emission associated with the southern galaxy has no
well-defined peak and shows a patchy brightness distribution. An
inspection of the X-ray and optical images suggests that the X-ray
emission in this galaxy may be composed of multiple X-ray sources
associated with individual star clusters or giant HII regions within
this galaxy.

\smallskip

\noindent {\bf VV 705 (= I Zw 107) :}\ 
%z = 0.0402
This system contains a close pair of galaxies with clear tidal
tails. The northern galaxy dominates the IRAS flux. While X-ray
emission is detected from both galaxies, the northern source is also
much brighter in X-rays (N:S = 8:1).  
A bright core is found at the northern nucleus with a faint arc to the NE that has a flat
brightness distribution. There are two faint sources to the NW at (RA,
Dec) = (15h18m05.22s, +42d44m 58.23s)$_{\rm J2000}$ and to the W (RA, Dec)
= (15h18m05.05s, 42d44m47.53s)$_{\rm J2000}$, which are probably
background sources.

\smallskip

\noindent {\bf IRAS F15250+3608 :}\ 
% z = 0.0552
With only a 10 ks exposure, a weak soft X-ray source is detected. No
emission is detected above 2 keV, which implies an extremely hard X-ray
limit compared to the IR luminosity, log $(L_{\rm HX}/L_{\rm IR})< -5$. The
Chandra result has been reported in Teng et al. (2005) while the
XMM-Newton results are published in Franceschini et al. (2003). 

% teng05
%xmm fra03 sb

\smallskip

\noindent {\bf UGC 9913 (= Arp 220) :}\ 
%z = 0.018
This object, the nearest ULIRG, has been studied extensively at all
wavelengths. The hard X-ray emission is elongated across the two
nuclei with a separation of $1^{\prime\prime}$, and is peaked at the western
nucleus (Clements et al. 2002; Iwasawa et al. 2005). The strong, Fe XXV
emission, detected in XMM-Newton (Iwasawa et al. 2005) and Suzaku
(Teng et al. 2009) spectra, should come from this region. The line
detection in the Chandra data is barely $2 \sigma $ (Clements et al.
2002, and Table 10). The region around the western nucleus within $2.7^{\prime\prime}$ 
shows a hard X-ray colour, $HR = -0.16$, which is in the range
observed for AGN. However, absorption, rather than an intrinsicaly hard spectrum of
an AGN, plays an important role in making the spectrum appear hard.

The soft X-ray peak is displaced to the NW from the western nucleus where the
hard X-ray emission is peaked. This is primarily due to absorption in
the nuclear region, which leads to the dip at the innermost bin of the
soft X-ray radial profile. The abrupt drop of the surface brightness
at $3-4^{\prime\prime}$ NW from the nucleus is also due to absorption which is
caused by the dust lane running across the nucleus. Towards the SE, a relatively
flat soft X-ray tail is seen up to $10^{\prime\prime}$, where the brightness
drops sharply.

The low surface brightness emission extending beyond the inner bright
part emits little emission above 2 keV. The soft X-ray nebula's extension is up
to $\sim 45^{\prime\prime}$ arcsec in the E-W direction. Both ends seem to show
a looped morphology with a well-defined edge. The same data have appeared in
Ptak et al. (2003), McDowell et al. (2003) and Grimes et al. (2005).

\smallskip

\noindent {\bf ESO 69-IG6 :}\ 
This system is a widely separated ($\sim 60^{\prime\prime}$) pair, and the
northern galaxy likely accounts for almost all the IRAS flux, as it
is the only source in the MIPS image (Mazzarella et al. 2011). The X-ray source properties
described in this paper are only for  this northern galaxy. The northern
galaxy shows an intriguing X-ray morphology. The soft X-ray image shows a
double morphology with diffuse emission, while the hard band shows three
compact sources aligned linearly. One of the hard X-ray sources is
located inbetween the soft X-ray double peaks. This suggests that the
soft X-ray depression between the brightness peaks may be due to
absorption. This complex X-ray morphology has no correspondence in the
images in the other bands (UV/optical/near-IR). The optical nucleus
imaged by the HST-ACS is located at the northern blob seen in the soft
X-ray band, and this is the apex of the X-ray radial profiles. The
X-ray spectrum is soft ($HR = -0.67$) and shows strong Si XIII (Table 9).

The southern galaxy hosts a point-like, strongly absorbed X-ray source
with \nH~$\sim 4\times 10^{22}$\psqcm. The absorption-corrected 2-10
keV luminosity is estimated to be $1.5\times 10^{42}$\ergps. It is
likely a moderately luminous Seyfert 2 nucleus. 

\smallskip

\noindent {\bf NGC 6240 :}\ 
% z= 0.025
This nearby LIRG has often been considered to be starburst dominated. The
presence of an AGN was suggested by the detection of a strong Fe K
complex in the ASCA spectrum (Iwasawa \& Comastri 1998), followed by
the detection of an hard X-ray excess with BeppoSAX (Vignati et al.
1999) and RXTE (Ikebe et al. 2000), indicating a strongly absorbed (\nH
$\sim 2\times 10^{24}$ \psqcm) AGN.  Additional evidence for an AGN was 
also obtained from infrared spectroscopy (Armus et al. 2006; Risaliti et al.
2006a).  Chandra observations resolved the X-ray source into two
nuclei and revealed that both nuclei host Compton-thick AGN through
the detection of reflection-dominated spectra (Komossa et al.
2003). The same Chandra data are included in Ptak et al. (2003) and
Grimes et al. (2005). A detailed analysis of the soft X-ray emission
spectrum, using the X-ray grating spectrometer data, is presented by
Netzer et al. (2005).

The low surface brightness emission in soft X-rays extends to large
radii corresponding to a projected diameter of $\sim 70-80$ kpc. This
soft X-ray nebula has a similar morphology to that of  other galactic wind
signatures like H$\alpha $ (Heckman et al. 1996; Veilleux et al.
2003). The hard X-ray emission is concentrated at the two nuclei with
a projected separation of $1.6^{\prime\prime}$, which are barely
separated at the Chandra resolution. The southern nucleus is the 
brighter hard X-ray source. The X-ray spectrum presented in
this paper is taken from the brighter, butterfly-shaped, inner part.

%Grimes 05 Ptak 03

\smallskip

\noindent {\bf IRAS F17132+5313 :}\  Two merging galaxies aligned in the E-W
direction make up this system, and both contribute to the IRAS
flux. 
%The eastern galaxy is classified as HII (V95) and with $D_{\rm
%  agn}=0$ (YKS10). The western galaxy is unclassified. 
In X-rays, the eastern galaxy is the main X-ray source and a faint
extension is seen towards the western galaxy both in the soft and hard
X-ray bands. The Chandra spectrum shows strong emission lines of Ne~IX
(0.9 keV) and Mg~XI (1.3 keV). The Si~XIII spectral region is noisy.
No AGN signatures are detected in optical (Veilleux et al 1995; Yun et
al 2010) or in X-ray, while the mid-IR [Ne V] is detected in the Spitzer
IRS spectrum (Petric et al 2010).

% Image smoothing kernel is big (3pix)

\smallskip

\noindent {\bf IRAS F17207--0014 :}\ 
% z = 0.0428
This galaxy has been observed with Chandra twice, one snapshot with 9
ks (ObsID 4114) and another longer (49 ks) exposure (ObsID
2035). Here, we used the long exposure data. The same data
have been published in Ptak et al. (2003), Grimes et al. (2005) and Teng
et al. (2005). 

The X-ray source has two peaks separated by $\sim 2^{\prime\prime}$. The
southern peak is due to hard X-ray emission and coincides with the NED
nuclear position. The northern peak is mainly due to soft X-ray
emission. The weak soft X-ray emission at the nuclear position means
that an excess of absorption is responsible for the suppression. The
dip of the soft X-ray surface brightness profile at the innermost $1^{\prime\prime}$, 
similar to Arp 220, is likely an effect of absorption.

The optical and infrared diagnostics (Veilleux et al. 1999; Lutz, Veilleux,
Genzel 1999; Risaliti et al. 2006b), the XMM-Newton observation (Franceschini
et al. 2003) and previous Chandra publications all point to
a starburst classification. Our analysis shows the possible presence of
strong, high-ionization Fe~K (Table 10) on a hard continuum (Table 8),
which is, again, similar to Arp 220.

% Teng 0
%XMM Fra03 SB

\smallskip

\noindent {\bf IRAS F18293--3413 :}\ 
%This galaxy is located at a relatively low Galactic latitude ($b =
%-11.3$) and has been less studied. 
X-ray emission is detected only at the brighter face-on galaxy in this
pair system. The observed X-ray flux is consistent with the BeppoSAX
value reported in Risaliti et al. (2000). There is a possible 2-7 keV
excess at the position of SN 2004ip (Perez-Torres et al. 2007, RA =
18h32m41.2s, Dec -34d11m26.8s). However, the position is too close to
the nuclear source ($\sim 1^{\prime\prime}$ away) to be resolved,
given the small detected counts (6 cts). The X-ray source in this
galaxy is one of the brightest sources among our  26 Cycle-8 targets. The
X-ray emission is resolved both in the soft and hard bands (up to
$12^{\prime\prime}$ and $7^{\prime\prime}$, respectively). The soft
X-ray emission extends towards the south beyond the main body of the
galaxy, suggesting a possible outflow in this direction. The X-ray
spectrum shows a strong Si XIII line as well as Mg XI and Fe~L blends
around 0.9 keV.

\smallskip

\noindent {\bf ESO 593-IG8 :}\ 
A cross-shaped double galaxy system aligned perpendicularly. Dudley
(1999) has classified this source as a ``PAH galaxy" based on near-IR
spectroscopy. The mid-IR spectroscopy detected [Ne V] (Petric et al 2010).
% while the optical classification is LINER (V95). The SDSS class is
%composite $D_{\rm agn} = 0.2$ for the northern galaxy and
%``ambiguous'' ($D_{\rm agn}=0.4$) for the souther galaxy. 
The X-ray emission peaks near the intersection of the two
galaxies and shows a curious morphology. The bright core is elongated
in the NE-SW direction and soft X-ray emission extends primarily 
towards the south. The hard X-ray source has a narrow $\sim 10^{\prime\prime}$ 
 tail extending to the south. There is a sharp peak, $r< 1^{\prime\prime}$, 
in the soft X-ray band. 
%The hard X-ray profile isinstead in a power-law form with slope $-1.8\pm 0.4$
The spectrum shows distinct emission-line peaks at 0.95, 1.8, and
possibly at 0.6 keV. Possible identifications for these lines are
Ne~IX, Si~XIII and O~VIII, respectively. If the presence of O~VIII is
real, this soft X-ray spectrum would be difficult to account for by a
keV-temperature thermal spectrum. The Si~XIII is strong (Table 9).

\smallskip

\noindent {\bf IRAS F19297--0406 :}\ 
%This galaxy is also at low Galactic latitude ($b = -10.9$) and less
%well-studied. 
This is the third most luminous object in the C-GOALS sample but the
observed X-ray emission is faint. While the soft X-ray emission shows
a slight extension, the hard X-ray emission is point-like.

\smallskip 

\noindent {\bf IRAS 19542+1110 :}\ 
%Also low Galactic latitude ($b = -8.9$). 
This face-on galaxy has a compact X-ray source with a very hard
spectrum. The hard X-ray emission is point-like. The hard X-ray
spectrum is consistent with a transmitted AGN continuum absorbed by
\nH $\sim 5\times 10^{22}$ \psqcm. The absorption-corrected 2-10 keV
luminosity is estimated to be $6\times 10^{42}$ \ergps. No information
on the optical classification is available, most likely due to the low
Galactic lattitude $b=-8.9^{\circ}$.
%The 4-panel image is in logarithmic scale.

\smallskip

\noindent {\bf CGCG 448-020 (= II Zw 96) :}\ This V-shaped interacting
pair has been suggested to host obscured star forming sites. Four
principal near-IR peaks are identified by Zenner \& Lenzen (1993) and
denoted as A, B, C, and D (also see Fig. 2 in Goldader et al.
1997). The Spitzer-MIPS image has revealed that the more than 70 per cent of
the total 24 $\mu$m emission comes from D (Inami et al. 2010), which is
almost invisible in the B-band image. Source D is the closest to the
IRAS peak and is likely to dominate the IRAS flux. There are two compact
hard X-ray sources, one of which is slightly elongated encompassing
Sources C+D and the other is coincident with Source A.  Source C+D has a X-ray
colour, $HR = +0.05\pm 0.14$, which is in the range of AGN, but
like the nuclear region of Arp 220, absorption within these red knots
is likely the major reason for the hard X-ray colour. There is a spur
extending from Source A to the north, which is identified as the base
of a soft X-ray filament bending over to the NW. There is another soft
X-ray filament extending towards the NW.  Both filaments appear to
connect to a northern soft X-ray blob, which altogether form an
elongated loop. The location of the NW soft X-ray blob is displaced
from the NW galaxy, and the relationship between the X-ray source and
optical galaxy is not clear.

There are three faint sources to the east, aligned linearly and only
seen in soft X-rays (\S 5.2). Two of these sources have clear optical
counterparts in the HST-ACS image. They are luminous star
clusters or dwarf galaxies residing within the tidal tail of the
merger.

The radial surface brightness profiles were constructed with Source A
as the apex. The soft X-ray profile has a power-law form with a 
slope of $-1.5\pm 0.1$ out to $17^{\prime\prime}$. The hard X-ray profile has a
bump at $4-10^{\prime\prime}$ due to Source C+D. The X-ray colour map shows
that Source D is the hardest X-ray source, which is consistent with
an absorbed X-ray source associated with an obscured far-IR source. No
clear evidence for an obscured AGN can be found in the X-ray
spectrum. A clear line feature at 1.6 keV, which is, however not at the
energy of Si XIII, remains unidentified. There is a marginally
significant line feature at 4.1 keV, which also has no obvious
identification.

%>>>>
In terms of the $L_{\rm x}-IR$ correlation (Ranalli et al. 2003), a comparison between Source
A and Source C+D may be illuminating. While the hard X-ray luminosities 
of  Source C+D and Source A are comparable, the IR luminosity of
Source C+D is larger than Source A by factor of $\sim 2$ at 24 $\mu$m
and factor of $\sim 7$ at 70 $\mu$m (Inami et al. 2010), i.e., within this single object, 
it is obvious that the $L_{\rm x}-IR$ correlation 
does not hold for individual ``knots". The powerful,
obscured IR Source C+D is X-ray quiet (at a given IR luminosity),
compared with the less powerful, unobscured Source A. This is
reminiscent of the hard X-ray queiet nature of the HXQ galaxies relative
to the nearby star-forming galaxies (Iwasawa et al 2009). 
%>>>

%% Ptak et al and Grimes et al identical ULIRG sample (9)

\smallskip

\noindent {\bf ESO 286-IG19 :}\ 
%z = 0.0430
A bright elliptical soft X-ray core, elongated along the NE-SW
direction, is seen around the nucleus of the galaxy. A faint spur
towards the NW is extended up to $10^{\prime\prime}$. The hard X-ray
image shows a point like source with a resolved faint extension to the
NE up to $3^{\prime\prime}$.  The Chandra data have been published in
Ptak et al. (2003) and Grimes et al. (2005). Franceschini et
al. (2003) classified this object as an AGN based on the XMM-Newton
data. No AGN signatures were detected in the optical and mid-IR
spectra (Table 6).

There is a faint point-like source, visible both in the soft and hard
bands, at $7^{\prime\prime}$ to the SE from the nucleus. A brighter compact source
with a rather hard spectrum, located at $8.5^{\prime\prime}$ to the SSW from the
nucleus, is also found. An optical counterpart of this X-ray source is
found in the HST-ACS I-band image (see the Chandra overlay in Fig. 2), and
is probaly a background AGN.

The X-ray spectrum shows a clear hard X-ray tail with $\Gamma\sim
0.7$. While the X-ray colour of the total emission ({\sl HR}$=-0.77$)
does not qualify as AGN, this hard X-ray emission is the one identifed
as an absorbed AGN component in the XMM-Newton spectrum (Franceschini
et al 2003). An iron K line is not detected either in the Chandra or
XMM-Newton spectra. As shown in Fig. 12., the relatively good quality
soft X-ray spectrum suggests that the X-ray emitting gas is $\alpha $
element rich relative to iron and the metallicity pattern deviates
from solar. Strong Si~XIII is detected, suggesting a further
enhancement of Si metallicity (Table 11), similar to the same line
detected in Compton thick AGN (Sect. 5.1.4).

%Ptak 03 Grimes 05
%XMM (Fra 03) AGN

\noindent {\bf IRAS 21101+5810 :}\ 
%Also at low Galactic latitude at ($b = +6.9$) with large Galactic absorption. 
A faint, compact X-ray source is detected at the nuclear position. A
western extension is seen only in the soft band where no obvious
optical counterpart is visible in the ACS image.

\smallskip

\noindent {\bf ESO 239-IG2 :}\ 
A point-like source is detected at the nucleus in the hard band while
the soft band shows a sharp peak at the nucleus with
low-brightness emission extending to $10^{\prime\prime}$. The soft
X-ray extension is primarily in the N-S direction. The spectrum is softer than the
 sample average ($HR = -0.75$) with the possible presence of a weak hard
 X-ray tail.

\smallskip

\noindent {\bf IRAS F22491--1808 :}\  This is a well-known merger system with
spectacular tidal tails. Farrah et al. (2003) conclude that more than half
of the bolometric luminosity originates from an AGN, based on an SED
fit while optical data favour a starburst. The observed X-ray source
is faint and most of the counts are detected in the soft band, thus it has
a very soft spectrum. No X-ray signature of an AGN can be found in the
data. The XMM-Newton data have been presented in Franceschini et al.
(2003).

%xmm fra03 SB

\smallskip 

\noindent {\bf ESO 148-IG2 :}\ 
% z= 0.0446
Two galaxy nuclei are aligned in the N-S direction with a
$5^{\prime\prime}$ separation, and the optical image shows a classic
double tidal tail. The XMM-Newton data show that this object contains
an absorbed AGN based on the X-ray spectrum (Franceschini et al.
2003), with an estimated absorbing column density, $N_{\rm H}\sim
7\times 10^{22}$ \psqcm, although the broad PSF of XMM-Newton did not
identify the location of the AGN. Many pieces of evidence show that
the southern nucleus, which is a much brighter source in the Spitzer-MIPS 24$\mu$m 
image (we assumed IR flux ratio of 1:3 for N:S, based on the
mid-IR flux ratio estimated by Charmandaris et al. 2002), contains the
AGN. The HST-ACS image shows a point-like optical source at the
southern nucleus. Risaliti et al. (2006) classified this nucleus as an
AGN based on their analysis of the L-band spectrum.

Chandra results have been reported previously in Ptak et al. (2003)
and Grimes et al. (2005). The Chandra observations clearly demonstrate that
the southern nucleus is the hard X-ray source, and the location of the
absorbed AGN. When the intrinsic spectral slope is assumed to have
$\Gamma = 1.8$, the absorbing column density is estimated to be \nH =
$(5\pm 2) \times 10^{22}$ \psqcm. The absorption corrected 2-10 keV
luminosity is then $0.8\times 10^{42}$ \ergps.

%Grimes 05
% xmm fra SB/AGN

\smallskip

\noindent {\bf ESO 77-IG14 :}\  This galaxy pair has two similar disk galaxies 
separated by $\sim 17^{\prime\prime}$,  and two Spitzer-MIPS sources with comparable
fluxes are detected at the respective nuclei; thus both members are likely
to contribute to the IRAS flux. The optical spectral type listed in NED is 
HII. Two X-ray sources, also with comparable fluxes (1:0.6 
in the 0.4-7 keV count ratio for the NE and SW galaxies). The NE
source is harder in X-ray colour. Both X-ray sources are faint but
they are not point-like. There is another faint X-ray source ($7.8\pm
2.8$ cts in the 0.4-7 keV band) located further south, which has a very
faint optical counterpart in the ACS image.

% not enough count for spectral analysis

\smallskip

\noindent {\bf IRAS F23365+3604 :}\ 
%z= 0.0645   10ks Wilson  Teng et al
This galaxy is generally considered to be a heavily obscured object
(Burston et al 2001) and the optical class is LINER (Veilleux et al
1995). A faint X-ray source is detected at the nucleus. The quality of
the spectrum of the short exposure (10~ks) observation is poor, but
the hard X-ray colour, $HR = -0.22$, classifies this object as an AGN,
which could be Compton thick, based on the small HX/IR ratio (Table
6). The Chandra results are also reported in Teng et al. (2005).

\smallskip

\section{Image extension of the very soft X-ray band}

The 0.4-0.7 keV images of Arp 220 (UGC 9913), ESO 286-IG19, NGC 6240
and Mrk 231 (UGC 8058) are shown as a supporting material for 
footnote 4, which argues against a spectral model in which a power-law
component accounts for the very soft band (e.g., 0.4-0.7 keV) spectrum
as well as the hard band (3-7 keV) in a starburst galaxy. The four
objects were selected for their brightness in the 0.4-0.7 keV
band. The upper panels for Arp 220 and ESO 286-IG19 are objects in
which no clear AGN signature is found in the hard band, and thus
relevant to the above argument. The bottom two panels are for NGC 6240
and Mrk 231 in which an AGN component is found in their hard band
spectra. In all cases, the soft band (0.4-0.7 keV) emission is significantly
extended, while the hard band (3-7 keV) emission is much more compact,
as shown in contours. Furthermore, the soft band emission is
displaced from the 3-7 keV hard band emission apart from Mrk 231. It
should be noted that the unresolved nuclear component of Mrk 231 is
always dominant over the whole energy band. In Arp 220 and NGC 6240,
it is evident that the soft X-ray emission is suppressed in their
nuclear regions where obscuration is substantial and only the hard
band emission comes through. As shown in the above examples, the
distinct morphology between the 0.4-0.7 keV and 3-7 keV bands means
that the origins of the emission in the two bands are different and
cannot be represented by a single component in spectral modelling, even
if it can provide a good fit. 

\begin{figure}
\centerline{\includegraphics[width=0.48\textwidth,angle=0]{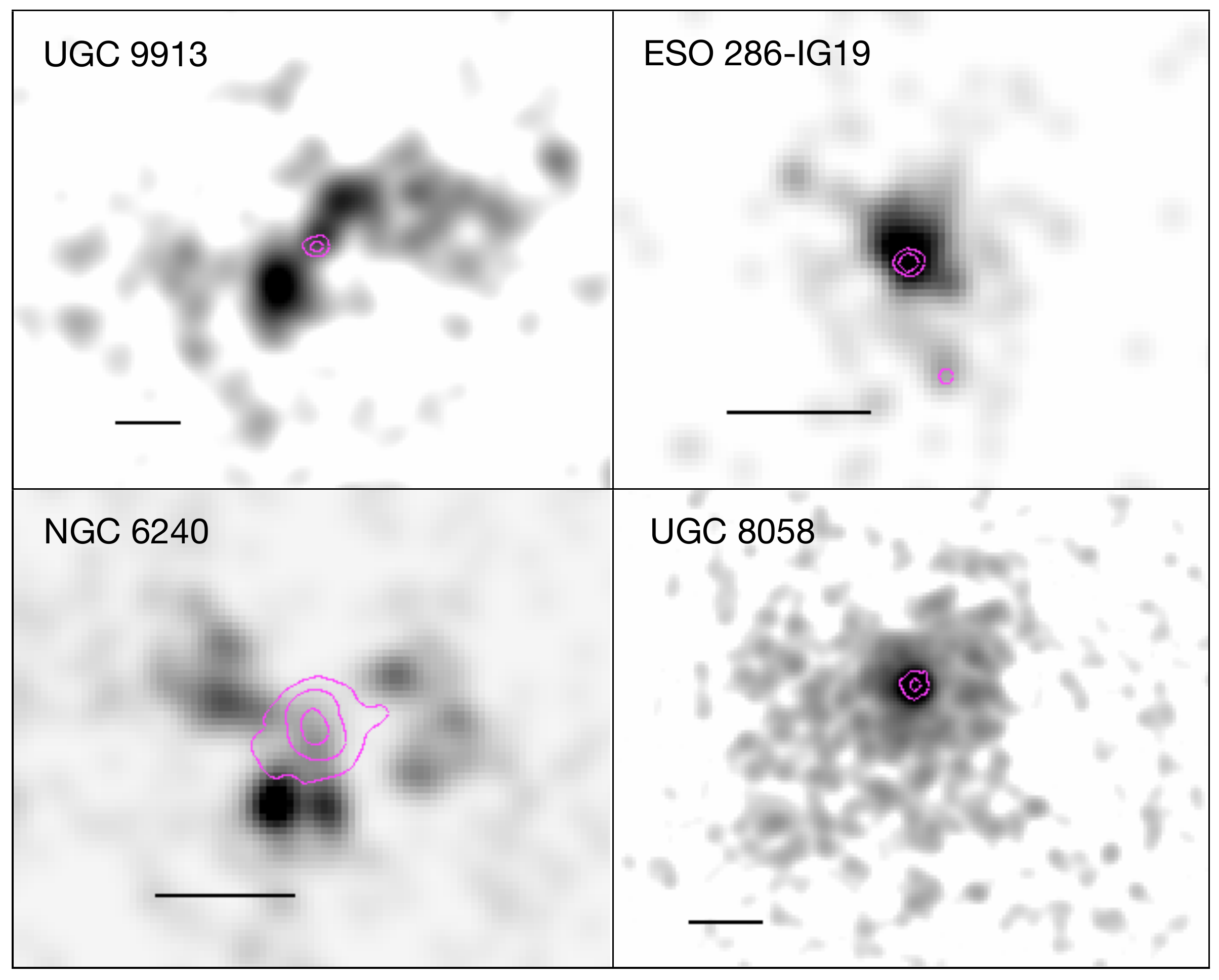}}
\caption{The soft band (0.4-0.7 keV), grey scale images of Arp 220, ESO 286-IG19, NGC 6240 and
  Mrk 231. The hard band (3-7 keV) source in each object is indicated by the purple 
  contours. The scale bar in each panel indicates 5 arcsec. }
\end{figure}

\section{X-ray images of the C-GOALS sample}

\newpage

All the image figures of the 44 individual objects, similar to Fig. 3,
are available in a separate file.

\end{document}